%% file: Article.tex
\documentclass[copyright,creativecommons]{eptcs}

\input{Format}

\input{Macros}

\input{Figures}

\input{Algorithms}

\hypersetup
  {
  pdftitle = {Robust Exponential Worst Cases for Divide-et-Impera Algorithms
    for Parity Games},
  pdfauthor = {M. Benerecetti, D. Dell'Erba, F. Mogavero}
  }

\begin{document}

  \title{Robust Exponential Worst Cases for Divide-et-Impera Algorithms for
    Parity Games}
  \def\titlerunning{Robust Exponential Worst Cases for Divide-et-Impera
    Algorithms for Parity Games}

  \author{Massimo Benerecetti \& Daniele Dell'Erba \institute{Universit\`a degli
    Studi di Napoli Federico II} \and Fabio Mogavero \institute{Universit\`a
    degli Studi di Verona}}

  \def\authorrunning{M. Benerecetti, D. Dell'Erba, \& F. Mogavero}

  \maketitle

\input{Abstract}

\input{Introduction}

\input{Preliminaries}

\input{SectionI}

\input{SectionII}

\input{SectionIII}

\input{SectionIV}

  % \input{SectionV}

  % \input{SectionVI}

  % \input{SectionVII}

  % \input{SectionVIII}

  % \input{Discussion}

  % \input{Conclusion}

  % \input{Acknowledgments}

  \bibliographystyle{eptcs}
  \bibliography{References}

  \newpage
  \appendix

\end{document}

%% file: Format.tex
\usepackage[font=small,labelfont=bf]{caption}

\usepackage{amsthm}

\usepackage{hyperref}

\usepackage{cite}

\usepackage[final]{microtype}

\usepackage{times}

\hypersetup
  {
  pdfsubject    = {},
  pdfkeywords   = {},
  pdfproducer   = {},
  pdfcreator    = {},
  pdfpagemode   = {UseNone},
  pdfstartview  = {FitH}
  }

\usepackage{etoolbox}

\AtEndPreamble
  {

  \usepackage{fixltx2e}

  }

%% file: Macros.tex
\input{Macro/Macro}

\theoremstyle{plain}

\newtheorem{definition}{Definition}[section]

\newtheorem{theorem}{Theorem}[section]
\newtheorem{lemma}{Lemma}[section]

\newtheorem{corollary}{Corollary}[section]

\newcounter{flushenumerate}
  {\end{list}}

\renewcommand{\iff}
  {\txtabr{iff}}

\newcommand{\atrfun}{atr}
\newcommandx{\atrFun}[4][1=, 2=, 3=, 4=]
	{\mthargfun{\atrfun#4}[#1][#2]{#3}}

\newcommand{\prtset}{Pr}
\newcommandx{\PrtSet}[3][1=, 2=, 3=]
	{\mthset{\prtset#3}[#1][#2]}

\newcommand{\prtfun}{pr}
\newcommandx{\prtFun}[4][1=, 2=, 3=, 4=]
	{\mthargfun{\prtfun#4}[#1][#2]{#3}}

\newcommand{\prefun}{pre}
\newcommandx{\preFun}[4][1=, 2=, 3=, 4=]
	{\mthargfun{\prefun#4}[#1][#2]{#3}}

\newcommand{\solfun}{sol}
\newcommandx{\solFun}[4][1=, 2=, 3=, 4=]
  {\mthargfun{\solfun#4}[#1][#2]{#3}}

\newcommand{\IndSubTre}
  {\mthset[1]{G}}

\newcommand{\PGC}
  {\PG_{\CSym}}

%% file: Macro/Macro.tex
\usepackage{xargs}
\usepackage{xspace}
\usepackage{xstring}
\usepackage{boolexpr}

\usepackage[cmex10]{mathtools}
\usepackage{amsxtra}
\usepackage{amssymb}
\usepackage{amsfonts}
\usepackage{mathrsfs}
\usepackage{mathdots}
\usepackage{stmaryrd}
\usepackage{latexsym}

\usepackage{textcomp}

\input{Macro/Tools}

\input{Macro/Style}

\input{Macro/Text}

\input{Macro/Math}
\input{Macro/Models}
\input{Macro/Games}

\input{Macro/Logics}

\input{Macro/Complexity}

%% file: Macro/Tools.tex
\newcommand{\argemp}[2]
  {\if&#1&\else#2\fi}

\newcommand{\argdef}[2]
  {\if&#1&#2\else#1\fi}

\newcommand{\argint}[3]
  {\if&#2&\else#1#2#3\fi}

\newcommand{\argext}[3]
  {\if&#1&#3\else#1\if&#3&\else#2#3\fi\fi}

\newcommandx{\argsubsup}[3][2=, 3=]
  {\def\argsubscript{{#2}}\def\argsuperscript{{#3}}#1}

\newcommandx{\argind}[9][2=, 3=, 4=, 5=, 6=, 7=, 8=, 9=]
  {%
  \switch[#1=]%
    \case{0}#2%
    \case{1}#3%
    \case{2}#4%
    \case{3}#5%
    \case{4}#6%
    \case{5}#7%
    \case{6}#8%
    \case{7}#9%
    \otherwise\ensuremath{\clubsuit}%
  \endswitch%
  }

\newcommand{\arga}[1]
  {#1}

\newcommand{\argb}[2]
  {\argext{\arga{#1}}{, \allowbreak}{#2}}

\newcommand{\argc}[3]
  {\argext{\argb{#1}{#2}}{, \allowbreak}{#3}}

\newcommand{\argd}[4]
  {\argext{\argc{#1}{#2}{#3}}{, \allowbreak}{#4}}

\newcommand{\arge}[5]
  {\argext{\argd{#1}{#2}{#3}{#4}}{, \allowbreak}{#5}}

\newcommand{\argf}[6]
  {\argext{\arge{#1}{#2}{#3}{#4}{#5}}{, \allowbreak}{#6}}

\newcommand{\argg}[7]
  {\argext{\argf{#1}{#2}{#3}{#4}{#5}{#6}}{, \allowbreak}{#7}}

\newcommand{\argh}[8]
  {\argext{\argg{#1}{#2}{#3}{#4}{#5}{#6}{#7}}{, \allowbreak}{#8}}

\newcommand{\argi}[9]
  {\argext{\argh{#1}{#2}{#3}{#4}{#5}{#6}{#7}{#8}}{, \allowbreak}{#9}}

\newcommand{\argj}[9]
  {%
  \def\valarga{#1}%
  \def\valargb{#2}%
  \def\valargc{#3}%
  \def\valargd{#4}%
  \def\valarge{#5}%
  \def\valargf{#6}%
  \def\valargg{#7}%
  \def\valargh{#8}%
  \def\valargi{#9}%
  \argauxj%
  }
\newcommand{\argauxj}[1]
  {%
  \argext%
    {
    \argi
      {\valarga} {\valargb} {\valargc} {\valargd} {\valarge} {\valargf}
      {\valargg} {\valargh} {\valargi}
    }
    {, \allowbreak}{#1}%
  }

\newcommand{\argk}[9]
  {%
  \def\valarga{#1}%
  \def\valargb{#2}%
  \def\valargc{#3}%
  \def\valargd{#4}%
  \def\valarge{#5}%
  \def\valargf{#6}%
  \def\valargg{#7}%
  \def\valargh{#8}%
  \def\valargi{#9}%
  \argauxk%
  }
\newcommand{\argauxk}[2]
  {\argext{\argauxj{#1}}{, \allowbreak}{#2}}

\newcommand{\argl}[9]
  {%
  \def\valarga{#1}%
  \def\valargb{#2}%
  \def\valargc{#3}%
  \def\valargd{#4}%
  \def\valarge{#5}%
  \def\valargf{#6}%
  \def\valargg{#7}%
  \def\valargh{#8}%
  \def\valargi{#9}%
  \argauxl%
  }
\newcommand{\argauxl}[3]
  {\argext{\argauxk{#1}{#2}}{, \allowbreak}{#3}}

\newcommand{\argm}[9]
  {%
  \def\valarga{#1}%
  \def\valargb{#2}%
  \def\valargc{#3}%
  \def\valargd{#4}%
  \def\valarge{#5}%
  \def\valargf{#6}%
  \def\valargg{#7}%
  \def\valargh{#8}%
  \def\valargi{#9}%
  \argauxm%
  }
\newcommand{\argauxm}[4]
  {\argext{\argauxl{#1}{#2}{#3}}{, \allowbreak}{#4}}

\newcommand{\argn}[9]
  {%
  \def\valarga{#1}%
  \def\valargb{#2}%
  \def\valargc{#3}%
  \def\valargd{#4}%
  \def\valarge{#5}%
  \def\valargf{#6}%
  \def\valargg{#7}%
  \def\valargh{#8}%
  \def\valargi{#9}%
  \argauxn%
  }
\newcommand{\argauxn}[5]
  {\argext{\argauxm{#1}{#2}{#3}{#4}}{, \allowbreak}{#5}}

\newcommand{\argo}[9]
  {%
  \def\valarga{#1}%
  \def\valargb{#2}%
  \def\valargc{#3}%
  \def\valargd{#4}%
  \def\valarge{#5}%
  \def\valargf{#6}%
  \def\valargg{#7}%
  \def\valargh{#8}%
  \def\valargi{#9}%
  \argauxo%
  }
\newcommand{\argauxo}[6]
  {\argext{\argauxn{#1}{#2}{#3}{#4}{#5}}{, \allowbreak}{#6}}

\newcommand{\argp}[9]
  {%
  \def\valarga{#1}%
  \def\valargb{#2}%
  \def\valargc{#3}%
  \def\valargd{#4}%
  \def\valarge{#5}%
  \def\valargf{#6}%
  \def\valargg{#7}%
  \def\valargh{#8}%
  \def\valargi{#9}%
  \argauxp%
  }
\newcommand{\argauxp}[7]
  {\argext{\argauxo{#1}{#2}{#3}{#4}{#5}{#6}}{, \allowbreak}{#7}}

\newcommand{\argq}[9]
  {%
  \def\valarga{#1}%
  \def\valargb{#2}%
  \def\valargc{#3}%
  \def\valargd{#4}%
  \def\valarge{#5}%
  \def\valargf{#6}%
  \def\valargg{#7}%
  \def\valargh{#8}%
  \def\valargi{#9}%
  \argauxq%
  }
\newcommand{\argauxq}[8]
  {\argext{\argauxp{#1}{#2}{#3}{#4}{#5}{#6}{#7}}{, \allowbreak}{#8}}

\newcommand{\argr}[9]
  {%
  \def\valarga{#1}%
  \def\valargb{#2}%
  \def\valargc{#3}%
  \def\valargd{#4}%
  \def\valarge{#5}%
  \def\valargf{#6}%
  \def\valargg{#7}%
  \def\valargh{#8}%
  \def\valargi{#9}%
  \argauxr%
  }
\newcommand{\argauxr}[9]
  {\argext{\argauxq{#1}{#2}{#3}{#4}{#5}{#6}{#7}{#8}}{, \allowbreak}{#9}}

%% file: Macro/Style.tex
\newcommand{\txtfnt}[2][]
  {{%
  \IfStrEq{#1}{}
    {#2}
    {%
    \StrLeft{#1}{2}[\optbgn]%
    \StrGobbleLeft{#1}{2}[\optend]%
    \IfStrEqCase{\optbgn}
      {%
      {Rm}{\rmfamily\txtfnt[\optend]{#2}}%
      {Sf}{\sffamily\txtfnt[\optend]{#2}}%
      {Tt}{\ttfamily\txtfnt[\optend]{#2}}%
      {Up}{\upshape\txtfnt[\optend]{#2}}%
      {It}{\itshape\txtfnt[\optend]{#2}}%
      {Sl}{\slshape\txtfnt[\optend]{#2}}%
      {Sc}{\scshape\txtfnt[\optend]{#2}}%
      {Md}{\mdseries\txtfnt[\optend]{#2}}%
      {Bf}{\bfseries\txtfnt[\optend]{#2}}%
      {Em}{\emph{\txtfnt[\optend]{#2}}}%
      }
      [\ensuremath{\clubsuit}]%
    }%
  }}

\newcommand{\txtsubsup}[2]
  {\txtfnt[#1]{#2}}

\newcommand{\txtsub}[2][]
  {\argemp{#2}{\ensuremath{_{\text{\txtsubsup{#1}{#2}}}}}}

\newcommand{\txtsup}[2][]
  {\argemp{#2}{\ensuremath{^{\text{\txtsubsup{#1}{#2}}}}}}

\newcommandx{\txt}[4][1=, 3=, 4=]
  {\text{\txtfnt[#1]{#2}\ensuremath{\txtsub[#1]{#3}\txtsup[#1]{#4}}}\xspace}

\newcommandx{\txtarg}[5][1=, 3=, 4=]
  {{\txt[#1]{#2}[#3][#4]\argint{(}{#5}{)}}\xspace}

\newcommand{\txtstyname}{RmScMd}
\newcommand{\txtname}[1][]
  {\txt[\argdef{#1}{\txtstyname}]}
\newcommand{\txtargname}[1][]
  {\txtarg[\argdef{#1}{\txtstyname}]}

\newcommand{\txtstyabr}{Em}
\newcommand{\txtabr}[1][]
  {\txt[\argdef{#1}{\txtstyabr}]}

\newcommandx{\mthfnt}[3][1=, 2=0]
  {{%
  \IfStrEqCase{#1}
    {%
    {}%
      {#3}%
    {Name}%
      {%
      \IfStrEqCase{#2}
        {%
        {0}{\mathcal{#3}}%
        {1}{\mathscr{#3}}%
        {2}{\mathfrak{#3}}%
        {3}{\mathbf{#3}}%
        {4}{\mathnormal{#3}}%
        }
        [\ensuremath{\clubsuit}]%
      }%
    {Set}%
      {%
      \IfStrEqCase{#2}
        {%
        {0}{\mathrm{#3}}%
        {1}{\mathbb{#3}}%
        {2}{\mathsf{#3}}%
        {3}{\mathtt{#3}}%
        {4}{\mathnormal{#3}}%
        }
        [\ensuremath{\clubsuit}]%
      }%
    {Fun}%
      {%
      \IfStrEqCase{#2}
        {%
        {0}{\mathsf{#3}}%
        {1}{\mathrm{#3}}%
        {2}{\mathnormal{#3}}%
        }
        [\ensuremath{\clubsuit}]%
      }%
    {Rel}%
      {%
      \IfStrEqCase{#2}
        {%
        {0}{\mathit{#3}}%
        {1}{\mathtt{#3}}%
        {2}{\mathnormal{#3}}%
        }
        [\ensuremath{\clubsuit}]%
      }%
    {Tup}%
      {%
      \IfStrEqCase{#2}
        {%
        {0}{\mathbf{#3}}%
        {1}{\mathtt{#3}}%
        {2}{\mathnormal{#3}}%
        }
        [\ensuremath{\clubsuit}]%
      }%
    {Sym}%
      {%
      \IfStrEqCase{#2}
        {%
        {0}{\mathtt{#3}}%
        {1}{\mathbf{#3}}%
        {2}{\mathnormal{#3}}%
        }
        [\ensuremath{\clubsuit}]%
      }%
    {Elm}%
      {\mathnormal{#3}}
    }
    [\ensuremath{\clubsuit}]%
  }}

\newcommand{\mthsubsup}[1]
  {\mathnormal{#1}}

\newcommand{\mthsub}[1]
  {\argemp{#1}{\ensuremath{_{\mthsubsup{#1}}}}}

\newcommand{\mthsup}[1]
  {\argemp{#1}{\ensuremath{^{\mthsubsup{#1}}}}}

\newcommandx{\mth}[5][1=, 2=0, 4=, 5=]
  {{\ensuremath{\mthfnt[#1][#2]{#3}\mthsub{#4}\mthsup{#5}}}}

\newcommandx{\mtharg}[6][1=, 2=0, 4=, 5=]
  {{\mth[#1][#2]{#3}[#4][#5]\ensuremath{\argint{\!\left(}{#6}{\right)}}}}

\newcommand{\mthempty}
  {\mth[][]}

\newcommand{\mthstyname}{0}
\newcommand{\mthname}[1][]
  {\mth[Name][\argdef{#1}{\mthstyname}]}

\newcommand{\mthstyset}{0}
\newcommand{\mthset}[1][]
  {\mth[Set][\argdef{#1}{\mthstyset}]}
\newcommand{\mthargset}[1][]
  {\mtharg[Set][\argdef{#1}{\mthstyset}]}

\newcommand{\mthstyfun}{0}
\newcommand{\mthfun}[1][]
  {\mth[Fun][\argdef{#1}{\mthstyfun}]}
\newcommand{\mthargfun}[1][]
  {\mtharg[Fun][\argdef{#1}{\mthstyfun}]}

\newcommand{\mthstyrel}{0}
\newcommand{\mthrel}[1][]
  {\mth[Rel][\argdef{#1}{\mthstyrel}]}

\newcommand{\mthstytup}{0}
\newcommand{\mthtup}[1][]
  {\mth[Tup][\argdef{#1}{\mthstytup}]}

\newcommand{\mthstysym}{0}
\newcommand{\mthsym}[1][]
  {\mth[Sym][\argdef{#1}{\mthstysym}]}

\newcommand{\mthstyelm}{0}
\newcommand{\mthelm}[1][]
  {\mth[Elm][\argdef{#1}{\mthstyelm}]}

\newcommand{\mthstylettername}{0}

\newcommandx{\AName}[4][1=, 2=, 3=, 4=]
  {\mthname[\argdef{#4}{\mthstylettername}]{A#3}[#1][#2]}
\newcommandx{\BName}[4][1=, 2=, 3=, 4=]
  {\mthname[\argdef{#4}{\mthstylettername}]{B#3}[#1][#2]}
\newcommandx{\CName}[4][1=, 2=, 3=, 4=]
  {\mthname[\argdef{#4}{\mthstylettername}]{C#3}[#1][#2]}
\newcommandx{\DName}[4][1=, 2=, 3=, 4=]
  {\mthname[\argdef{#4}{\mthstylettername}]{D#3}[#1][#2]}
\newcommandx{\EName}[4][1=, 2=, 3=, 4=]
  {\mthname[\argdef{#4}{\mthstylettername}]{E#3}[#1][#2]}
\newcommandx{\FName}[4][1=, 2=, 3=, 4=]
  {\mthname[\argdef{#4}{\mthstylettername}]{F#3}[#1][#2]}
\newcommandx{\GName}[4][1=, 2=, 3=, 4=]
  {\mthname[\argdef{#4}{\mthstylettername}]{G#3}[#1][#2]}
\newcommandx{\HName}[4][1=, 2=, 3=, 4=]
  {\mthname[\argdef{#4}{\mthstylettername}]{H#3}[#1][#2]}
\newcommandx{\IName}[4][1=, 2=, 3=, 4=]
  {\mthname[\argdef{#4}{\mthstylettername}]{I#3}[#1][#2]}
\newcommandx{\JName}[4][1=, 2=, 3=, 4=]
  {\mthname[\argdef{#4}{\mthstylettername}]{J#3}[#1][#2]}
\newcommandx{\KName}[4][1=, 2=, 3=, 4=]
  {\mthname[\argdef{#4}{\mthstylettername}]{K#3}[#1][#2]}
\newcommandx{\LName}[4][1=, 2=, 3=, 4=]
  {\mthname[\argdef{#4}{\mthstylettername}]{L#3}[#1][#2]}
\newcommandx{\MName}[4][1=, 2=, 3=, 4=]
  {\mthname[\argdef{#4}{\mthstylettername}]{M#3}[#1][#2]}
\newcommandx{\NName}[4][1=, 2=, 3=, 4=]
  {\mthname[\argdef{#4}{\mthstylettername}]{N#3}[#1][#2]}
\newcommandx{\OName}[4][1=, 2=, 3=, 4=]
  {\mthname[\argdef{#4}{\mthstylettername}]{O#3}[#1][#2]}
\newcommandx{\PName}[4][1=, 2=, 3=, 4=]
  {\mthname[\argdef{#4}{\mthstylettername}]{P#3}[#1][#2]}
\newcommandx{\QName}[4][1=, 2=, 3=, 4=]
  {\mthname[\argdef{#4}{\mthstylettername}]{Q#3}[#1][#2]}
\newcommandx{\RName}[4][1=, 2=, 3=, 4=]
  {\mthname[\argdef{#4}{\mthstylettername}]{R#3}[#1][#2]}
\newcommandx{\SName}[4][1=, 2=, 3=, 4=]
  {\mthname[\argdef{#4}{\mthstylettername}]{S#3}[#1][#2]}
\newcommandx{\TName}[4][1=, 2=, 3=, 4=]
  {\mthname[\argdef{#4}{\mthstylettername}]{T#3}[#1][#2]}
\newcommandx{\UName}[4][1=, 2=, 3=, 4=]
  {\mthname[\argdef{#4}{\mthstylettername}]{U#3}[#1][#2]}
\newcommandx{\VName}[4][1=, 2=, 3=, 4=]
  {\mthname[\argdef{#4}{\mthstylettername}]{V#3}[#1][#2]}
\newcommandx{\WName}[4][1=, 2=, 3=, 4=]
  {\mthname[\argdef{#4}{\mthstylettername}]{W#3}[#1][#2]}
\newcommandx{\XName}[4][1=, 2=, 3=, 4=]
  {\mthname[\argdef{#4}{\mthstylettername}]{X#3}[#1][#2]}
\newcommandx{\YName}[4][1=, 2=, 3=, 4=]
  {\mthname[\argdef{#4}{\mthstylettername}]{Y#3}[#1][#2]}
\newcommandx{\ZName}[4][1=, 2=, 3=, 4=]
  {\mthname[\argdef{#4}{\mthstylettername}]{Z#3}[#1][#2]}

\newcommand{\mthstyletterset}{0}

\newcommandx{\ASet}[4][1=, 2=, 3=, 4=]
  {\mthset[\argdef{#4}{\mthstyletterset}]{A#3}[#1][#2]}
\newcommandx{\BSet}[4][1=, 2=, 3=, 4=]
  {\mthset[\argdef{#4}{\mthstyletterset}]{B#3}[#1][#2]}
\newcommandx{\CSet}[4][1=, 2=, 3=, 4=]
  {\mthset[\argdef{#4}{\mthstyletterset}]{C#3}[#1][#2]}
\newcommandx{\DSet}[4][1=, 2=, 3=, 4=]
  {\mthset[\argdef{#4}{\mthstyletterset}]{D#3}[#1][#2]}
\newcommandx{\ESet}[4][1=, 2=, 3=, 4=]
  {\mthset[\argdef{#4}{\mthstyletterset}]{E#3}[#1][#2]}
\newcommandx{\FSet}[4][1=, 2=, 3=, 4=]
  {\mthset[\argdef{#4}{\mthstyletterset}]{F#3}[#1][#2]}
\newcommandx{\GSet}[4][1=, 2=, 3=, 4=]
  {\mthset[\argdef{#4}{\mthstyletterset}]{G#3}[#1][#2]}
\newcommandx{\HSet}[4][1=, 2=, 3=, 4=]
  {\mthset[\argdef{#4}{\mthstyletterset}]{H#3}[#1][#2]}
\newcommandx{\ISet}[4][1=, 2=, 3=, 4=]
  {\mthset[\argdef{#4}{\mthstyletterset}]{I#3}[#1][#2]}
\newcommandx{\JSet}[4][1=, 2=, 3=, 4=]
  {\mthset[\argdef{#4}{\mthstyletterset}]{J#3}[#1][#2]}
\newcommandx{\KSet}[4][1=, 2=, 3=, 4=]
  {\mthset[\argdef{#4}{\mthstyletterset}]{K#3}[#1][#2]}
\newcommandx{\LSet}[4][1=, 2=, 3=, 4=]
  {\mthset[\argdef{#4}{\mthstyletterset}]{L#3}[#1][#2]}
\newcommandx{\MSet}[4][1=, 2=, 3=, 4=]
  {\mthset[\argdef{#4}{\mthstyletterset}]{M#3}[#1][#2]}
\newcommandx{\NSet}[4][1=, 2=, 3=, 4=]
  {\mthset[\argdef{#4}{\mthstyletterset}]{N#3}[#1][#2]}
\newcommandx{\OSet}[4][1=, 2=, 3=, 4=]
  {\mthset[\argdef{#4}{\mthstyletterset}]{O#3}[#1][#2]}
\newcommandx{\PSet}[4][1=, 2=, 3=, 4=]
  {\mthset[\argdef{#4}{\mthstyletterset}]{P#3}[#1][#2]}
\newcommandx{\QSet}[4][1=, 2=, 3=, 4=]
  {\mthset[\argdef{#4}{\mthstyletterset}]{Q#3}[#1][#2]}
\newcommandx{\RSet}[4][1=, 2=, 3=, 4=]
  {\mthset[\argdef{#4}{\mthstyletterset}]{R#3}[#1][#2]}
\newcommandx{\SSet}[4][1=, 2=, 3=, 4=]
  {\mthset[\argdef{#4}{\mthstyletterset}]{S#3}[#1][#2]}
\newcommandx{\TSet}[4][1=, 2=, 3=, 4=]
  {\mthset[\argdef{#4}{\mthstyletterset}]{T#3}[#1][#2]}
\newcommandx{\USet}[4][1=, 2=, 3=, 4=]
  {\mthset[\argdef{#4}{\mthstyletterset}]{U#3}[#1][#2]}
\newcommandx{\VSet}[4][1=, 2=, 3=, 4=]
  {\mthset[\argdef{#4}{\mthstyletterset}]{V#3}[#1][#2]}
\newcommandx{\WSet}[4][1=, 2=, 3=, 4=]
  {\mthset[\argdef{#4}{\mthstyletterset}]{W#3}[#1][#2]}
\newcommandx{\XSet}[4][1=, 2=, 3=, 4=]
  {\mthset[\argdef{#4}{\mthstyletterset}]{X#3}[#1][#2]}
\newcommandx{\YSet}[4][1=, 2=, 3=, 4=]
  {\mthset[\argdef{#4}{\mthstyletterset}]{Y#3}[#1][#2]}
\newcommandx{\ZSet}[4][1=, 2=, 3=, 4=]
  {\mthset[\argdef{#4}{\mthstyletterset}]{Z#3}[#1][#2]}

\newcommandx{\aSet}[4][1=, 2=, 3=, 4=]
  {\mthset[\argdef{#4}{\mthstyletterset}]{a#3}[#1][#2]}
\newcommandx{\bSet}[4][1=, 2=, 3=, 4=]
  {\mthset[\argdef{#4}{\mthstyletterset}]{b#3}[#1][#2]}
\newcommandx{\cSet}[4][1=, 2=, 3=, 4=]
  {\mthset[\argdef{#4}{\mthstyletterset}]{c#3}[#1][#2]}
\newcommandx{\dSet}[4][1=, 2=, 3=, 4=]
  {\mthset[\argdef{#4}{\mthstyletterset}]{d#3}[#1][#2]}
\newcommandx{\eSet}[4][1=, 2=, 3=, 4=]
  {\mthset[\argdef{#4}{\mthstyletterset}]{e#3}[#1][#2]}
\newcommandx{\fSet}[4][1=, 2=, 3=, 4=]
  {\mthset[\argdef{#4}{\mthstyletterset}]{f#3}[#1][#2]}
\newcommandx{\gSet}[4][1=, 2=, 3=, 4=]
  {\mthset[\argdef{#4}{\mthstyletterset}]{g#3}[#1][#2]}
\newcommandx{\hSet}[4][1=, 2=, 3=, 4=]
  {\mthset[\argdef{#4}{\mthstyletterset}]{h#3}[#1][#2]}
\newcommandx{\iSet}[4][1=, 2=, 3=, 4=]
  {\mthset[\argdef{#4}{\mthstyletterset}]{i#3}[#1][#2]}
\newcommandx{\jSet}[4][1=, 2=, 3=, 4=]
  {\mthset[\argdef{#4}{\mthstyletterset}]{j#3}[#1][#2]}
\newcommandx{\kSet}[4][1=, 2=, 3=, 4=]
  {\mthset[\argdef{#4}{\mthstyletterset}]{k#3}[#1][#2]}
\newcommandx{\lSet}[4][1=, 2=, 3=, 4=]
  {\mthset[\argdef{#4}{\mthstyletterset}]{l#3}[#1][#2]}
\newcommandx{\mSet}[4][1=, 2=, 3=, 4=]
  {\mthset[\argdef{#4}{\mthstyletterset}]{m#3}[#1][#2]}
\newcommandx{\nSet}[4][1=, 2=, 3=, 4=]
  {\mthset[\argdef{#4}{\mthstyletterset}]{n#3}[#1][#2]}
\newcommandx{\oSet}[4][1=, 2=, 3=, 4=]
  {\mthset[\argdef{#4}{\mthstyletterset}]{o#3}[#1][#2]}
\newcommandx{\pSet}[4][1=, 2=, 3=, 4=]
  {\mthset[\argdef{#4}{\mthstyletterset}]{p#3}[#1][#2]}
\newcommandx{\qSet}[4][1=, 2=, 3=, 4=]
  {\mthset[\argdef{#4}{\mthstyletterset}]{q#3}[#1][#2]}
\newcommandx{\rSet}[4][1=, 2=, 3=, 4=]
  {\mthset[\argdef{#4}{\mthstyletterset}]{r#3}[#1][#2]}
\newcommandx{\sSet}[4][1=, 2=, 3=, 4=]
  {\mthset[\argdef{#4}{\mthstyletterset}]{s#3}[#1][#2]}
\newcommandx{\tSet}[4][1=, 2=, 3=, 4=]
  {\mthset[\argdef{#4}{\mthstyletterset}]{t#3}[#1][#2]}
\newcommandx{\uSet}[4][1=, 2=, 3=, 4=]
  {\mthset[\argdef{#4}{\mthstyletterset}]{u#3}[#1][#2]}
\newcommandx{\vSet}[4][1=, 2=, 3=, 4=]
  {\mthset[\argdef{#4}{\mthstyletterset}]{v#3}[#1][#2]}
\newcommandx{\wSet}[4][1=, 2=, 3=, 4=]
  {\mthset[\argdef{#4}{\mthstyletterset}]{w#3}[#1][#2]}
\newcommandx{\xSet}[4][1=, 2=, 3=, 4=]
  {\mthset[\argdef{#4}{\mthstyletterset}]{x#3}[#1][#2]}
\newcommandx{\ySet}[4][1=, 2=, 3=, 4=]
  {\mthset[\argdef{#4}{\mthstyletterset}]{y#3}[#1][#2]}
\newcommandx{\zSet}[4][1=, 2=, 3=, 4=]
  {\mthset[\argdef{#4}{\mthstyletterset}]{z#3}[#1][#2]}

\newcommand{\mthstyletterfun}{0}

\newcommandx{\AFun}[4][1=, 2=, 3=, 4=]
  {\mthfun[\argdef{#4}{\mthstyletterfun}]{A#3}[#1][#2]}
\newcommandx{\BFun}[4][1=, 2=, 3=, 4=]
  {\mthfun[\argdef{#4}{\mthstyletterfun}]{B#3}[#1][#2]}
\newcommandx{\CFun}[4][1=, 2=, 3=, 4=]
  {\mthfun[\argdef{#4}{\mthstyletterfun}]{C#3}[#1][#2]}
\newcommandx{\DFun}[4][1=, 2=, 3=, 4=]
  {\mthfun[\argdef{#4}{\mthstyletterfun}]{D#3}[#1][#2]}
\newcommandx{\EFun}[4][1=, 2=, 3=, 4=]
  {\mthfun[\argdef{#4}{\mthstyletterfun}]{E#3}[#1][#2]}
\newcommandx{\FFun}[4][1=, 2=, 3=, 4=]
  {\mthfun[\argdef{#4}{\mthstyletterfun}]{F#3}[#1][#2]}
\newcommandx{\GFun}[4][1=, 2=, 3=, 4=]
  {\mthfun[\argdef{#4}{\mthstyletterfun}]{G#3}[#1][#2]}
\newcommandx{\HFun}[4][1=, 2=, 3=, 4=]
  {\mthfun[\argdef{#4}{\mthstyletterfun}]{H#3}[#1][#2]}
\newcommandx{\IFun}[4][1=, 2=, 3=, 4=]
  {\mthfun[\argdef{#4}{\mthstyletterfun}]{I#3}[#1][#2]}
\newcommandx{\JFun}[4][1=, 2=, 3=, 4=]
  {\mthfun[\argdef{#4}{\mthstyletterfun}]{J#3}[#1][#2]}
\newcommandx{\KFun}[4][1=, 2=, 3=, 4=]
  {\mthfun[\argdef{#4}{\mthstyletterfun}]{K#3}[#1][#2]}
\newcommandx{\LFun}[4][1=, 2=, 3=, 4=]
  {\mthfun[\argdef{#4}{\mthstyletterfun}]{L#3}[#1][#2]}
\newcommandx{\MFun}[4][1=, 2=, 3=, 4=]
  {\mthfun[\argdef{#4}{\mthstyletterfun}]{M#3}[#1][#2]}
\newcommandx{\NFun}[4][1=, 2=, 3=, 4=]
  {\mthfun[\argdef{#4}{\mthstyletterfun}]{N#3}[#1][#2]}
\newcommandx{\OFun}[4][1=, 2=, 3=, 4=]
  {\mthfun[\argdef{#4}{\mthstyletterfun}]{O#3}[#1][#2]}
\newcommandx{\PFun}[4][1=, 2=, 3=, 4=]
  {\mthfun[\argdef{#4}{\mthstyletterfun}]{P#3}[#1][#2]}
\newcommandx{\QFun}[4][1=, 2=, 3=, 4=]
  {\mthfun[\argdef{#4}{\mthstyletterfun}]{Q#3}[#1][#2]}
\newcommandx{\RFun}[4][1=, 2=, 3=, 4=]
  {\mthfun[\argdef{#4}{\mthstyletterfun}]{R#3}[#1][#2]}
\newcommandx{\SFun}[4][1=, 2=, 3=, 4=]
  {\mthfun[\argdef{#4}{\mthstyletterfun}]{S#3}[#1][#2]}
\newcommandx{\TFun}[4][1=, 2=, 3=, 4=]
  {\mthfun[\argdef{#4}{\mthstyletterfun}]{T#3}[#1][#2]}
\newcommandx{\UFun}[4][1=, 2=, 3=, 4=]
  {\mthfun[\argdef{#4}{\mthstyletterfun}]{U#3}[#1][#2]}
\newcommandx{\VFun}[4][1=, 2=, 3=, 4=]
  {\mthfun[\argdef{#4}{\mthstyletterfun}]{V#3}[#1][#2]}
\newcommandx{\WFun}[4][1=, 2=, 3=, 4=]
  {\mthfun[\argdef{#4}{\mthstyletterfun}]{W#3}[#1][#2]}
\newcommandx{\XFun}[4][1=, 2=, 3=, 4=]
  {\mthfun[\argdef{#4}{\mthstyletterfun}]{X#3}[#1][#2]}
\newcommandx{\YFun}[4][1=, 2=, 3=, 4=]
  {\mthfun[\argdef{#4}{\mthstyletterfun}]{Y#3}[#1][#2]}
\newcommandx{\ZFun}[4][1=, 2=, 3=, 4=]
  {\mthfun[\argdef{#4}{\mthstyletterfun}]{Z#3}[#1][#2]}

\newcommandx{\aFun}[4][1=, 2=, 3=, 4=]
  {\mthfun[\argdef{#4}{\mthstyletterfun}]{a#3}[#1][#2]}
\newcommandx{\bFun}[4][1=, 2=, 3=, 4=]
  {\mthfun[\argdef{#4}{\mthstyletterfun}]{b#3}[#1][#2]}
\newcommandx{\cFun}[4][1=, 2=, 3=, 4=]
  {\mthfun[\argdef{#4}{\mthstyletterfun}]{c#3}[#1][#2]}
\newcommandx{\dFun}[4][1=, 2=, 3=, 4=]
  {\mthfun[\argdef{#4}{\mthstyletterfun}]{d#3}[#1][#2]}
\newcommandx{\eFun}[4][1=, 2=, 3=, 4=]
  {\mthfun[\argdef{#4}{\mthstyletterfun}]{e#3}[#1][#2]}
\newcommandx{\fFun}[4][1=, 2=, 3=, 4=]
  {\mthfun[\argdef{#4}{\mthstyletterfun}]{f#3}[#1][#2]}
\newcommandx{\gFun}[4][1=, 2=, 3=, 4=]
  {\mthfun[\argdef{#4}{\mthstyletterfun}]{g#3}[#1][#2]}
\newcommandx{\hFun}[4][1=, 2=, 3=, 4=]
  {\mthfun[\argdef{#4}{\mthstyletterfun}]{h#3}[#1][#2]}
\newcommandx{\iFun}[4][1=, 2=, 3=, 4=]
  {\mthfun[\argdef{#4}{\mthstyletterfun}]{i#3}[#1][#2]}
\newcommandx{\jFun}[4][1=, 2=, 3=, 4=]
  {\mthfun[\argdef{#4}{\mthstyletterfun}]{j#3}[#1][#2]}
\newcommandx{\kFun}[4][1=, 2=, 3=, 4=]
  {\mthfun[\argdef{#4}{\mthstyletterfun}]{k#3}[#1][#2]}
\newcommandx{\lFun}[4][1=, 2=, 3=, 4=]
  {\mthfun[\argdef{#4}{\mthstyletterfun}]{l#3}[#1][#2]}
\newcommandx{\mFun}[4][1=, 2=, 3=, 4=]
  {\mthfun[\argdef{#4}{\mthstyletterfun}]{m#3}[#1][#2]}
\newcommandx{\nFun}[4][1=, 2=, 3=, 4=]
  {\mthfun[\argdef{#4}{\mthstyletterfun}]{n#3}[#1][#2]}
\newcommandx{\oFun}[4][1=, 2=, 3=, 4=]
  {\mthfun[\argdef{#4}{\mthstyletterfun}]{o#3}[#1][#2]}
\newcommandx{\pFun}[4][1=, 2=, 3=, 4=]
  {\mthfun[\argdef{#4}{\mthstyletterfun}]{p#3}[#1][#2]}
\newcommandx{\qFun}[4][1=, 2=, 3=, 4=]
  {\mthfun[\argdef{#4}{\mthstyletterfun}]{q#3}[#1][#2]}
\newcommandx{\rFun}[4][1=, 2=, 3=, 4=]
  {\mthfun[\argdef{#4}{\mthstyletterfun}]{r#3}[#1][#2]}
\newcommandx{\sFun}[4][1=, 2=, 3=, 4=]
  {\mthfun[\argdef{#4}{\mthstyletterfun}]{s#3}[#1][#2]}
\newcommandx{\tFun}[4][1=, 2=, 3=, 4=]
  {\mthfun[\argdef{#4}{\mthstyletterfun}]{t#3}[#1][#2]}
\newcommandx{\uFun}[4][1=, 2=, 3=, 4=]
  {\mthfun[\argdef{#4}{\mthstyletterfun}]{u#3}[#1][#2]}
\newcommandx{\vFun}[4][1=, 2=, 3=, 4=]
  {\mthfun[\argdef{#4}{\mthstyletterfun}]{v#3}[#1][#2]}
\newcommandx{\wFun}[4][1=, 2=, 3=, 4=]
  {\mthfun[\argdef{#4}{\mthstyletterfun}]{w#3}[#1][#2]}
\newcommandx{\xFun}[4][1=, 2=, 3=, 4=]
  {\mthfun[\argdef{#4}{\mthstyletterfun}]{x#3}[#1][#2]}
\newcommandx{\yFun}[4][1=, 2=, 3=, 4=]
  {\mthfun[\argdef{#4}{\mthstyletterfun}]{y#3}[#1][#2]}
\newcommandx{\zFun}[4][1=, 2=, 3=, 4=]
  {\mthfun[\argdef{#4}{\mthstyletterfun}]{z#3}[#1][#2]}

\newcommand{\mthstyletterrel}{0}

\newcommandx{\ARel}[4][1=, 2=, 3=, 4=]
  {\mthrel[\argdef{#4}{\mthstyletterrel}]{A#3}[#1][#2]}
\newcommandx{\BRel}[4][1=, 2=, 3=, 4=]
  {\mthrel[\argdef{#4}{\mthstyletterrel}]{B#3}[#1][#2]}
\newcommandx{\CRel}[4][1=, 2=, 3=, 4=]
  {\mthrel[\argdef{#4}{\mthstyletterrel}]{C#3}[#1][#2]}
\newcommandx{\DRel}[4][1=, 2=, 3=, 4=]
  {\mthrel[\argdef{#4}{\mthstyletterrel}]{D#3}[#1][#2]}
\newcommandx{\ERel}[4][1=, 2=, 3=, 4=]
  {\mthrel[\argdef{#4}{\mthstyletterrel}]{E#3}[#1][#2]}
\newcommandx{\FRel}[4][1=, 2=, 3=, 4=]
  {\mthrel[\argdef{#4}{\mthstyletterrel}]{F#3}[#1][#2]}
\newcommandx{\GRel}[4][1=, 2=, 3=, 4=]
  {\mthrel[\argdef{#4}{\mthstyletterrel}]{G#3}[#1][#2]}
\newcommandx{\HRel}[4][1=, 2=, 3=, 4=]
  {\mthrel[\argdef{#4}{\mthstyletterrel}]{H#3}[#1][#2]}
\newcommandx{\IRel}[4][1=, 2=, 3=, 4=]
  {\mthrel[\argdef{#4}{\mthstyletterrel}]{I#3}[#1][#2]}
\newcommandx{\JRel}[4][1=, 2=, 3=, 4=]
  {\mthrel[\argdef{#4}{\mthstyletterrel}]{J#3}[#1][#2]}
\newcommandx{\KRel}[4][1=, 2=, 3=, 4=]
  {\mthrel[\argdef{#4}{\mthstyletterrel}]{K#3}[#1][#2]}
\newcommandx{\LRel}[4][1=, 2=, 3=, 4=]
  {\mthrel[\argdef{#4}{\mthstyletterrel}]{L#3}[#1][#2]}
\newcommandx{\MRel}[4][1=, 2=, 3=, 4=]
  {\mthrel[\argdef{#4}{\mthstyletterrel}]{M#3}[#1][#2]}
\newcommandx{\NRel}[4][1=, 2=, 3=, 4=]
  {\mthrel[\argdef{#4}{\mthstyletterrel}]{N#3}[#1][#2]}
\newcommandx{\ORel}[4][1=, 2=, 3=, 4=]
  {\mthrel[\argdef{#4}{\mthstyletterrel}]{O#3}[#1][#2]}
\newcommandx{\PRel}[4][1=, 2=, 3=, 4=]
  {\mthrel[\argdef{#4}{\mthstyletterrel}]{P#3}[#1][#2]}
\newcommandx{\QRel}[4][1=, 2=, 3=, 4=]
  {\mthrel[\argdef{#4}{\mthstyletterrel}]{Q#3}[#1][#2]}
\newcommandx{\RRel}[4][1=, 2=, 3=, 4=]
  {\mthrel[\argdef{#4}{\mthstyletterrel}]{R#3}[#1][#2]}
\newcommandx{\SRel}[4][1=, 2=, 3=, 4=]
  {\mthrel[\argdef{#4}{\mthstyletterrel}]{S#3}[#1][#2]}
\newcommandx{\TRel}[4][1=, 2=, 3=, 4=]
  {\mthrel[\argdef{#4}{\mthstyletterrel}]{T#3}[#1][#2]}
\newcommandx{\URel}[4][1=, 2=, 3=, 4=]
  {\mthrel[\argdef{#4}{\mthstyletterrel}]{U#3}[#1][#2]}
\newcommandx{\VRel}[4][1=, 2=, 3=, 4=]
  {\mthrel[\argdef{#4}{\mthstyletterrel}]{V#3}[#1][#2]}
\newcommandx{\WRel}[4][1=, 2=, 3=, 4=]
  {\mthrel[\argdef{#4}{\mthstyletterrel}]{W#3}[#1][#2]}
\newcommandx{\XRel}[4][1=, 2=, 3=, 4=]
  {\mthrel[\argdef{#4}{\mthstyletterrel}]{X#3}[#1][#2]}
\newcommandx{\YRel}[4][1=, 2=, 3=, 4=]
  {\mthrel[\argdef{#4}{\mthstyletterrel}]{Y#3}[#1][#2]}
\newcommandx{\ZRel}[4][1=, 2=, 3=, 4=]
  {\mthrel[\argdef{#4}{\mthstyletterrel}]{Z#3}[#1][#2]}

\newcommandx{\aRel}[4][1=, 2=, 3=, 4=]
  {\mthrel[\argdef{#4}{\mthstyletterrel}]{a#3}[#1][#2]}
\newcommandx{\bRel}[4][1=, 2=, 3=, 4=]
  {\mthrel[\argdef{#4}{\mthstyletterrel}]{b#3}[#1][#2]}
\newcommandx{\cRel}[4][1=, 2=, 3=, 4=]
  {\mthrel[\argdef{#4}{\mthstyletterrel}]{c#3}[#1][#2]}
\newcommandx{\dRel}[4][1=, 2=, 3=, 4=]
  {\mthrel[\argdef{#4}{\mthstyletterrel}]{d#3}[#1][#2]}
\newcommandx{\eRel}[4][1=, 2=, 3=, 4=]
  {\mthrel[\argdef{#4}{\mthstyletterrel}]{e#3}[#1][#2]}
\newcommandx{\fRel}[4][1=, 2=, 3=, 4=]
  {\mthrel[\argdef{#4}{\mthstyletterrel}]{f#3}[#1][#2]}
\newcommandx{\gRel}[4][1=, 2=, 3=, 4=]
  {\mthrel[\argdef{#4}{\mthstyletterrel}]{g#3}[#1][#2]}
\newcommandx{\hRel}[4][1=, 2=, 3=, 4=]
  {\mthrel[\argdef{#4}{\mthstyletterrel}]{h#3}[#1][#2]}
\newcommandx{\iRel}[4][1=, 2=, 3=, 4=]
  {\mthrel[\argdef{#4}{\mthstyletterrel}]{i#3}[#1][#2]}
\newcommandx{\jRel}[4][1=, 2=, 3=, 4=]
  {\mthrel[\argdef{#4}{\mthstyletterrel}]{j#3}[#1][#2]}
\newcommandx{\kRel}[4][1=, 2=, 3=, 4=]
  {\mthrel[\argdef{#4}{\mthstyletterrel}]{k#3}[#1][#2]}
\newcommandx{\lRel}[4][1=, 2=, 3=, 4=]
  {\mthrel[\argdef{#4}{\mthstyletterrel}]{l#3}[#1][#2]}
\newcommandx{\mRel}[4][1=, 2=, 3=, 4=]
  {\mthrel[\argdef{#4}{\mthstyletterrel}]{m#3}[#1][#2]}
\newcommandx{\nRel}[4][1=, 2=, 3=, 4=]
  {\mthrel[\argdef{#4}{\mthstyletterrel}]{n#3}[#1][#2]}
\newcommandx{\oRel}[4][1=, 2=, 3=, 4=]
  {\mthrel[\argdef{#4}{\mthstyletterrel}]{o#3}[#1][#2]}
\newcommandx{\pRel}[4][1=, 2=, 3=, 4=]
  {\mthrel[\argdef{#4}{\mthstyletterrel}]{p#3}[#1][#2]}
\newcommandx{\qRel}[4][1=, 2=, 3=, 4=]
  {\mthrel[\argdef{#4}{\mthstyletterrel}]{q#3}[#1][#2]}
\newcommandx{\rRel}[4][1=, 2=, 3=, 4=]
  {\mthrel[\argdef{#4}{\mthstyletterrel}]{r#3}[#1][#2]}
\newcommandx{\sRel}[4][1=, 2=, 3=, 4=]
  {\mthrel[\argdef{#4}{\mthstyletterrel}]{s#3}[#1][#2]}
\newcommandx{\tRel}[4][1=, 2=, 3=, 4=]
  {\mthrel[\argdef{#4}{\mthstyletterrel}]{t#3}[#1][#2]}
\newcommandx{\uRel}[4][1=, 2=, 3=, 4=]
  {\mthrel[\argdef{#4}{\mthstyletterrel}]{u#3}[#1][#2]}
\newcommandx{\vRel}[4][1=, 2=, 3=, 4=]
  {\mthrel[\argdef{#4}{\mthstyletterrel}]{v#3}[#1][#2]}
\newcommandx{\wRel}[4][1=, 2=, 3=, 4=]
  {\mthrel[\argdef{#4}{\mthstyletterrel}]{w#3}[#1][#2]}
\newcommandx{\xRel}[4][1=, 2=, 3=, 4=]
  {\mthrel[\argdef{#4}{\mthstyletterrel}]{x#3}[#1][#2]}
\newcommandx{\yRel}[4][1=, 2=, 3=, 4=]
  {\mthrel[\argdef{#4}{\mthstyletterrel}]{y#3}[#1][#2]}
\newcommandx{\zRel}[4][1=, 2=, 3=, 4=]
  {\mthrel[\argdef{#4}{\mthstyletterrel}]{z#3}[#1][#2]}

\newcommand{\mthstylettertup}{0}

\newcommandx{\ATup}[4][1=, 2=, 3=, 4=]
  {\mthtup[\argdef{#4}{\mthstylettertup}]{A#3}[#1][#2]}
\newcommandx{\BTup}[4][1=, 2=, 3=, 4=]
  {\mthtup[\argdef{#4}{\mthstylettertup}]{B#3}[#1][#2]}
\newcommandx{\CTup}[4][1=, 2=, 3=, 4=]
  {\mthtup[\argdef{#4}{\mthstylettertup}]{C#3}[#1][#2]}
\newcommandx{\DTup}[4][1=, 2=, 3=, 4=]
  {\mthtup[\argdef{#4}{\mthstylettertup}]{D#3}[#1][#2]}
\newcommandx{\ETup}[4][1=, 2=, 3=, 4=]
  {\mthtup[\argdef{#4}{\mthstylettertup}]{E#3}[#1][#2]}
\newcommandx{\FTup}[4][1=, 2=, 3=, 4=]
  {\mthtup[\argdef{#4}{\mthstylettertup}]{F#3}[#1][#2]}
\newcommandx{\GTup}[4][1=, 2=, 3=, 4=]
  {\mthtup[\argdef{#4}{\mthstylettertup}]{G#3}[#1][#2]}
\newcommandx{\HTup}[4][1=, 2=, 3=, 4=]
  {\mthtup[\argdef{#4}{\mthstylettertup}]{H#3}[#1][#2]}
\newcommandx{\ITup}[4][1=, 2=, 3=, 4=]
  {\mthtup[\argdef{#4}{\mthstylettertup}]{I#3}[#1][#2]}
\newcommandx{\JTup}[4][1=, 2=, 3=, 4=]
  {\mthtup[\argdef{#4}{\mthstylettertup}]{J#3}[#1][#2]}
\newcommandx{\KTup}[4][1=, 2=, 3=, 4=]
  {\mthtup[\argdef{#4}{\mthstylettertup}]{K#3}[#1][#2]}
\newcommandx{\LTup}[4][1=, 2=, 3=, 4=]
  {\mthtup[\argdef{#4}{\mthstylettertup}]{L#3}[#1][#2]}
\newcommandx{\MTup}[4][1=, 2=, 3=, 4=]
  {\mthtup[\argdef{#4}{\mthstylettertup}]{M#3}[#1][#2]}
\newcommandx{\NTup}[4][1=, 2=, 3=, 4=]
  {\mthtup[\argdef{#4}{\mthstylettertup}]{N#3}[#1][#2]}
\newcommandx{\OTup}[4][1=, 2=, 3=, 4=]
  {\mthtup[\argdef{#4}{\mthstylettertup}]{O#3}[#1][#2]}
\newcommandx{\PTup}[4][1=, 2=, 3=, 4=]
  {\mthtup[\argdef{#4}{\mthstylettertup}]{P#3}[#1][#2]}
\newcommandx{\QTup}[4][1=, 2=, 3=, 4=]
  {\mthtup[\argdef{#4}{\mthstylettertup}]{Q#3}[#1][#2]}
\newcommandx{\RTup}[4][1=, 2=, 3=, 4=]
  {\mthtup[\argdef{#4}{\mthstylettertup}]{R#3}[#1][#2]}
\newcommandx{\STup}[4][1=, 2=, 3=, 4=]
  {\mthtup[\argdef{#4}{\mthstylettertup}]{S#3}[#1][#2]}
\newcommandx{\TTup}[4][1=, 2=, 3=, 4=]
  {\mthtup[\argdef{#4}{\mthstylettertup}]{T#3}[#1][#2]}
\newcommandx{\UTup}[4][1=, 2=, 3=, 4=]
  {\mthtup[\argdef{#4}{\mthstylettertup}]{U#3}[#1][#2]}
\newcommandx{\VTup}[4][1=, 2=, 3=, 4=]
  {\mthtup[\argdef{#4}{\mthstylettertup}]{V#3}[#1][#2]}
\newcommandx{\WTup}[4][1=, 2=, 3=, 4=]
  {\mthtup[\argdef{#4}{\mthstylettertup}]{W#3}[#1][#2]}
\newcommandx{\XTup}[4][1=, 2=, 3=, 4=]
  {\mthtup[\argdef{#4}{\mthstylettertup}]{X#3}[#1][#2]}
\newcommandx{\YTup}[4][1=, 2=, 3=, 4=]
  {\mthtup[\argdef{#4}{\mthstylettertup}]{Y#3}[#1][#2]}
\newcommandx{\ZTup}[4][1=, 2=, 3=, 4=]
  {\mthtup[\argdef{#4}{\mthstylettertup}]{Z#3}[#1][#2]}

\newcommandx{\aTup}[4][1=, 2=, 3=, 4=]
  {\mthtup[\argdef{#4}{\mthstylettertup}]{a#3}[#1][#2]}
\newcommandx{\bTup}[4][1=, 2=, 3=, 4=]
  {\mthtup[\argdef{#4}{\mthstylettertup}]{b#3}[#1][#2]}
\newcommandx{\cTup}[4][1=, 2=, 3=, 4=]
  {\mthtup[\argdef{#4}{\mthstylettertup}]{c#3}[#1][#2]}
\newcommandx{\dTup}[4][1=, 2=, 3=, 4=]
  {\mthtup[\argdef{#4}{\mthstylettertup}]{d#3}[#1][#2]}
\newcommandx{\eTup}[4][1=, 2=, 3=, 4=]
  {\mthtup[\argdef{#4}{\mthstylettertup}]{e#3}[#1][#2]}
\newcommandx{\fTup}[4][1=, 2=, 3=, 4=]
  {\mthtup[\argdef{#4}{\mthstylettertup}]{f#3}[#1][#2]}
\newcommandx{\gTup}[4][1=, 2=, 3=, 4=]
  {\mthtup[\argdef{#4}{\mthstylettertup}]{g#3}[#1][#2]}
\newcommandx{\hTup}[4][1=, 2=, 3=, 4=]
  {\mthtup[\argdef{#4}{\mthstylettertup}]{h#3}[#1][#2]}
\newcommandx{\iTup}[4][1=, 2=, 3=, 4=]
  {\mthtup[\argdef{#4}{\mthstylettertup}]{i#3}[#1][#2]}
\newcommandx{\jTup}[4][1=, 2=, 3=, 4=]
  {\mthtup[\argdef{#4}{\mthstylettertup}]{j#3}[#1][#2]}
\newcommandx{\kTup}[4][1=, 2=, 3=, 4=]
  {\mthtup[\argdef{#4}{\mthstylettertup}]{k#3}[#1][#2]}
\newcommandx{\lTup}[4][1=, 2=, 3=, 4=]
  {\mthtup[\argdef{#4}{\mthstylettertup}]{l#3}[#1][#2]}
\newcommandx{\mTup}[4][1=, 2=, 3=, 4=]
  {\mthtup[\argdef{#4}{\mthstylettertup}]{m#3}[#1][#2]}
\newcommandx{\nTup}[4][1=, 2=, 3=, 4=]
  {\mthtup[\argdef{#4}{\mthstylettertup}]{n#3}[#1][#2]}
\newcommandx{\oTup}[4][1=, 2=, 3=, 4=]
  {\mthtup[\argdef{#4}{\mthstylettertup}]{o#3}[#1][#2]}
\newcommandx{\pTup}[4][1=, 2=, 3=, 4=]
  {\mthtup[\argdef{#4}{\mthstylettertup}]{p#3}[#1][#2]}
\newcommandx{\qTup}[4][1=, 2=, 3=, 4=]
  {\mthtup[\argdef{#4}{\mthstylettertup}]{q#3}[#1][#2]}
\newcommandx{\rTup}[4][1=, 2=, 3=, 4=]
  {\mthtup[\argdef{#4}{\mthstylettertup}]{r#3}[#1][#2]}
\newcommandx{\sTup}[4][1=, 2=, 3=, 4=]
  {\mthtup[\argdef{#4}{\mthstylettertup}]{s#3}[#1][#2]}
\newcommandx{\tTup}[4][1=, 2=, 3=, 4=]
  {\mthtup[\argdef{#4}{\mthstylettertup}]{t#3}[#1][#2]}
\newcommandx{\uTup}[4][1=, 2=, 3=, 4=]
  {\mthtup[\argdef{#4}{\mthstylettertup}]{u#3}[#1][#2]}
\newcommandx{\vTup}[4][1=, 2=, 3=, 4=]
  {\mthtup[\argdef{#4}{\mthstylettertup}]{v#3}[#1][#2]}
\newcommandx{\wTup}[4][1=, 2=, 3=, 4=]
  {\mthtup[\argdef{#4}{\mthstylettertup}]{w#3}[#1][#2]}
\newcommandx{\xTup}[4][1=, 2=, 3=, 4=]
  {\mthtup[\argdef{#4}{\mthstylettertup}]{x#3}[#1][#2]}
\newcommandx{\yTup}[4][1=, 2=, 3=, 4=]
  {\mthtup[\argdef{#4}{\mthstylettertup}]{y#3}[#1][#2]}
\newcommandx{\zTup}[4][1=, 2=, 3=, 4=]
  {\mthtup[\argdef{#4}{\mthstylettertup}]{z#3}[#1][#2]}

\newcommand{\mthstylettersym}{0}

\newcommandx{\ASym}[4][1=, 2=, 3=, 4=]
  {\mthsym[\argdef{#4}{\mthstylettersym}]{A#3}[#1][#2]}
\newcommandx{\BSym}[4][1=, 2=, 3=, 4=]
  {\mthsym[\argdef{#4}{\mthstylettersym}]{B#3}[#1][#2]}
\newcommandx{\CSym}[4][1=, 2=, 3=, 4=]
  {\mthsym[\argdef{#4}{\mthstylettersym}]{C#3}[#1][#2]}
\newcommandx{\DSym}[4][1=, 2=, 3=, 4=]
  {\mthsym[\argdef{#4}{\mthstylettersym}]{D#3}[#1][#2]}
\newcommandx{\ESym}[4][1=, 2=, 3=, 4=]
  {\mthsym[\argdef{#4}{\mthstylettersym}]{E#3}[#1][#2]}
\newcommandx{\FSym}[4][1=, 2=, 3=, 4=]
  {\mthsym[\argdef{#4}{\mthstylettersym}]{F#3}[#1][#2]}
\newcommandx{\GSym}[4][1=, 2=, 3=, 4=]
  {\mthsym[\argdef{#4}{\mthstylettersym}]{G#3}[#1][#2]}
\newcommandx{\HSym}[4][1=, 2=, 3=, 4=]
  {\mthsym[\argdef{#4}{\mthstylettersym}]{H#3}[#1][#2]}
\newcommandx{\ISym}[4][1=, 2=, 3=, 4=]
  {\mthsym[\argdef{#4}{\mthstylettersym}]{I#3}[#1][#2]}
\newcommandx{\JSym}[4][1=, 2=, 3=, 4=]
  {\mthsym[\argdef{#4}{\mthstylettersym}]{J#3}[#1][#2]}
\newcommandx{\KSym}[4][1=, 2=, 3=, 4=]
  {\mthsym[\argdef{#4}{\mthstylettersym}]{K#3}[#1][#2]}
\newcommandx{\LSym}[4][1=, 2=, 3=, 4=]
  {\mthsym[\argdef{#4}{\mthstylettersym}]{L#3}[#1][#2]}
\newcommandx{\MSym}[4][1=, 2=, 3=, 4=]
  {\mthsym[\argdef{#4}{\mthstylettersym}]{M#3}[#1][#2]}
\newcommandx{\NSym}[4][1=, 2=, 3=, 4=]
  {\mthsym[\argdef{#4}{\mthstylettersym}]{N#3}[#1][#2]}
\newcommandx{\OSym}[4][1=, 2=, 3=, 4=]
  {\mthsym[\argdef{#4}{\mthstylettersym}]{O#3}[#1][#2]}
\newcommandx{\PSym}[4][1=, 2=, 3=, 4=]
  {\mthsym[\argdef{#4}{\mthstylettersym}]{P#3}[#1][#2]}
\newcommandx{\QSym}[4][1=, 2=, 3=, 4=]
  {\mthsym[\argdef{#4}{\mthstylettersym}]{Q#3}[#1][#2]}
\newcommandx{\RSym}[4][1=, 2=, 3=, 4=]
  {\mthsym[\argdef{#4}{\mthstylettersym}]{R#3}[#1][#2]}
\newcommandx{\SSym}[4][1=, 2=, 3=, 4=]
  {\mthsym[\argdef{#4}{\mthstylettersym}]{S#3}[#1][#2]}
\newcommandx{\TSym}[4][1=, 2=, 3=, 4=]
  {\mthsym[\argdef{#4}{\mthstylettersym}]{T#3}[#1][#2]}
\newcommandx{\USym}[4][1=, 2=, 3=, 4=]
  {\mthsym[\argdef{#4}{\mthstylettersym}]{U#3}[#1][#2]}
\newcommandx{\VSym}[4][1=, 2=, 3=, 4=]
  {\mthsym[\argdef{#4}{\mthstylettersym}]{V#3}[#1][#2]}
\newcommandx{\WSym}[4][1=, 2=, 3=, 4=]
  {\mthsym[\argdef{#4}{\mthstylettersym}]{W#3}[#1][#2]}
\newcommandx{\XSym}[4][1=, 2=, 3=, 4=]
  {\mthsym[\argdef{#4}{\mthstylettersym}]{X#3}[#1][#2]}
\newcommandx{\YSym}[4][1=, 2=, 3=, 4=]
  {\mthsym[\argdef{#4}{\mthstylettersym}]{Y#3}[#1][#2]}
\newcommandx{\ZSym}[4][1=, 2=, 3=, 4=]
  {\mthsym[\argdef{#4}{\mthstylettersym}]{Z#3}[#1][#2]}

\newcommandx{\aSym}[4][1=, 2=, 3=, 4=]
  {\mthsym[\argdef{#4}{\mthstylettersym}]{a#3}[#1][#2]}
\newcommandx{\bSym}[4][1=, 2=, 3=, 4=]
  {\mthsym[\argdef{#4}{\mthstylettersym}]{b#3}[#1][#2]}
\newcommandx{\cSym}[4][1=, 2=, 3=, 4=]
  {\mthsym[\argdef{#4}{\mthstylettersym}]{c#3}[#1][#2]}
\newcommandx{\dSym}[4][1=, 2=, 3=, 4=]
  {\mthsym[\argdef{#4}{\mthstylettersym}]{d#3}[#1][#2]}
\newcommandx{\eSym}[4][1=, 2=, 3=, 4=]
  {\mthsym[\argdef{#4}{\mthstylettersym}]{e#3}[#1][#2]}
\newcommandx{\fSym}[4][1=, 2=, 3=, 4=]
  {\mthsym[\argdef{#4}{\mthstylettersym}]{f#3}[#1][#2]}
\newcommandx{\gSym}[4][1=, 2=, 3=, 4=]
  {\mthsym[\argdef{#4}{\mthstylettersym}]{g#3}[#1][#2]}
\newcommandx{\hSym}[4][1=, 2=, 3=, 4=]
  {\mthsym[\argdef{#4}{\mthstylettersym}]{h#3}[#1][#2]}
\newcommandx{\iSym}[4][1=, 2=, 3=, 4=]
  {\mthsym[\argdef{#4}{\mthstylettersym}]{i#3}[#1][#2]}
\newcommandx{\jSym}[4][1=, 2=, 3=, 4=]
  {\mthsym[\argdef{#4}{\mthstylettersym}]{j#3}[#1][#2]}
\newcommandx{\kSym}[4][1=, 2=, 3=, 4=]
  {\mthsym[\argdef{#4}{\mthstylettersym}]{k#3}[#1][#2]}
\newcommandx{\lSym}[4][1=, 2=, 3=, 4=]
  {\mthsym[\argdef{#4}{\mthstylettersym}]{l#3}[#1][#2]}
\newcommandx{\mSym}[4][1=, 2=, 3=, 4=]
  {\mthsym[\argdef{#4}{\mthstylettersym}]{m#3}[#1][#2]}
\newcommandx{\nSym}[4][1=, 2=, 3=, 4=]
  {\mthsym[\argdef{#4}{\mthstylettersym}]{n#3}[#1][#2]}
\newcommandx{\oSym}[4][1=, 2=, 3=, 4=]
  {\mthsym[\argdef{#4}{\mthstylettersym}]{o#3}[#1][#2]}
\newcommandx{\pSym}[4][1=, 2=, 3=, 4=]
  {\mthsym[\argdef{#4}{\mthstylettersym}]{p#3}[#1][#2]}
\newcommandx{\qSym}[4][1=, 2=, 3=, 4=]
  {\mthsym[\argdef{#4}{\mthstylettersym}]{q#3}[#1][#2]}
\newcommandx{\rSym}[4][1=, 2=, 3=, 4=]
  {\mthsym[\argdef{#4}{\mthstylettersym}]{r#3}[#1][#2]}
\newcommandx{\sSym}[4][1=, 2=, 3=, 4=]
  {\mthsym[\argdef{#4}{\mthstylettersym}]{s#3}[#1][#2]}
\newcommandx{\tSym}[4][1=, 2=, 3=, 4=]
  {\mthsym[\argdef{#4}{\mthstylettersym}]{t#3}[#1][#2]}
\newcommandx{\uSym}[4][1=, 2=, 3=, 4=]
  {\mthsym[\argdef{#4}{\mthstylettersym}]{u#3}[#1][#2]}
\newcommandx{\vSym}[4][1=, 2=, 3=, 4=]
  {\mthsym[\argdef{#4}{\mthstylettersym}]{v#3}[#1][#2]}
\newcommandx{\wSym}[4][1=, 2=, 3=, 4=]
  {\mthsym[\argdef{#4}{\mthstylettersym}]{w#3}[#1][#2]}
\newcommandx{\xSym}[4][1=, 2=, 3=, 4=]
  {\mthsym[\argdef{#4}{\mthstylettersym}]{x#3}[#1][#2]}
\newcommandx{\ySym}[4][1=, 2=, 3=, 4=]
  {\mthsym[\argdef{#4}{\mthstylettersym}]{y#3}[#1][#2]}
\newcommandx{\zSym}[4][1=, 2=, 3=, 4=]
  {\mthsym[\argdef{#4}{\mthstylettersym}]{z#3}[#1][#2]}

\newcommand{\mthstyletterelm}{0}

\newcommandx{\AElm}[4][1=, 2=, 3=, 4=]
  {\mthelm[\argdef{#4}{\mthstyletterelm}]{A#3}[#1][#2]}
\newcommandx{\BElm}[4][1=, 2=, 3=, 4=]
  {\mthelm[\argdef{#4}{\mthstyletterelm}]{B#3}[#1][#2]}
\newcommandx{\CElm}[4][1=, 2=, 3=, 4=]
  {\mthelm[\argdef{#4}{\mthstyletterelm}]{C#3}[#1][#2]}
\newcommandx{\DElm}[4][1=, 2=, 3=, 4=]
  {\mthelm[\argdef{#4}{\mthstyletterelm}]{D#3}[#1][#2]}
\newcommandx{\EElm}[4][1=, 2=, 3=, 4=]
  {\mthelm[\argdef{#4}{\mthstyletterelm}]{E#3}[#1][#2]}
\newcommandx{\FElm}[4][1=, 2=, 3=, 4=]
  {\mthelm[\argdef{#4}{\mthstyletterelm}]{F#3}[#1][#2]}
\newcommandx{\GElm}[4][1=, 2=, 3=, 4=]
  {\mthelm[\argdef{#4}{\mthstyletterelm}]{G#3}[#1][#2]}
\newcommandx{\HElm}[4][1=, 2=, 3=, 4=]
  {\mthelm[\argdef{#4}{\mthstyletterelm}]{H#3}[#1][#2]}
\newcommandx{\IElm}[4][1=, 2=, 3=, 4=]
  {\mthelm[\argdef{#4}{\mthstyletterelm}]{I#3}[#1][#2]}
\newcommandx{\JElm}[4][1=, 2=, 3=, 4=]
  {\mthelm[\argdef{#4}{\mthstyletterelm}]{J#3}[#1][#2]}
\newcommandx{\KElm}[4][1=, 2=, 3=, 4=]
  {\mthelm[\argdef{#4}{\mthstyletterelm}]{K#3}[#1][#2]}
\newcommandx{\LElm}[4][1=, 2=, 3=, 4=]
  {\mthelm[\argdef{#4}{\mthstyletterelm}]{L#3}[#1][#2]}
\newcommandx{\MElm}[4][1=, 2=, 3=, 4=]
  {\mthelm[\argdef{#4}{\mthstyletterelm}]{M#3}[#1][#2]}
\newcommandx{\NElm}[4][1=, 2=, 3=, 4=]
  {\mthelm[\argdef{#4}{\mthstyletterelm}]{N#3}[#1][#2]}
\newcommandx{\OElm}[4][1=, 2=, 3=, 4=]
  {\mthelm[\argdef{#4}{\mthstyletterelm}]{O#3}[#1][#2]}
\newcommandx{\PElm}[4][1=, 2=, 3=, 4=]
  {\mthelm[\argdef{#4}{\mthstyletterelm}]{P#3}[#1][#2]}
\newcommandx{\QElm}[4][1=, 2=, 3=, 4=]
  {\mthelm[\argdef{#4}{\mthstyletterelm}]{Q#3}[#1][#2]}
\newcommandx{\RElm}[4][1=, 2=, 3=, 4=]
  {\mthelm[\argdef{#4}{\mthstyletterelm}]{R#3}[#1][#2]}
\newcommandx{\SElm}[4][1=, 2=, 3=, 4=]
  {\mthelm[\argdef{#4}{\mthstyletterelm}]{S#3}[#1][#2]}
\newcommandx{\TElm}[4][1=, 2=, 3=, 4=]
  {\mthelm[\argdef{#4}{\mthstyletterelm}]{T#3}[#1][#2]}
\newcommandx{\UElm}[4][1=, 2=, 3=, 4=]
  {\mthelm[\argdef{#4}{\mthstyletterelm}]{U#3}[#1][#2]}
\newcommandx{\VElm}[4][1=, 2=, 3=, 4=]
  {\mthelm[\argdef{#4}{\mthstyletterelm}]{V#3}[#1][#2]}
\newcommandx{\WElm}[4][1=, 2=, 3=, 4=]
  {\mthelm[\argdef{#4}{\mthstyletterelm}]{W#3}[#1][#2]}
\newcommandx{\XElm}[4][1=, 2=, 3=, 4=]
  {\mthelm[\argdef{#4}{\mthstyletterelm}]{X#3}[#1][#2]}
\newcommandx{\YElm}[4][1=, 2=, 3=, 4=]
  {\mthelm[\argdef{#4}{\mthstyletterelm}]{Y#3}[#1][#2]}
\newcommandx{\ZElm}[4][1=, 2=, 3=, 4=]
  {\mthelm[\argdef{#4}{\mthstyletterelm}]{Z#3}[#1][#2]}

\newcommandx{\aElm}[4][1=, 2=, 3=, 4=]
  {\mthelm[\argdef{#4}{\mthstyletterelm}]{a#3}[#1][#2]}
\newcommandx{\bElm}[4][1=, 2=, 3=, 4=]
  {\mthelm[\argdef{#4}{\mthstyletterelm}]{b#3}[#1][#2]}
\newcommandx{\cElm}[4][1=, 2=, 3=, 4=]
  {\mthelm[\argdef{#4}{\mthstyletterelm}]{c#3}[#1][#2]}
\newcommandx{\dElm}[4][1=, 2=, 3=, 4=]
  {\mthelm[\argdef{#4}{\mthstyletterelm}]{d#3}[#1][#2]}
\newcommandx{\eElm}[4][1=, 2=, 3=, 4=]
  {\mthelm[\argdef{#4}{\mthstyletterelm}]{e#3}[#1][#2]}
\newcommandx{\fElm}[4][1=, 2=, 3=, 4=]
  {\mthelm[\argdef{#4}{\mthstyletterelm}]{f#3}[#1][#2]}
\newcommandx{\gElm}[4][1=, 2=, 3=, 4=]
  {\mthelm[\argdef{#4}{\mthstyletterelm}]{g#3}[#1][#2]}
\newcommandx{\hElm}[4][1=, 2=, 3=, 4=]
  {\mthelm[\argdef{#4}{\mthstyletterelm}]{h#3}[#1][#2]}
\newcommandx{\iElm}[4][1=, 2=, 3=, 4=]
  {\mthelm[\argdef{#4}{\mthstyletterelm}]{i#3}[#1][#2]}
\newcommandx{\jElm}[4][1=, 2=, 3=, 4=]
  {\mthelm[\argdef{#4}{\mthstyletterelm}]{j#3}[#1][#2]}
\newcommandx{\kElm}[4][1=, 2=, 3=, 4=]
  {\mthelm[\argdef{#4}{\mthstyletterelm}]{k#3}[#1][#2]}
\newcommandx{\lElm}[4][1=, 2=, 3=, 4=]
  {\mthelm[\argdef{#4}{\mthstyletterelm}]{l#3}[#1][#2]}
\newcommandx{\mElm}[4][1=, 2=, 3=, 4=]
  {\mthelm[\argdef{#4}{\mthstyletterelm}]{m#3}[#1][#2]}
\newcommandx{\nElm}[4][1=, 2=, 3=, 4=]
  {\mthelm[\argdef{#4}{\mthstyletterelm}]{n#3}[#1][#2]}
\newcommandx{\oElm}[4][1=, 2=, 3=, 4=]
  {\mthelm[\argdef{#4}{\mthstyletterelm}]{o#3}[#1][#2]}
\newcommandx{\pElm}[4][1=, 2=, 3=, 4=]
  {\mthelm[\argdef{#4}{\mthstyletterelm}]{p#3}[#1][#2]}
\newcommandx{\qElm}[4][1=, 2=, 3=, 4=]
  {\mthelm[\argdef{#4}{\mthstyletterelm}]{q#3}[#1][#2]}
\newcommandx{\rElm}[4][1=, 2=, 3=, 4=]
  {\mthelm[\argdef{#4}{\mthstyletterelm}]{r#3}[#1][#2]}
\newcommandx{\sElm}[4][1=, 2=, 3=, 4=]
  {\mthelm[\argdef{#4}{\mthstyletterelm}]{s#3}[#1][#2]}
\newcommandx{\tElm}[4][1=, 2=, 3=, 4=]
  {\mthelm[\argdef{#4}{\mthstyletterelm}]{t#3}[#1][#2]}
\newcommandx{\uElm}[4][1=, 2=, 3=, 4=]
  {\mthelm[\argdef{#4}{\mthstyletterelm}]{u#3}[#1][#2]}
\newcommandx{\vElm}[4][1=, 2=, 3=, 4=]
  {\mthelm[\argdef{#4}{\mthstyletterelm}]{v#3}[#1][#2]}
\newcommandx{\wElm}[4][1=, 2=, 3=, 4=]
  {\mthelm[\argdef{#4}{\mthstyletterelm}]{w#3}[#1][#2]}
\newcommandx{\xElm}[4][1=, 2=, 3=, 4=]
  {\mthelm[\argdef{#4}{\mthstyletterelm}]{x#3}[#1][#2]}
\newcommandx{\yElm}[4][1=, 2=, 3=, 4=]
  {\mthelm[\argdef{#4}{\mthstyletterelm}]{y#3}[#1][#2]}
\newcommandx{\zElm}[4][1=, 2=, 3=, 4=]
  {\mthelm[\argdef{#4}{\mthstyletterelm}]{z#3}[#1][#2]}

%% file: Macro/Text.tex
\newcommand{\afortiori}
  {\txtabr{a fortiori}}

\newcommand{\divideetimpera}
  {\txtabr{divide et impera}}

\newcommand{\eg}
  {\txtabr{e.g.}}

\newcommand{\etal}
  {\txtabr{et al.}}

\newcommand{\ie}
  {\txtabr{i.e.}}

\newcommand{\viceversa}
  {\txtabr{vice versa}}

\renewcommand{\iff}
  {\txtabr{iff}}

\newcommand{\resp}
  {\txtabr{resp.}}

\newcommand{\wrt}
  {\txtabr{w.r.t.}}

%% file: Macro/Math.tex
\newcommand{\defeq}
  {\ensuremath{\triangleq}}

\newcommand{\lst}
  {\mthargfun{lst}}

\newcommand{\dual}[1]
  {\mthempty{\overline{#1}}}

\newcommand{\der}[1]
  {\mthempty{\widehat{#1}}}
\newcommand{\trn}[1]
  {\mthempty{\overline{#1}}}

\newcommand{\tuple}[1]
  {\ensuremath{\!\argint{\langle}{#1}{\rangle}}}

\newcommand{\tupleb}[2]
  {\tuple{\argb{#1}{#2}}}
\newcommand{\tuplec}[3]
  {\tuple{\argc{#1}{#2}{#3}}}
\newcommand{\tupled}[4]
  {\tuple{\argd{#1}{#2}{#3}{#4}}}
\newcommand{\tuplee}[5]
  {\tuple{\arge{#1}{#2}{#3}{#4}{#5}}}
\newcommand{\tuplef}[6]
  {\tuple{\argf{#1}{#2}{#3}{#4}{#5}{#6}}}
\newcommand{\tupleg}[7]
  {\tuple{\argg{#1}{#2}{#3}{#4}{#5}{#6}{#7}}}
\newcommand{\tupleh}[8]
  {\tuple{\argh{#1}{#2}{#3}{#4}{#5}{#6}{#7}{#8}}}
\newcommand{\tuplei}[9]
  {\tuple{\argi{#1}{#2}{#3}{#4}{#5}{#6}{#7}{#8}{#9}}}

\newcommand{\tuplej}[9]
  {%
  \def\defarga{#1}%
  \def\defargb{#2}%
  \def\defargc{#3}%
  \def\defargd{#4}%
  \def\defarge{#5}%
  \def\defargf{#6}%
  \def\defargg{#7}%
  \def\defargh{#8}%
  \def\defargi{#9}%
  \tupleauxj%
  }
\newcommand{\tuplek}[9]
  {%
  \def\defarga{#1}%
  \def\defargb{#2}%
  \def\defargc{#3}%
  \def\defargd{#4}%
  \def\defarge{#5}%
  \def\defargf{#6}%
  \def\defargg{#7}%
  \def\defargh{#8}%
  \def\defargi{#9}%
  \tupleauxk%
  }
\newcommand{\tuplel}[9]
  {%
  \def\defarga{#1}%
  \def\defargb{#2}%
  \def\defargc{#3}%
  \def\defargd{#4}%
  \def\defarge{#5}%
  \def\defargf{#6}%
  \def\defargg{#7}%
  \def\defargh{#8}%
  \def\defargi{#9}%
  \tupleauxl%
  }
\newcommand{\tuplem}[9]
  {%
  \def\defarga{#1}%
  \def\defargb{#2}%
  \def\defargc{#3}%
  \def\defargd{#4}%
  \def\defarge{#5}%
  \def\defargf{#6}%
  \def\defargg{#7}%
  \def\defargh{#8}%
  \def\defargi{#9}%
  \tupleauxm%
  }
\newcommand{\tuplen}[9]
  {%
  \def\defarga{#1}%
  \def\defargb{#2}%
  \def\defargc{#3}%
  \def\defargd{#4}%
  \def\defarge{#5}%
  \def\defargf{#6}%
  \def\defargg{#7}%
  \def\defargh{#8}%
  \def\defargi{#9}%
  \tupleauxn%
  }
\newcommand{\tupleo}[9]
  {%
  \def\defarga{#1}%
  \def\defargb{#2}%
  \def\defargc{#3}%
  \def\defargd{#4}%
  \def\defarge{#5}%
  \def\defargf{#6}%
  \def\defargg{#7}%
  \def\defargh{#8}%
  \def\defargi{#9}%
  \tupleauxo%
  }
\newcommand{\tuplep}[9]
  {%
  \def\defarga{#1}%
  \def\defargb{#2}%
  \def\defargc{#3}%
  \def\defargd{#4}%
  \def\defarge{#5}%
  \def\defargf{#6}%
  \def\defargg{#7}%
  \def\defargh{#8}%
  \def\defargi{#9}%
  \tupleauxp%
  }
\newcommand{\tupleq}[9]
  {%
  \def\defarga{#1}%
  \def\defargb{#2}%
  \def\defargc{#3}%
  \def\defargd{#4}%
  \def\defarge{#5}%
  \def\defargf{#6}%
  \def\defargg{#7}%
  \def\defargh{#8}%
  \def\defargi{#9}%
  \tupleauxq%
  }
\newcommand{\tupler}[9]
  {%
  \def\defarga{#1}%
  \def\defargb{#2}%
  \def\defargc{#3}%
  \def\defargd{#4}%
  \def\defarge{#5}%
  \def\defargf{#6}%
  \def\defargg{#7}%
  \def\defargh{#8}%
  \def\defargi{#9}%
  \tupleauxr%
  }

\newcommand{\tupleauxj}[1]
  {%
  \tuple{\argj{\defarga}{\defargb}{\defargc}{\defargd}{\defarge}{\defargf}%
    {\defargg}{\defargh}{\defargi}{#1}}%
  }
\newcommand{\tupleauxk}[2]
  {%
  \tuple{\argk{\defarga}{\defargb}{\defargc}{\defargd}{\defarge}{\defargf}%
    {\defargg}{\defargh}{\defargi}{#1}{#2}}%
  }
\newcommand{\tupleauxl}[3]
  {%
  \tuple{\argl{\defarga}{\defargb}{\defargc}{\defargd}{\defarge}{\defargf}%
    {\defargg}{\defargh}{\defargi}{#1}{#2}{#3}}%
  }
\newcommand{\tupleauxm}[4]
  {%
  \tuple{\argm{\defarga}{\defargb}{\defargc}{\defargd}{\defarge}{\defargf}%
    {\defargg}{\defargh}{\defargi}{#1}{#2}{#3}{#4}}%
  }
\newcommand{\tupleauxn}[5]
  {%
  \tuple{\argn{\defarga}{\defargb}{\defargc}{\defargd}{\defarge}{\defargf}%
    {\defargg}{\defargh}{\defargi}{#1}{#2}{#3}{#4}{#5}}%
  }
\newcommand{\tupleauxo}[6]
  {%
  \tuple{\argo{\defarga}{\defargb}{\defargc}{\defargd}{\defarge}{\defargf}%
    {\defargg}{\defargh}{\defargi}{#1}{#2}{#3}{#4}{#5}{#6}}%
  }
\newcommand{\tupleauxp}[7]
  {%
  \tuple{\argp{\defarga}{\defargb}{\defargc}{\defargd}{\defarge}{\defargf}%
    {\defargg}{\defargh}{\defargi}{#1}{#2}{#3}{#4}{#5}{#6}{#7}}%
  }
\newcommand{\tupleauxq}[8]
  {%
  \tuple{\argq{\defarga}{\defargb}{\defargc}{\defargd}{\defarge}{\defargf}%
    {\defargg}{\defargh}{\defargi}{#1}{#2}{#3}{#4}{#5}{#6}{#7}{#8}}%
  }
\newcommand{\tupleauxr}[9]
  {%
  \tuple{\argr{\defarga}{\defargb}{\defargc}{\defargd}{\defarge}{\defargf}%
    {\defargg}{\defargh}{\defargi}{#1}{#2}{#3}{#4}{#5}{#6}{#7}{#8}{#9}}%
  }

\newcommand{\tuplecx}[3]
  {%
  \def\defarga{#1}%
  \def\defargb{#2}%
  \def\defargc{#3}%
  \argsubsup{\tupleauxcx}%
  }
\newcommand{\tupledx}[4]
  {%
  \def\defarga{#1}%
  \def\defargb{#2}%
  \def\defargc{#3}%
  \def\defargd{#4}%
  \argsubsup{\tupleauxdx}%
  }
\newcommand{\tupleex}[5]
  {%
  \def\defarga{#1}%
  \def\defargb{#2}%
  \def\defargc{#3}%
  \def\defargd{#4}%
  \def\defarge{#5}%
  \argsubsup{\tupleauxex}%
  }
\newcommand{\tuplefx}[6]
  {%
  \def\defarga{#1}%
  \def\defargb{#2}%
  \def\defargc{#3}%
  \def\defargd{#4}%
  \def\defarge{#5}%
  \def\defargf{#6}%
  \argsubsup{\tupleauxfx}%
  }
\newcommand{\tuplegx}[7]
  {%
  \def\defarga{#1}%
  \def\defargb{#2}%
  \def\defargc{#3}%
  \def\defargd{#4}%
  \def\defarge{#5}%
  \def\defargf{#6}%
  \def\defargg{#7}%
  \argsubsup{\tupleauxgx}%
  }
\newcommand{\tuplehx}[8]
  {%
  \def\defarga{#1}%
  \def\defargb{#2}%
  \def\defargc{#3}%
  \def\defargd{#4}%
  \def\defarge{#5}%
  \def\defargf{#6}%
  \def\defargg{#7}%
  \def\defargh{#8}%
  \argsubsup{\tupleauxhx}%
  }
\newcommand{\tupleix}[9]
  {%
  \def\defarga{#1}%
  \def\defargb{#2}%
  \def\defargc{#3}%
  \def\defargd{#4}%
  \def\defarge{#5}%
  \def\defargf{#6}%
  \def\defargg{#7}%
  \def\defargh{#8}%
  \def\defargi{#9}%
  \argsubsup{\tupleauxix}%
  }

\newcommandx{\tupleauxbx}[2][1=, 2=]
  {%
  \tupleb
    {\argdef{#1}{\defarga[\argsubscript][\argsuperscript]}}
    {\argdef{#2}{\defargb[\argsubscript][\argsuperscript]}}%
  }
\newcommandx{\tupleauxcx}[3][1=, 2=, 3=]
  {%
  \tuplec
    {\argdef{#1}{\defarga[\argsubscript][\argsuperscript]}}
    {\argdef{#2}{\defargb[\argsubscript][\argsuperscript]}}
    {\argdef{#3}{\defargc[\argsubscript][\argsuperscript]}}%
  }
\newcommandx{\tupleauxdx}[4][1=, 2=, 3=, 4=]
  {%
  \tupled
    {\argdef{#1}{\defarga[\argsubscript][\argsuperscript]}}
    {\argdef{#2}{\defargb[\argsubscript][\argsuperscript]}}
    {\argdef{#3}{\defargc[\argsubscript][\argsuperscript]}}
    {\argdef{#4}{\defargd[\argsubscript][\argsuperscript]}}%
  }
\newcommandx{\tupleauxex}[5][1=, 2=, 3=, 4=, 5=]
  {%
  \tuplee
    {\argdef{#1}{\defarga[\argsubscript][\argsuperscript]}}
    {\argdef{#2}{\defargb[\argsubscript][\argsuperscript]}}
    {\argdef{#3}{\defargc[\argsubscript][\argsuperscript]}}
    {\argdef{#4}{\defargd[\argsubscript][\argsuperscript]}}
    {\argdef{#5}{\defarge[\argsubscript][\argsuperscript]}}%
  }
\newcommandx{\tupleauxfx}[6][1=, 2=, 3=, 4=, 5=, 6=]
  {%
  \tuplef
    {\argdef{#1}{\defarga[\argsubscript][\argsuperscript]}}
    {\argdef{#2}{\defargb[\argsubscript][\argsuperscript]}}
    {\argdef{#3}{\defargc[\argsubscript][\argsuperscript]}}
    {\argdef{#4}{\defargd[\argsubscript][\argsuperscript]}}
    {\argdef{#5}{\defarge[\argsubscript][\argsuperscript]}}
    {\argdef{#6}{\defargf[\argsubscript][\argsuperscript]}}%
  }
\newcommandx{\tupleauxgx}[7][1=, 2=, 3=, 4=, 5=, 6=, 7=]
  {%
  \tupleg
    {\argdef{#1}{\defarga[\argsubscript][\argsuperscript]}}
    {\argdef{#2}{\defargb[\argsubscript][\argsuperscript]}}
    {\argdef{#3}{\defargc[\argsubscript][\argsuperscript]}}
    {\argdef{#4}{\defargd[\argsubscript][\argsuperscript]}}
    {\argdef{#5}{\defarge[\argsubscript][\argsuperscript]}}
    {\argdef{#6}{\defargf[\argsubscript][\argsuperscript]}}
    {\argdef{#7}{\defargg[\argsubscript][\argsuperscript]}}%
  }
\newcommandx{\tupleauxhx}[8][1=, 2=, 3=, 4=, 5=, 6=, 7=, 8=]
  {%
  \tupleh
    {\argdef{#1}{\defarga[\argsubscript][\argsuperscript]}}
    {\argdef{#2}{\defargb[\argsubscript][\argsuperscript]}}
    {\argdef{#3}{\defargc[\argsubscript][\argsuperscript]}}
    {\argdef{#4}{\defargd[\argsubscript][\argsuperscript]}}
    {\argdef{#5}{\defarge[\argsubscript][\argsuperscript]}}
    {\argdef{#6}{\defargf[\argsubscript][\argsuperscript]}}
    {\argdef{#7}{\defargg[\argsubscript][\argsuperscript]}}
    {\argdef{#8}{\defargh[\argsubscript][\argsuperscript]}}%
  }
\newcommandx{\tupleauxix}[9][1=, 2=, 3=, 4=, 5=, 6=, 7=, 8=, 9=]
  {%
  \tuplei
    {\argdef{#1}{\defarga[\argsubscript][\argsuperscript]}}
    {\argdef{#2}{\defargb[\argsubscript][\argsuperscript]}}
    {\argdef{#3}{\defargc[\argsubscript][\argsuperscript]}}
    {\argdef{#4}{\defargd[\argsubscript][\argsuperscript]}}
    {\argdef{#5}{\defarge[\argsubscript][\argsuperscript]}}
    {\argdef{#6}{\defargf[\argsubscript][\argsuperscript]}}
    {\argdef{#7}{\defargg[\argsubscript][\argsuperscript]}}
    {\argdef{#8}{\defargh[\argsubscript][\argsuperscript]}}
    {\argdef{#9}{\defargi[\argsubscript][\argsuperscript]}}%
  }

\newcommand{\tuplejx}[9]
  {%
  \def\tuplearga{#1}%
  \def\tupleargb{#2}%
  \def\tupleargc{#3}%
  \def\tupleargd{#4}%
  \def\tuplearge{#5}%
  \def\tupleargf{#6}%
  \def\tupleargg{#7}%
  \def\tupleargh{#8}%
  \def\tupleargi{#9}%
  \argsubsup{\tupleauxjx}%
  }

\newcommand{\tupleauxjx}[1]
  {%
  \def\tupleargj{#1}%
  \argsubsup{\tupleauxxjx}%
  }

\newcommandx{\tupleauxxjx}[9][1=, 2=, 3=, 4=, 5=, 6=, 7=, 8=, 9=]
  {%
  \def\optarga{#1}%
  \def\optargb{#2}%
  \def\optargc{#3}%
  \def\optargd{#4}%
  \def\optarge{#5}%
  \def\optargf{#6}%
  \def\optargg{#7}%
  \def\optargh{#8}%
  \def\optargi{#9}%
  \tupleauxxxjx%
  }
\newcommandx{\tupleauxxkx}[9][1=, 2=, 3=, 4=, 5=, 6=, 7=, 8=, 9=]
  {%
  \def\optarga{#1}%
  \def\optargb{#2}%
  \def\optargc{#3}%
  \def\optargd{#4}%
  \def\optarge{#5}%
  \def\optargf{#6}%
  \def\optargg{#7}%
  \def\optargh{#8}%
  \def\optargi{#9}%
  \tupleauxxxkx%
  }
\newcommandx{\tupleauxxlx}[9][1=, 2=, 3=, 4=, 5=, 6=, 7=, 8=, 9=]
  {%
  \def\optarga{#1}%
  \def\optargb{#2}%
  \def\optargc{#3}%
  \def\optargd{#4}%
  \def\optarge{#5}%
  \def\optargf{#6}%
  \def\optargg{#7}%
  \def\optargh{#8}%
  \def\optargi{#9}%
  \tupleauxxxlx%
  }
\newcommandx{\tupleauxxmx}[9][1=, 2=, 3=, 4=, 5=, 6=, 7=, 8=, 9=]
  {%
  \def\optarga{#1}%
  \def\optargb{#2}%
  \def\optargc{#3}%
  \def\optargd{#4}%
  \def\optarge{#5}%
  \def\optargf{#6}%
  \def\optargg{#7}%
  \def\optargh{#8}%
  \def\optargi{#9}%
  \tupleauxxxmx%
  }
\newcommandx{\tupleauxxnx}[9][1=, 2=, 3=, 4=, 5=, 6=, 7=, 8=, 9=]
  {%
  \def\optarga{#1}%
  \def\optargb{#2}%
  \def\optargc{#3}%
  \def\optargd{#4}%
  \def\optarge{#5}%
  \def\optargf{#6}%
  \def\optargg{#7}%
  \def\optargh{#8}%
  \def\optargi{#9}%
  \tupleauxxxnx%
  }
\newcommandx{\tupleauxxox}[9][1=, 2=, 3=, 4=, 5=, 6=, 7=, 8=, 9=]
  {%
  \def\optarga{#1}%
  \def\optargb{#2}%
  \def\optargc{#3}%
  \def\optargd{#4}%
  \def\optarge{#5}%
  \def\optargf{#6}%
  \def\optargg{#7}%
  \def\optargh{#8}%
  \def\optargi{#9}%
  \tupleauxxxox%
  }
\newcommandx{\tupleauxxpx}[9][1=, 2=, 3=, 4=, 5=, 6=, 7=, 8=, 9=]
  {%
  \def\optarga{#1}%
  \def\optargb{#2}%
  \def\optargc{#3}%
  \def\optargd{#4}%
  \def\optarge{#5}%
  \def\optargf{#6}%
  \def\optargg{#7}%
  \def\optargh{#8}%
  \def\optargi{#9}%
  \tupleauxxxpx%
  }
\newcommandx{\tupleauxxqx}[9][1=, 2=, 3=, 4=, 5=, 6=, 7=, 8=, 9=]
  {%
  \def\optarga{#1}%
  \def\optargb{#2}%
  \def\optargc{#3}%
  \def\optargd{#4}%
  \def\optarge{#5}%
  \def\optargf{#6}%
  \def\optargg{#7}%
  \def\optargh{#8}%
  \def\optargi{#9}%
  \tupleauxxxqx%
  }
\newcommandx{\tupleauxxrx}[9][1=, 2=, 3=, 4=, 5=, 6=, 7=, 8=, 9=]
  {%
  \def\optarga{#1}%
  \def\optargb{#2}%
  \def\optargc{#3}%
  \def\optargd{#4}%
  \def\optarge{#5}%
  \def\optargf{#6}%
  \def\optargg{#7}%
  \def\optargh{#8}%
  \def\optargi{#9}%
  \tupleauxxxrx%
  }

\newcommandx{\tupleauxxxjx}[1][1=]
  {%
  \tuplej
    {\argdef{\optarga}{\tuplearga[\argsubscript][\argsuperscript]}}
    {\argdef{\optargb}{\tupleargb[\argsubscript][\argsuperscript]}}
    {\argdef{\optargc}{\tupleargc[\argsubscript][\argsuperscript]}}
    {\argdef{\optargd}{\tupleargd[\argsubscript][\argsuperscript]}}
    {\argdef{\optarge}{\tuplearge[\argsubscript][\argsuperscript]}}
    {\argdef{\optargf}{\tupleargf[\argsubscript][\argsuperscript]}}
    {\argdef{\optargg}{\tupleargg[\argsubscript][\argsuperscript]}}
    {\argdef{\optargh}{\tupleargh[\argsubscript][\argsuperscript]}}
    {\argdef{\optargi}{\tupleargi[\argsubscript][\argsuperscript]}}
    {\argdef{#1}{\tupleargj[\argsubscript][\argsuperscript]}}%
  }
\newcommandx{\tupleauxxxkx}[2][1=, 2=]
  {%
  \tuplek
    {\argdef{\optarga}{\tuplearga[\argsubscript][\argsuperscript]}}
    {\argdef{\optargb}{\tupleargb[\argsubscript][\argsuperscript]}}
    {\argdef{\optargc}{\tupleargc[\argsubscript][\argsuperscript]}}
    {\argdef{\optargd}{\tupleargd[\argsubscript][\argsuperscript]}}
    {\argdef{\optarge}{\tuplearge[\argsubscript][\argsuperscript]}}
    {\argdef{\optargf}{\tupleargf[\argsubscript][\argsuperscript]}}
    {\argdef{\optargg}{\tupleargg[\argsubscript][\argsuperscript]}}
    {\argdef{\optargh}{\tupleargh[\argsubscript][\argsuperscript]}}
    {\argdef{\optargi}{\tupleargi[\argsubscript][\argsuperscript]}}
    {\argdef{#1}{\tupleargj[\argsubscript][\argsuperscript]}}
    {\argdef{#2}{\tupleargk[\argsubscript][\argsuperscript]}}
  }
\newcommandx{\tupleauxxxlx}[3][1=, 2=, 3=]
  {%
  \tuplel
    {\argdef{\optarga}{\tuplearga[\argsubscript][\argsuperscript]}}
    {\argdef{\optargb}{\tupleargb[\argsubscript][\argsuperscript]}}
    {\argdef{\optargc}{\tupleargc[\argsubscript][\argsuperscript]}}
    {\argdef{\optargd}{\tupleargd[\argsubscript][\argsuperscript]}}
    {\argdef{\optarge}{\tuplearge[\argsubscript][\argsuperscript]}}
    {\argdef{\optargf}{\tupleargf[\argsubscript][\argsuperscript]}}
    {\argdef{\optargg}{\tupleargg[\argsubscript][\argsuperscript]}}
    {\argdef{\optargh}{\tupleargh[\argsubscript][\argsuperscript]}}
    {\argdef{\optargi}{\tupleargi[\argsubscript][\argsuperscript]}}
    {\argdef{#1}{\tupleargj[\argsubscript][\argsuperscript]}}
    {\argdef{#2}{\tupleargk[\argsubscript][\argsuperscript]}}
    {\argdef{#3}{\tupleargl[\argsubscript][\argsuperscript]}}
  }
\newcommandx{\tupleauxxxmx}[4][1=, 2=, 3=, 4=]
  {%
  \tuplem
    {\argdef{\optarga}{\tuplearga[\argsubscript][\argsuperscript]}}
    {\argdef{\optargb}{\tupleargb[\argsubscript][\argsuperscript]}}
    {\argdef{\optargc}{\tupleargc[\argsubscript][\argsuperscript]}}
    {\argdef{\optargd}{\tupleargd[\argsubscript][\argsuperscript]}}
    {\argdef{\optarge}{\tuplearge[\argsubscript][\argsuperscript]}}
    {\argdef{\optargf}{\tupleargf[\argsubscript][\argsuperscript]}}
    {\argdef{\optargg}{\tupleargg[\argsubscript][\argsuperscript]}}
    {\argdef{\optargh}{\tupleargh[\argsubscript][\argsuperscript]}}
    {\argdef{\optargi}{\tupleargi[\argsubscript][\argsuperscript]}}
    {\argdef{#1}{\tupleargj[\argsubscript][\argsuperscript]}}
    {\argdef{#2}{\tupleargk[\argsubscript][\argsuperscript]}}
    {\argdef{#3}{\tupleargl[\argsubscript][\argsuperscript]}}
    {\argdef{#4}{\tupleargm[\argsubscript][\argsuperscript]}}
  }
\newcommandx{\tupleauxxxnx}[5][1=, 2=, 3=, 4=, 5=]
  {%
  \tuplen
    {\argdef{\optarga}{\tuplearga[\argsubscript][\argsuperscript]}}
    {\argdef{\optargb}{\tupleargb[\argsubscript][\argsuperscript]}}
    {\argdef{\optargc}{\tupleargc[\argsubscript][\argsuperscript]}}
    {\argdef{\optargd}{\tupleargd[\argsubscript][\argsuperscript]}}
    {\argdef{\optarge}{\tuplearge[\argsubscript][\argsuperscript]}}
    {\argdef{\optargf}{\tupleargf[\argsubscript][\argsuperscript]}}
    {\argdef{\optargg}{\tupleargg[\argsubscript][\argsuperscript]}}
    {\argdef{\optargh}{\tupleargh[\argsubscript][\argsuperscript]}}
    {\argdef{\optargi}{\tupleargi[\argsubscript][\argsuperscript]}}
    {\argdef{#1}{\tupleargj[\argsubscript][\argsuperscript]}}
    {\argdef{#2}{\tupleargk[\argsubscript][\argsuperscript]}}
    {\argdef{#3}{\tupleargl[\argsubscript][\argsuperscript]}}
    {\argdef{#4}{\tupleargm[\argsubscript][\argsuperscript]}}
    {\argdef{#5}{\tupleargn[\argsubscript][\argsuperscript]}}
  }
\newcommandx{\tupleauxxxox}[6][1=, 2=, 3=, 4=, 5=, 6=]
  {%
  \tupleo
    {\argdef{\optarga}{\tuplearga[\argsubscript][\argsuperscript]}}
    {\argdef{\optargb}{\tupleargb[\argsubscript][\argsuperscript]}}
    {\argdef{\optargc}{\tupleargc[\argsubscript][\argsuperscript]}}
    {\argdef{\optargd}{\tupleargd[\argsubscript][\argsuperscript]}}
    {\argdef{\optarge}{\tuplearge[\argsubscript][\argsuperscript]}}
    {\argdef{\optargf}{\tupleargf[\argsubscript][\argsuperscript]}}
    {\argdef{\optargg}{\tupleargg[\argsubscript][\argsuperscript]}}
    {\argdef{\optargh}{\tupleargh[\argsubscript][\argsuperscript]}}
    {\argdef{\optargi}{\tupleargi[\argsubscript][\argsuperscript]}}
    {\argdef{#1}{\tupleargj[\argsubscript][\argsuperscript]}}
    {\argdef{#2}{\tupleargk[\argsubscript][\argsuperscript]}}
    {\argdef{#3}{\tupleargl[\argsubscript][\argsuperscript]}}
    {\argdef{#4}{\tupleargm[\argsubscript][\argsuperscript]}}
    {\argdef{#5}{\tupleargn[\argsubscript][\argsuperscript]}}
    {\argdef{#6}{\tupleargo[\argsubscript][\argsuperscript]}}
  }
\newcommandx{\tupleauxxxpx}[7][1=, 2=, 3=, 4=, 5=, 6=, 7=]
  {%
  \tuplep
    {\argdef{\optarga}{\tuplearga[\argsubscript][\argsuperscript]}}
    {\argdef{\optargb}{\tupleargb[\argsubscript][\argsuperscript]}}
    {\argdef{\optargc}{\tupleargc[\argsubscript][\argsuperscript]}}
    {\argdef{\optargd}{\tupleargd[\argsubscript][\argsuperscript]}}
    {\argdef{\optarge}{\tuplearge[\argsubscript][\argsuperscript]}}
    {\argdef{\optargf}{\tupleargf[\argsubscript][\argsuperscript]}}
    {\argdef{\optargg}{\tupleargg[\argsubscript][\argsuperscript]}}
    {\argdef{\optargh}{\tupleargh[\argsubscript][\argsuperscript]}}
    {\argdef{\optargi}{\tupleargi[\argsubscript][\argsuperscript]}}
    {\argdef{#1}{\tupleargj[\argsubscript][\argsuperscript]}}
    {\argdef{#2}{\tupleargk[\argsubscript][\argsuperscript]}}
    {\argdef{#3}{\tupleargl[\argsubscript][\argsuperscript]}}
    {\argdef{#4}{\tupleargm[\argsubscript][\argsuperscript]}}
    {\argdef{#5}{\tupleargn[\argsubscript][\argsuperscript]}}
    {\argdef{#6}{\tupleargo[\argsubscript][\argsuperscript]}}
    {\argdef{#7}{\tupleargp[\argsubscript][\argsuperscript]}}
  }
\newcommandx{\tupleauxxxqx}[8][1=, 2=, 3=, 4=, 5=, 6=, 7=, 8=]
  {%
  \tupleq
    {\argdef{\optarga}{\tuplearga[\argsubscript][\argsuperscript]}}
    {\argdef{\optargb}{\tupleargb[\argsubscript][\argsuperscript]}}
    {\argdef{\optargc}{\tupleargc[\argsubscript][\argsuperscript]}}
    {\argdef{\optargd}{\tupleargd[\argsubscript][\argsuperscript]}}
    {\argdef{\optarge}{\tuplearge[\argsubscript][\argsuperscript]}}
    {\argdef{\optargf}{\tupleargf[\argsubscript][\argsuperscript]}}
    {\argdef{\optargg}{\tupleargg[\argsubscript][\argsuperscript]}}
    {\argdef{\optargh}{\tupleargh[\argsubscript][\argsuperscript]}}
    {\argdef{\optargi}{\tupleargi[\argsubscript][\argsuperscript]}}
    {\argdef{#1}{\tupleargj[\argsubscript][\argsuperscript]}}
    {\argdef{#2}{\tupleargk[\argsubscript][\argsuperscript]}}
    {\argdef{#3}{\tupleargl[\argsubscript][\argsuperscript]}}
    {\argdef{#4}{\tupleargm[\argsubscript][\argsuperscript]}}
    {\argdef{#5}{\tupleargn[\argsubscript][\argsuperscript]}}
    {\argdef{#6}{\tupleargo[\argsubscript][\argsuperscript]}}
    {\argdef{#7}{\tupleargp[\argsubscript][\argsuperscript]}}
    {\argdef{#8}{\tupleargq[\argsubscript][\argsuperscript]}}
  }
\newcommandx{\tupleauxxxrx}[9][1=, 2=, 3=, 4=, 5=, 6=, 7=, 8=, 9=]
  {%
  \tupler
    {\argdef{\optarga}{\tuplearga[\argsubscript][\argsuperscript]}}
    {\argdef{\optargb}{\tupleargb[\argsubscript][\argsuperscript]}}
    {\argdef{\optargc}{\tupleargc[\argsubscript][\argsuperscript]}}
    {\argdef{\optargd}{\tupleargd[\argsubscript][\argsuperscript]}}
    {\argdef{\optarge}{\tuplearge[\argsubscript][\argsuperscript]}}
    {\argdef{\optargf}{\tupleargf[\argsubscript][\argsuperscript]}}
    {\argdef{\optargg}{\tupleargg[\argsubscript][\argsuperscript]}}
    {\argdef{\optargh}{\tupleargh[\argsubscript][\argsuperscript]}}
    {\argdef{\optargi}{\tupleargi[\argsubscript][\argsuperscript]}}
    {\argdef{#1}{\tupleargj[\argsubscript][\argsuperscript]}}
    {\argdef{#2}{\tupleargk[\argsubscript][\argsuperscript]}}
    {\argdef{#3}{\tupleargl[\argsubscript][\argsuperscript]}}
    {\argdef{#4}{\tupleargm[\argsubscript][\argsuperscript]}}
    {\argdef{#5}{\tupleargn[\argsubscript][\argsuperscript]}}
    {\argdef{#6}{\tupleargo[\argsubscript][\argsuperscript]}}
    {\argdef{#7}{\tupleargp[\argsubscript][\argsuperscript]}}
    {\argdef{#8}{\tupleargq[\argsubscript][\argsuperscript]}}
    {\argdef{#9}{\tupleargr[\argsubscript][\argsuperscript]}}%
  }

\newcommand{\set}[2]
  {
  \argint
    {\{}
    {\argext{#1}{\!\allowbreak\,:\allowbreak}{#2}}
    {\}}
  }

\newcommand{\pow}[1]
  {\ensuremath{2^{#1}}}

\newcommand{\card}[1]
  {\mthempty{\argint{\left\vert}{#1}{\right\vert}}}

\newcommand{\ctimes}
  {\mthempty{\times}}

\newcommandx{\pto}[2][1=, 2=]
  {\ensuremath{\rightharpoonup}}

\newcommandx{\cto}[2][1=, 2=]
  {\:\mthempty{\to}[#1][#2]\:}
\newcommandx{\cpto}[2][1=, 2=]
  {\:\mthempty{\pto}[#1][#2]\:}

\newcommand{\AOmega}
  {\mthargset{\Omega}}

\newcommand{\SetN}
  {\mthset[1]{N}}

\newcommand{\numcc}[2]
  {\mthempty{[\argb{#1}{#2}]}}

\DeclareRobustCommand{\ceil}[1]
  {\mthempty{\left\lceil{#1}\right\rceil}}

\DeclareRobustCommand{\max}
  {\mthfun{max}}

%% file: Macro/Models.tex
\input{Macro/Models/Structures}

\input{Macro/Models/Graphs}

\input{Macro/Models/Games}

\input{Macro/Models/KS}

\input{Macro/Models/CGS}

\input{Macro/Models/Machines}
\input{Macro/Models/Dominoes}

%% file: Macro/Models/Structures.tex
\newcommand{\argset}{Ar}
\newcommandx{\ArgSet}[3][1=, 2=, 3=]
	{\mthset{\argset#3}[#1][#2]}

\newcommand{\argsym}{a}
\newcommandx{\argSym}[3][1=, 2=, 3=]
	{\mthsym{\argsym#3}[#1][#2]}

\newcommand{\argelm}{a}
\newcommandx{\argElm}[3][1=, 2=, 3=]
	{\mthelm{\argelm#3}[#1][#2]}

\newcommand{\relset}{Rl}
\newcommandx{\RelSet}[3][1=, 2=, 3=]
	{\mthset{\relset#3}[#1][#2]}

\newcommand{\relsym}{r}
\newcommandx{\relSym}[3][1=, 2=, 3=]
	{\mthsym{\relsym#3}[#1][#2]}

\newcommand{\relelm}{r}
\newcommandx{\relElm}[3][1=, 2=, 3=]
	{\mthelm{\relelm#3}[#1][#2]}

\newcommand{\argfun}{ar}
\newcommandx{\argFun}[4][1=, 2=, 3=, 4=]
	{\mthargfun{\argfun#4}[#1][#2]{#3}}

\newcommand{\lansig}{LS}
\newcommandx{\LanSig}[5][1=, 2=, 3=, 4=, 5=]
	{\txtargname{\lansig#5{\small\argint{$[$}{#1}{$]$}}}[#2][#3]{#4}\xspace}

\newcommand{\lansigcls}{LS}
\newcommandx{\LanSigCls}[5][1=, 2=, 3=, 4=, 5=]
	{\mthset[#5]{\lansigcls#4\text{\txtname{\small\argint{$[$}{#1}{$]$}}}}[#2]%
	[#3]}

\newcommand{\domset}{Dm}
\newcommandx{\DomSet}[3][1=, 2=, 3=]
	{\mthset{\domset#3}[#1][#2]}

\newcommand{\domsym}{d}
\newcommandx{\domSym}[3][1=, 2=, 3=]
	{\mthsym{\domsym#3}[#1][#2]}

\newcommand{\domelm}{d}
\newcommandx{\domElm}[3][1=, 2=, 3=]
	{\mthelm{\domelm#3}[#1][#2]}

\newcommand{\relfun}{rl}
\newcommandx{\relFun}[4][1=, 2=, 3=, 4=]
	{\mthargfun{\relfun#4}[#1][#2]{#3}}

\newcommand{\relstr}{RS}
\newcommandx{\RelStr}[5][1=, 2=, 3=, 4=, 5=]
	{\txtargname{\relstr#5{\small\argint{$[$}{#1}{$]$}}}[#2][#3]{#4}\xspace}

\newcommand{\relstrcls}{RS}
\newcommandx{\RelStrCls}[5][1=, 2=, 3=, 4=, 5=]
	{\mthset[#5]{\relstrcls#4\text{\txtname{\small\argint{$[$}{#1}{$]$}}}}[#2]%
	[#3]}

\newcommandx{\ordFun}[3][1=, 2=, 3=]
	{\mthempty{\argint{\left\vert}{#3}{\right\vert}}[#1][#2]}

\newcommandx{\sizFun}[3][1=, 2=, 3=]
	{\mthempty{\argint{\left\Vert}{#3}{\right\Vert}}[#1][#2]}

%% file: Macro/Models/Graphs.tex
\newcommand{\verset}{Vr}
\newcommandx{\VerSet}[3][1=, 2=, 3=]
	{\mthset{\verset#3}[#1][#2]}

\newcommand{\versym}{v}
\newcommandx{\verSym}[3][1=, 2=, 3=]
	{\mthsym{\versym#3}[#1][#2]}

\newcommand{\verelm}{v}
\newcommandx{\verElm}[3][1=, 2=, 3=]
	{\mthelm{\verelm#3}[#1][#2]}

\newcommand{\edgrel}{Ed}
\newcommandx{\EdgRel}[3][1=, 2=, 3=]
	{\mthrel{\edgrel#3}[#1][#2]}

\newcommand{\edgsym}{e}
\newcommandx{\edgSym}[3][1=, 2=, 3=]
	{\mthsym{\edgsym#3}[#1][#2]}

\newcommand{\edgelm}{e}
\newcommandx{\edgElm}[3][1=, 2=, 3=]
	{\mthelm{\edgelm#3}[#1][#2]}

\newcommand{\orgfun}{or}
\newcommandx{\orgFun}[4][1=, 2=, 3=, 4=]
	{\mthargfun{\orgfun#4}[#1][#2]{#3}}

\newcommand{\desfun}{ds}
\newcommandx{\desFun}[4][1=, 2=, 3=, 4=]
	{\mthargfun{\desfun#4}[#1][#2]{#3}}

\newcommand{\grp}{Gr}
\newcommandx{\Grp}[5][1=, 2=, 3=, 4=, 5=]
	{\txtargname{\grp#5{\small\argint{$[$}{#1}{$]$}}}[#2][#3]{#4}\xspace}

\newcommand{\grpcls}{Gr}
\newcommandx{\GrpCls}[5][1=, 2=, 3=, 4=, 5=]
	{\mthset[#5]{\grpcls#4\text{\small\txtname{\argint{$[$}{#1}{$]$}}}}[#2][#3]}

\newcommand{\pthset}{Pth}
\newcommandx{\PthSet}[3][1=, 2=, 3=]
	{\mthset{\pthset#3}[#1][#2]}

\newcommand{\pthsym}{\pi}
\newcommandx{\pthSym}[3][1=, 2=, 3=]
	{\mthsym{\pthsym#3}[#1][#2]}

\newcommand{\pthelm}{\pi}
\newcommandx{\pthElm}[3][1=, 2=, 3=]
	{\mthelm{\pthelm#3}[#1][#2]}

\newcommand{\apset}{AP}
\newcommandx{\APSet}[3][1=, 2=, 3=]
	{\mthset{\apset#3}[#1][#2]}

\newcommand{\apsym}{p}
\newcommandx{\apSym}[3][1=, 2=, 3=]
	{\mthsym{\apsym#3}[#1][#2]}

\newcommand{\apelm}{p}
\newcommandx{\apElm}[3][1=, 2=, 3=]
	{\mthelm{\apelm#3}[#1][#2]}

\newcommand{\apfun}{ap}
\newcommandx{\apFun}[4][1=, 2=, 3=, 4=]
	{\mthargfun{\apfun#4}[#1][#2]{#3}}

\newcommand{\labgrp}{L\grp}
\newcommandx{\LabGrp}[5][1=, 2=, 3=, 4=, 5=]
	{\txtargname{\labgrp#5{\small\argint{$[$}{#1}{$]$}}}[#2][#3]{#4}\xspace}

\newcommand{\labgrpcls}{L\grpcls}
\newcommandx{\LabGrpCls}[5][1=, 2=, 3=, 4=, 5=]
	{\mthset[#5]{\labgrpcls#4\text{\small\txtname{\argint{$[$}{#1}{$]$}}}}[#2]%
	[#3]}

\newcommand{\trcset}{Trc}
\newcommandx{\TrcSet}[3][1=, 2=, 3=]
	{\mthset{\trcset#3}[#1][#2]}

\newcommand{\trcsym}{\varrho}
\newcommandx{\trcSym}[3][1=, 2=, 3=]
	{\mthsym{\trcsym#3}[#1][#2]}

\newcommand{\trcelm}{\varrho}
\newcommandx{\trcElm}[3][1=, 2=, 3=]
	{\mthelm{\trcelm#3}[#1][#2]}

\newcommand{\colset}{Cl}
\newcommandx{\ColSet}[3][1=, 2=, 3=]
	{\mthset{\colset#3}[#1][#2]}

\newcommand{\colsym}{c}
\newcommandx{\colSym}[3][1=, 2=, 3=]
	{\mthsym{\colsym#3}[#1][#2]}

\newcommand{\colelm}{c}
\newcommandx{\colElm}[3][1=, 2=, 3=]
	{\mthelm{\colelm#3}[#1][#2]}

\newcommand{\colfun}{cl}
\newcommandx{\colFun}[4][1=, 2=, 3=, 4=]
	{\mthargfun{\colfun#4}[#1][#2]{#3}}

\newcommand{\colgrp}{C\grp}
\newcommandx{\ColGrp}[5][1=, 2=, 3=, 4=, 5=]
	{\txtargname{\colgrp#5{\small\argint{$[$}{#1}{$]$}}}[#2][#3]{#4}\xspace}

\newcommand{\colgrpcls}{C\grpcls}
\newcommandx{\ColGrpCls}[5][1=, 2=, 3=, 4=, 5=]
	{\mthset[#5]{\colgrpcls#4\text{\small\txtname{\argint{$[$}{#1}{$]$}}}}[#2]%
	[#3]}

\newcommand{\wghset}{Wg}
\newcommandx{\WghSet}[3][1=, 2=, 3=]
	{\mthset{\wghset#3}[#1][#2]}

\newcommand{\wghsym}{w}
\newcommandx{\wghSym}[3][1=, 2=, 3=]
	{\mthsym{\wghsym#3}[#1][#2]}

\newcommand{\wghelm}{w}
\newcommandx{\wghElm}[3][1=, 2=, 3=]
	{\mthelm{\wghelm#3}[#1][#2]}

\newcommand{\wghfun}{wg}
\newcommandx{\wghFun}[4][1=, 2=, 3=, 4=]
	{\mthargfun{\wghfun#4}[#1][#2]{#3}}

\newcommand{\wghgrp}{W\grp}
\newcommandx{\WghGrp}[5][1=, 2=, 3=, 4=, 5=]
	{\txtargname{\wghgrp#5{\small\argint{$[$}{#1}{$]$}}}[#2][#3]{#4}\xspace}

\newcommand{\wghgrpcls}{W\grpcls}
\newcommandx{\WghGrpCls}[5][1=, 2=, 3=, 4=, 5=]
	{\mthset[#5]{\wghgrpcls#4\text{\small\txtname{\argint{$[$}{#1}{$]$}}}}[#2]%
	[#3]}

%% file: Macro/Models/Games.tex
\newcommand{\gamkin}{2PT}

\newcommand{\plrset}{Pl}
\newcommandx{\PlrSet}[3][1=, 2=, 3=]
	{\mthset{\plrset#3}[#1][#2]}

\newcommand{\plrsym}{p}
\newcommandx{\plrSym}[3][1=, 2=, 3=]
	{\mthsym{\plrsym#3}[#1][#2]}

\newcommand{\plrelm}{p}
\newcommandx{\plrElm}[3][1=, 2=, 3=]
	{\mthelm{\plrelm#3}[#1][#2]}

\newcommand{\agnset}{Ag}
\newcommandx{\AgnSet}[3][1=, 2=, 3=]
	{\mthset{\agnset#3}[#1][#2]}

\newcommand{\agnsym}{a}
\newcommandx{\agnSym}[3][1=, 2=, 3=]
	{\mthsym{\agnsym#3}[#1][#2]}

\newcommand{\agnelm}{a}
\newcommandx{\agnElm}[3][1=, 2=, 3=]
	{\mthelm{\agnelm#3}[#1][#2]}

\newcommand{\movset}{Mv}
\newcommandx{\MovSet}[3][1=, 2=, 3=]
	{\mthset{\movset#3}[#1][#2]}

\newcommand{\movrel}{Mv}
\newcommandx{\MovRel}[3][1=, 2=, 3=]
	{\mthrel{\movrel#3}[#1][#2]}

\newcommand{\movsym}{m}
\newcommandx{\movSym}[3][1=, 2=, 3=]
	{\mthsym{\movsym#3}[#1][#2]}

\newcommand{\movelm}{m}
\newcommandx{\movElm}[3][1=, 2=, 3=]
	{\mthelm{\movelm#3}[#1][#2]}

\newcommand{\actset}{Ac}
\newcommandx{\ActSet}[3][1=, 2=, 3=]
	{\mthset{\actset#3}[#1][#2]}

\newcommand{\actrel}{Ac}
\newcommandx{\ActRel}[3][1=, 2=, 3=]
	{\mthrel{\actrel#3}[#1][#2]}

\newcommand{\actsym}{c}
\newcommandx{\actSym}[3][1=, 2=, 3=]
	{\mthsym{\actsym#3}[#1][#2]}

\newcommand{\actelm}{c}
\newcommandx{\actElm}[3][1=, 2=, 3=]
	{\mthelm{\actelm#3}[#1][#2]}

\newcommand{\decset}{Dc}
\newcommandx{\DecSet}[3][1=, 2=, 3=]
	{\mthset{\decset#3}[#1][#2]}

\newcommand{\decsym}{\delta}
\newcommandx{\decSym}[4][1=, 2=, 3=, 4=]
	{\mthargfun{\decsym#4}[#1][#2]{#3}}

\newcommand{\decelm}{\delta}
\newcommandx{\decElm}[4][1=, 2=, 3=, 4=]
	{\mthargfun{\decelm#4}[#1][#2]{#3}}

\newcommand{\posset}{Ps}
\newcommandx{\PosSet}[3][1=, 2=, 3=]
	{\mthset{\posset#3}[#1][#2]}
\newcommand{\fpossub}{0}
\newcommandx{\FPosSet}[3][1=, 2=, 3=]
	{\mthset{\posset#3}[\fpossub#1][#2]}
\newcommand{\spossub}{1}
\newcommandx{\SPosSet}[3][1=, 2=, 3=]
	{\mthset{\posset#3}[\spossub#1][#2]}

\newcommand{\possym}{v}
\newcommandx{\posSym}[3][1=, 2=, 3=]
	{\mthsym{\possym#3}[#1][#2]}
\newcommandx{\fposSym}[1][1=]
	{\posSym[\fpossub#1]}
\newcommandx{\sposSym}[1][1=]
	{\posSym[\spossub#1]}
\newcommand{\ipossub}{I}
\newcommandx{\iposSym}[1][1=]
	{\posSym[\ipossub#1]}

\newcommand{\poselm}{v}
\newcommandx{\posElm}[3][1=, 2=, 3=]
	{\mthelm{\poselm#3}[#1][#2]}
\newcommandx{\fposElm}[1][1=]
	{\posElm[\fpossub#1]}
\newcommandx{\sposElm}[1][1=]
	{\posElm[\spossub#1]}
\newcommandx{\iposElm}[1][1=]
	{\posElm[\ipossub#1]}

\newcommand{\sttset}{St}
\newcommandx{\SttSet}[3][1=, 2=, 3=]
	{\mthset{\sttset#3}[#1][#2]}
\newcommand{\fsttsub}{0}
\newcommandx{\FSttSet}[3][1=, 2=, 3=]
	{\mthset{\sttset#3}[\fsttsub#1][#2]}
\newcommand{\ssttsub}{1}
\newcommandx{\SSttSet}[3][1=, 2=, 3=]
	{\mthset{\sttset#3}[\ssttsub#1][#2]}

\newcommand{\sttsym}{s}
\newcommandx{\sttSym}[3][1=, 2=, 3=]
	{\mthsym{\sttsym#3}[#1][#2]}
\newcommandx{\fsttSym}[1][1=]
	{\sttSym[\fsttsub#1]}
\newcommandx{\ssttSym}[1][1=]
	{\sttSym[\ssttsub#1]}
\newcommand{\isttsub}{I}
\newcommandx{\isttSym}[1][1=]
	{\sttSym[\isttsub#1]}

\newcommand{\sttelm}{s}
\newcommandx{\sttElm}[3][1=, 2=, 3=]
	{\mthelm{\sttelm#3}[#1][#2]}
\newcommandx{\fsttElm}[1][1=]
	{\sttElm[\fsttsub#1]}
\newcommandx{\ssttElm}[1][1=]
	{\sttElm[\ssttsub#1]}
\newcommandx{\isttElm}[1][1=]
	{\sttElm[\isttsub#1]}

\newcommand{\plrfun}{pl}
\newcommandx{\plrFun}[4][1=, 2=, 3=, 4=]
	{\mthargfun{\plrfun#4}[#1][#2]{#3}}

\newcommand{\agnfun}{ag}
\newcommandx{\agnFun}[4][1=, 2=, 3=, 4=]
	{\mthargfun{\agnfun#4}[#1][#2]{#3}}

\newcommand{\movfun}{mv}
\newcommandx{\movFun}[4][1=, 2=, 3=, 4=]
	{\mthargfun{\movfun#4}[#1][#2]{#3}}

\newcommand{\actfun}{ac}
\newcommandx{\actFun}[4][1=, 2=, 3=, 4=]
	{\mthargfun{\actfun#4}[#1][#2]{#3}}

\newcommand{\decfun}{dc}
\newcommandx{\decFun}[4][1=, 2=, 3=, 4=]
	{\mthargfun{\decfun#4}[#1][#2]{#3}}

\newcommand{\trnfun}{tr}
\newcommandx{\trnFun}[4][1=, 2=, 3=, 4=]
	{\mthargfun{\trnfun#4}[#1][#2]{#3}}

\newcommand{\arn}{Ar}
\newcommandx{\Arn}[5][1=, 2=, 3=, 4=, 5=]
	{\txtargname{\arn#5{\small\argint{$[$}{#1}{$]$}}}[#2][#3]{#4}\xspace}

\newcommand{\arnname}{A}
\newcommand{\ArnName}
	{\mthname{\arnname}}

\newcommand{\arncls}{Ar}
\newcommandx{\ArnCls}[5][1=, 2=, 3=, 4=, 5=]
	{\mthset[#5]{\arncls#4\text{\small\txtname{\argint{$[$}{#1}{$]$}}}}[#2][#3]}

\newcommand{\ArnStr}[1][]
	{%
	\IfStrEqCase{\argdef{#1}{\gamkin}}
		{%
		{2PT}
			{\tuplecx{\FPosSet}{\SPosSet}{\MovRel}}%
		{MPC0}
			{\tupledx{\PlrSet}{\MovSet}{\PosSet}{\trnFun}}%
		{MPC1}
			{\tupleex{\PlrSet}{\MovSet}{\PosSet}{\decFun}{\trnFun}}%
		{MPC2}
			{\tuplefx{\PlrSet}{\MovSet}{\PosSet}{\plrFun}{\movFun}{\trnFun}}%
		{MPC3}
			{\tuplegx{\PlrSet}{\MovSet}{\PosSet}{\plrFun}{\movFun}{\decFun}{\trnFun}}%
		{2AT}
			{\tuplecx{\FSttSet}{\SSttSet}{\ActRel}}%
		{MAC0}
			{\tupledx{\AgnSet}{\ActSet}{\SttSet}{\trnFun}}%
		{MAC1}
			{\tupleex{\AgnSet}{\ActSet}{\SttSet}{\decFun}{\trnFun}}%
		{MAC2}
			{\tuplefx{\AgnSet}{\ActSet}{\SttSet}{\agnFun}{\actFun}{\trnFun}}%
		{MAC3}
			{\tuplegx{\AgnSet}{\ActSet}{\SttSet}{\agnFun}{\actFun}{\decFun}{\trnFun}}%
		}
		[\ensuremath{\clubsuit}]%
	}

\newcommand{\hstset}{Hst}
\newcommandx{\HstSet}[3][1=, 2=, 3=]
	{\mthset{\hstset#3}[#1][#2]}

\newcommand{\hstsym}{\rho}
\newcommandx{\hstSym}[3][1=, 2=, 3=]
	{\mthsym{\hstsym#3}[#1][#2]}

\newcommand{\hstelm}{\rho}
\newcommandx{\hstElm}[3][1=, 2=, 3=]
	{\mthelm{\hstelm#3}[#1][#2]}

\newcommand{\strset}{Str}
\newcommandx{\StrSet}[3][1=, 2=, 3=]
	{\mthset{\strset#3}[#1][#2]}

\newcommand{\strsym}{\sigma}
\newcommandx{\strSym}[4][1=, 2=, 3=, 4=]
	{\mthargfun{\strsym#4}[#1][#2]{#3}}

\newcommand{\strelm}{\sigma}
\newcommandx{\strElm}[4][1=, 2=, 3=, 4=]
	{\mthargfun{\strelm#4}[#1][#2]{#3}}

\newcommand{\prfset}{Prf}
\newcommandx{\PrfSet}[3][1=, 2=, 3=]
	{\mthset{\prfset#3}[#1][#2]}

\newcommand{\prfsym}{\xi}
\newcommandx{\prfSym}[4][1=, 2=, 3=, 4=]
	{\mthargfun{\prfsym#4}[#1][#2]{#3}}

\newcommandx{\prfElm}[4][1=, 2=, 3=, 4=]
	{\mthargfun{\prfsym#4}[#1][#2]{#3}}

\newcommand{\playfun}{play}
\newcommandx{\playFun}[4][1=, 2=, 3=, 4=]
	{\mthargfun{\playfun#4}[#1][#2]{#3}}

\newcommand{\labarn}{L\arn}
\newcommandx{\LabArn}[5][1=, 2=, 3=, 4=, 5=]
	{\txtargname{\labarn#5{\small\argint{$[$}{#1}{$]$}}}[#2][#3]{#4}\xspace}

\newcommand{\labarncls}{L\arncls}
\newcommandx{\LabArnCls}[5][1=, 2=, 3=, 4=, 5=]
	{\mthset[#5]{\labarncls#4\text{\small\txtname{\argint{$[$}{#1}{$]$}}}}[#2]%
	[#3]}

\newcommand{\colarn}{C\arn}
\newcommandx{\ColArn}[5][1=, 2=, 3=, 4=, 5=]
	{\txtargname{\colarn#5{\small\argint{$[$}{#1}{$]$}}}[#2][#3]{#4}\xspace}

\newcommand{\colarncls}{C\arncls}
\newcommandx{\ColArnCls}[5][1=, 2=, 3=, 4=, 5=]
	{\mthset[#5]{\colarncls#4\text{\small\txtname{\argint{$[$}{#1}{$]$}}}}[#2]%
	[#3]}

\newcommand{\wgharn}{W\arn}
\newcommandx{\WghArn}[5][1=, 2=, 3=, 4=, 5=]
	{\txtargname{\wgharn#5{\small\argint{$[$}{#1}{$]$}}}[#2][#3]{#4}\xspace}

\newcommand{\wgharncls}{W\arncls}
\newcommandx{\WghArnCls}[5][1=, 2=, 3=, 4=, 5=]
	{\mthset[#5]{\wgharncls#4\text{\small\txtname{\argint{$[$}{#1}{$]$}}}}[#2]%
	[#3]}

\newcommand{\winset}{Wn}
\newcommandx{\WinSet}[3][1=, 2=, 3=]
	{\mthset{\winset#3}[#1][#2]}

\newcommand{\prdset}{Pr}
\newcommandx{\PrdSet}[3][1=, 2=, 3=]
	{\mthset{\prdset#3}[#1][#2]}

\newcommand{\prdsym}{p}
\newcommandx{\prdSym}[3][1=, 2=, 3=]
	{\mthsym{\prdsym#3}[#1][#2]}

\newcommand{\prdelm}{p}
\newcommandx{\prdElm}[3][1=, 2=, 3=]
	{\mthelm{\prdelm#3}[#1][#2]}

\newcommand{\prdfun}{pre}
\newcommandx{\prdFun}[4][1=, 2=, 3=, 4=]
	{\mthargfun{\prdfun#4}[#1][#2]{#3}}

\newcommand{\extname}{E}
\newcommand{\ExtName}
	{\mthname{\extname}}

\newcommand{\extcls}{Ex}
\newcommandx{\ExtCls}[5][1=, 2=, 3=, 4=, 5=]
	{\mthset[#5]{\extcls#4\text{\small\txtname{\argint{$[$}{#1}{$]$}}}}[#2][#3]}

\newcommand{\conset}{Cn}
\newcommandx{\ConSet}[3][1=, 2=, 3=]
	{\mthset{\conset#3}[#1][#2]}

\newcommand{\consym}{\varphi}
\newcommandx{\conSym}[3][1=, 2=, 3=]
	{\mthsym{\consym#3}[#1][#2]}

\newcommand{\conelm}{\varphi}
\newcommandx{\conElm}[3][1=, 2=, 3=]
	{\mthelm{\conelm#3}[#1][#2]}

\newcommand{\schrel}{\models}
\newcommandx{\schRel}[4][1=, 2=, 3=, 4=]
	{\mthrel{\schrel#3}[#1][#2]}

\newcommand{\schcls}{Sc}
\newcommandx{\SchCls}[5][1=, 2=, 3=, 4=, 5=]
	{\mthset[#5]{\schcls#4\text{\small\txtname{\argint{$[$}{#1}{$]$}}}}[#2][#3]}

\newcommand{\gamname}{\Game}
\newcommand{\GamName}
	{\mthname{\gamname}}

\newcommand{\gamcls}{Gm}
\newcommandx{\GamCls}[5][1=, 2=, 3=, 4=, 5=]
	{\mthset[#5]{\gamcls#4\text{\small\txtname{\argint{$[$}{#1}{$]$}}}}[#2][#3]}

\newcommandx{\GamStr}[1][1=]
	{%
	\StrLeft{\argdef{#1}{\gamkin}}{2}[\optgamkin]%
	\IfStrEqCase{\optgamkin}
		{%
		{2P}
			{\gamstrauxtp}%
		{MP}
			{\gamstrauxmp}%
		{2A}
			{\gamstrauxta}%
		{MA}
			{\gamstrauxma}%
		}
		[\ensuremath{\clubsuit}]%
	}
\newcommandx{\gamstrauxtp}[5][1=, 2=, 3=, 4=, 5=]
	{\tuplecx{\ArnName}{\iposElm}{\WinSet}[#3][#4][#5][#1][#2]}
\newcommandx{\gamstrauxmp}[5][1=, 2=, 3=, 4=, 5=]
	{\tuplecx{\ExtName}{\iposElm}{\conElm}[#3][#4][#5][#1][#2]}
\newcommandx{\gamstrauxta}[5][1=, 2=, 3=, 4=, 5=]
	{\tuplecx{\ArnName}{\isttElm}{\WinSet}[#3][#4][#5][#1][#2]}
\newcommandx{\gamstrauxma}[5][1=, 2=, 3=, 4=, 5=]
	{\tuplecx{\ExtName}{\isttElm}{\conElm}[#3][#4][#5][#1][#2]}

%% file: Macro/Models/KS.tex
\newcommand{\worset}{W}
\newcommandx{\WorSet}[3][1=, 2=, 3=]
	{\mthset{\worset#3}[#1][#2]}

\newcommand{\worsym}{w}
\newcommandx{\worSym}[3][1=, 2=, 3=]
	{\mthsym{\worsym#3}[#1][#2]}

\newcommand{\worelm}{w}
\newcommandx{\worElm}[3][1=, 2=, 3=]
	{\mthelm{\worelm#3}[#1][#2]}

\newcommand{\trnrel}{R}
\newcommandx{\TrnRel}[3][1=, 2=, 3=]
	{\mthrel{\trnrel#3}[#1][#2]}

\newcommand{\trnsym}{r}
\newcommandx{\trnSym}[3][1=, 2=, 3=]
	{\mthsym{\trnsym#3}[#1][#2]}

\newcommand{\trnelm}{r}
\newcommandx{\trnElm}[3][1=, 2=, 3=]
	{\mthelm{\trnelm#3}[#1][#2]}

\newcommand{\labfun}{L}
\newcommandx{\labFun}[4][1=, 2=, 3=, 4=]
	{\mthargfun{\labfun#4}[#1][#2]{#3}}

\newcommand{\krpstr}{KS}
\newcommandx{\KrpStr}[5][1=, 2=, 3=, 4=, 5=]
	{\txtargname{\krpstr#5{\small\argint{$[$}{#1}{$]$}}}[#2][#3]{#4}\xspace}

\newcommand{\krpstrcls}{KS}
\newcommandx{\KrpStrCls}[5][1=, 2=, 3=, 4=, 5=]
	{\mthset[#5]{\krpstrcls#4\text{\small\txtname{\argint{$[$}{#1}{$]$}}}}[#2]%
	[#3]}

\newcommand{\trkset}{Trk}
\newcommandx{\TrkSet}[3][1=, 2=, 3=]
	{\mthset{\trkset#3}[#1][#2]}

\newcommand{\trksym}{\rho}
\newcommandx{\trkSym}[3][1=, 2=, 3=]
	{\mthsym{\trksym#3}[#1][#2]}

\newcommand{\trkelm}{\rho}
\newcommandx{\trkElm}[3][1=, 2=, 3=]
	{\mthelm{\trkelm#3}[#1][#2]}

\newcommand{\krptree}{KT}
\newcommandx{\KrpTree}[5][1=, 2=, 3=, 4=, 5=]
	{\txtargname{\krptree#5{\small\argint{$[$}{#1}{$]$}}}[#2][#3]{#4}\xspace}

\newcommand{\krptreecls}{KT}
\newcommandx{\KrpTreeCls}[5][1=, 2=, 3=, 4=, 5=]
	{\mthset[#5]{\krptreecls#4\text{\small\txtname{\argint{$[$}{#1}{$]$}}}}[#2]%
	[#3]}

\newcommand{\dirset}{Dir}
\newcommandx{\DirSet}[3][1=, 2=, 3=]
	{\mthset{\dirset#3}[#1][#2]}

\newcommand{\dirsym}{d}
\newcommandx{\dirSym}[3][1=, 2=, 3=]
	{\mthsym{\dirsym#3}[#1][#2]}

\newcommand{\direlm}{d}
\newcommandx{\dirElm}[3][1=, 2=, 3=]
	{\mthelm{\direlm#3}[#1][#2]}

\newcommand{\unwfun}{unw}
\newcommandx{\unwFun}[4][1=, 2=, 3=, 4=]
	{\mthargfun{\unwfun#4}[#1][#2]{#3}}

%% file: Macro/Models/CGS.tex
\newcommand{\congamstrkin}{MAC0}

\newcommand{\congamstr}{CGS}
\newcommandx{\ConGamStr}[5][1=, 2=, 3=, 4=, 5=]
	{\txtargname{\congamstr#5{\small\argint{$[$}{#1}{$]$}}}[#2][#3]{#4}\xspace}

\newcommandx{\ConGamStrCls}[5][1=, 2=, 3=, 4=, 5=]
	{\mthset[#5]{\arncls#4\text{\small\txtname{\argint{$[$}{#1}{$]$}}}}[#2][#3]}

\newcommandx{\ConGamStrStr}[1][1=]
	{%
	\IfStrEqCase{\argdef{#1}{\congamstrkin}}
		{%
		{IP}
			{\congamstrstrauxip}%
		{2PT}
			{\congamstrstrauxpt}%
		{MPC0}
			{\congamstrstrauxpca}%
		{MPC1}
			{\congamstrstrauxpcb}%
		{MPC2}
			{\congamstrstrauxpcc}%
		{MPC3}
			{\congamstrstrauxpcd}%
		{IA}
			{\congamstrstrauxia}%
		{2AT}
			{\congamstrstrauxat}%
		{MAC0}
			{\congamstrstrauxaca}%
		{MAC1}
			{\congamstrstrauxacb}%
		{MAC2}
			{\congamstrstrauxacc}%
		{MAC3}
			{\congamstrstrauxacd}%
		}
		[\ensuremath{\clubsuit}]%
	}
\newcommandx{\congamstrstrauxip}[3][1=, 2=, 3=]
	{%
	\def\defini{#1}%
	\def\defsubscr{#2}%
	\def\defsupscr{#3}%
	\congamstrstrauxxip%
	}
\newcommandx{\congamstrstrauxxip}[3][1=, 2=, 3=]
	{%
	\tupledx{\ArnName}{\APSet}{\apFun}{\iposElm}%
		[\defsubscr][\defsupscr][#1][#2][#3][\defini]%
	}
\newcommandx{\congamstrstrauxpt}[3][1=, 2=, 3=]
	{%
	\def\defini{#1}%
	\def\defsubscr{#2}%
	\def\defsupscr{#3}%
	\congamstrstrauxxpt%
	}
\newcommandx{\congamstrstrauxxpt}[5][1=, 2=, 3=, 4=, 5=]
	{%
	\tuplefx{\APSet}{\FPosSet}{\SPosSet}{\MovRel}{\apFun}{\iposElm}%
		[\defsubscr][\defsupscr][#1][#2][#3][#4][#5][\defini]%
	}
\newcommandx{\congamstrstrauxpca}[3][1=, 2=, 3=]
	{%
	\def\defini{#1}%
	\def\defsubscr{#2}%
	\def\defsupscr{#3}%
	\congamstrstrauxxpca%
	}
\newcommandx{\congamstrstrauxxpca}[6][1=, 2=, 3=, 4=, 5=, 6=]
	{%
	\tuplegx{\APSet}{\PlrSet}{\MovSet}{\PosSet}{\trnFun}{\apFun}{\iposElm}%
		[\defsubscr][\defsupscr][#1][#2][#3][#4][#5][#6][\defini]%
	}
\newcommandx{\congamstrstrauxpcb}[3][1=, 2=, 3=]
	{%
	\def\defini{#1}%
	\def\defsubscr{#2}%
	\def\defsupscr{#3}%
	\congamstrstrauxxpcb%
	}
\newcommandx{\congamstrstrauxxpcb}[7][1=, 2=, 3=, 4=, 5=, 6=, 7=]
	{%
	\tuplehx{\APSet}{\PlrSet}{\MovSet}{\PosSet}{\decFun}{\trnFun}{\apFun}%
		{\iposElm}%
		[\defsubscr][\defsupscr][#1][#2][#3][#4][#5][#6][#7][\defini]%
	}
\newcommandx{\congamstrstrauxpcc}[3][1=, 2=, 3=]
	{%
	\def\defini{#1}%
	\def\defsubscr{#2}%
	\def\defsupscr{#3}%
	\congamstrstrauxxpcc%
	}
\newcommandx{\congamstrstrauxxpcc}[8][1=, 2=, 3=, 4=, 5=, 6=, 7=, 8=]
	{%
	\tupleix{\APSet}{\PlrSet}{\MovSet}{\PosSet}{\plrFun}{\movFun}{\trnFun}%
		{\apFun}{\iposElm}%
		[\defsubscr][\defsupscr][#1][#2][#3][#4][#5][#6][#7][#8][\defini]%
	}
\newcommandx{\congamstrstrauxpcd}[3][1=, 2=, 3=]
	{%
	\def\defini{#1}%
	\def\defsubscr{#2}%
	\def\defsupscr{#3}%
	\congamstrstrauxxpcd%
	}
\newcommandx{\congamstrstrauxxpcd}[9][1=, 2=, 3=, 4=, 5=, 6=, 7=, 8=, 9=]
	{%
	\tuplejx{\APSet}{\PlrSet}{\MovSet}{\PosSet}{\plrFun}{\movFun}{\decFun}%
	{\trnFun}{\apFun}{\iposElm}%
		[\defsubscr][\defsupscr][#1][#2][#3][#4][#5][#6][#7][#8][#9][\defini]%
	}
\newcommandx{\congamstrstrauxia}[3][1=, 2=, 3=]
	{%
	\def\defini{#1}%
	\def\defsubscr{#2}%
	\def\defsupscr{#3}%
	\congamstrstrauxxia%
	}
\newcommandx{\congamstrstrauxxia}[3][1=, 2=, 3=]
	{%
	\tupledx{\ArnName}{\APSet}{\apFun}{\isttElm}%
		[\defsubscr][\defsupscr][#1][#2][#3][\defini]%
	}
\newcommandx{\congamstrstrauxat}[3][1=, 2=, 3=]
	{%
	\def\defini{#1}%
	\def\defsubscr{#2}%
	\def\defsupscr{#3}%
	\congamstrstrauxxat%
	}
\newcommandx{\congamstrstrauxxat}[5][1=, 2=, 3=, 4=, 5=]
	{%
	\tuplefx{\APSet}{\FSttSet}{\SSttSet}{\ActRel}{\apFun}{\isttElm}%
		[\defsubscr][\defsupscr][#1][#2][#3][#4][#5][\defini]%
	}
\newcommandx{\congamstrstrauxaca}[3][1=, 2=, 3=]
	{%
	\def\defini{#1}%
	\def\defsubscr{#2}%
	\def\defsupscr{#3}%
	\congamstrstrauxxaca%
	}
\newcommandx{\congamstrstrauxxaca}[6][1=, 2=, 3=, 4=, 5=, 6=]
	{%
	\tuplegx{\APSet}{\AgnSet}{\ActSet}{\SttSet}{\trnFun}{\apFun}{\isttElm}%
		[\defsubscr][\defsupscr][#1][#2][#3][#4][#5][#6][\defini]%
	}
\newcommandx{\congamstrstrauxacb}[3][1=, 2=, 3=]
	{%
	\def\defini{#1}%
	\def\defsubscr{#2}%
	\def\defsupscr{#3}%
	\congamstrstrauxxacb%
	}
\newcommandx{\congamstrstrauxxacb}[7][1=, 2=, 3=, 4=, 5=, 6=, 7=]
	{%
	\tuplehx{\APSet}{\AgnSet}{\ActSet}{\SttSet}{\decFun}{\trnFun}{\apFun}%
		{\isttElm}%
		[\defsubscr][\defsupscr][#1][#2][#3][#4][#5][#6][#7][\defini]%
	}
\newcommandx{\congamstrstrauxacc}[3][1=, 2=, 3=]
	{%
	\def\defini{#1}%
	\def\defsubscr{#2}%
	\def\defsupscr{#3}%
	\congamstrstrauxxacc%
	}
\newcommandx{\congamstrstrauxxacc}[8][1=, 2=, 3=, 4=, 5=, 6=, 7=, 8=]
	{%
	\tupleix{\APSet}{\AgnSet}{\ActSet}{\SttSet}{\agnFun}{\actFun}{\trnFun}%
		{\apFun}{\isttElm}%
		[\defsubscr][\defsupscr][#1][#2][#3][#4][#5][#6][#7][#8][\defini]%
	}
\newcommandx{\congamstrstrauxacd}[3][1=, 2=, 3=]
	{%
	\def\defini{#1}%
	\def\defsubscr{#2}%
	\def\defsupscr{#3}%
	\congamstrstrauxxacd%
	}
\newcommandx{\congamstrstrauxxacd}[9][1=, 2=, 3=, 4=, 5=, 6=, 7=, 8=, 9=]
	{%
	\tuplejx{\APSet}{\AgnSet}{\ActSet}{\SttSet}{\agnFun}{\actFun}{\decFun}%
		{\trnFun}{\apFun}{\isttElm}%
		[\defsubscr][\defsupscr][#1][#2][#3][#4][#5][#6][#7][#8][#9][\defini]%
	}

%% file: Macro/Models/Machines.tex
\newcommand{\trntabkin}{D}

\newcommand{\symset}{Sm}
\newcommandx{\SymSet}[3][1=, 2=, 3=]
	{\mthset{\symset#3}[#1][#2]}

\newcommand{\symsym}{\ell}
\newcommandx{\symSym}[3][1=, 2=, 3=]
	{\mthsym{\symsym#3}[#1][#2]}

\newcommand{\symelm}{\ell}
\newcommandx{\symElm}[3][1=, 2=, 3=]
	{\mthelm{\symelm#3}[#1][#2]}

\newcommand{\DSttSet}[1][]
	{\SttSet[\Delta#1]}

\newcommand{\ESttSet}[1][]
	{\SttSet[\exists#1]}

\newcommand{\ASttSet}[1][]
	{\SttSet[\forall#1]}

\newcommand{\trntab}{tt}
\newcommandx{\TrnTab}[5][1=, 2=, 3=, 4=, 5=]
	{\txtargname{\trntab#5{\small\argint{$[$}{#1}{$]$}}}[#2][#3]{#4}\xspace}

\newcommand{\trntabcls}{TT}
\newcommandx{\TrnTabCls}[5][1=, 2=, 3=, 4=, 5=]
	{\mthset[#5]{\trntabcls#4\text{\txtname{\small\argint{$[$}{#1}{$]$}}}}[#2]%
	[#3]}

\newcommand{\TrnTabStr}[1][]
	{%
	\IfStrEqCase{\argdef{#1}{\trntabkin}}
		{%
		{D}{\tuplecx{\SymSet}{\SttSet}{\trnFun}}%
		{N}{\tupledx{\SymSet}{\DSttSet}{\ESttSet}{\trnFun}}%
		{U}{\tupledx{\SymSet}{\DSttSet}{\ASttSet}{\trnFun}}%
		{A}{\tupleex{\SymSet}{\DSttSet}{\ESttSet}{\ASttSet}{\trnFun}}%
		}
		[\ensuremath{\clubsuit}]%
	}

%% file: Macro/Games.tex
\newcommandx{\EF}[5][1=, 2=, 3=, 4=, 5=]
  {\txtargname{EF#5{\small\argint{$[$}{#1}{$]$}}}[#2][#3]{#4}}

\newcommandx{\BG}[5][1=, 2=, 3=, 4=, 5=]
  {\txtargname{BG#5{\small\argint{$[$}{#1}{$]$}}}[#2][#3]{#4}}

\newcommandx{\CG}[5][1=, 2=, 3=, 4=, 5=]
  {\txtargname{CG#5{\small\argint{$[$}{#1}{$]$}}}[#2][#3]{#4}}

\newcommandx{\PG}[5][1=, 2=, 3=, 4=, 5=]
  {\txtargname{PG#5{\small\argint{$[$}{#1}{$]$}}}[#2][#3]{#4}}

\newcommandx{\RG}[5][1=, 2=, 3=, 4=, 5=]
  {\txtargname{RG#5{\small\argint{$[$}{#1}{$]$}}}[#2][#3]{#4}}

\newcommandx{\SG}[5][1=, 2=, 3=, 4=, 5=]
  {\txtargname{SG#5{\small\argint{$[$}{#1}{$]$}}}[#2][#3]{#4}}

\newcommandx{\MG}[5][1=, 2=, 3=, 4=, 5=]
  {\txtargname{MG#5{\small\argint{$[$}{#1}{$]$}}}[#2][#3]{#4}}

\newcommandx{\MPG}[5][1=, 2=, 3=, 4=, 5=]
  {\txtargname{MPG#5{\small\argint{$[$}{#1}{$]$}}}[#2][#3]{#4}}

\newcommandx{\DPG}[5][1=, 2=, 3=, 4=, 5=]
  {\txtargname{DPG#5{\small\argint{$[$}{#1}{$]$}}}[#2][#3]{#4}}

\newcommandx{\SSG}[5][1=, 2=, 3=, 4=, 5=]
  {\txtargname{SSG#5{\small\argint{$[$}{#1}{$]$}}}[#2][#3]{#4}}

\renewcommandx{\SG}[5][1=, 2=, 3=, 4=, 5=]
  {\txtargname{SG#5{\small\argint{$[$}{#1}{$]$}}}[#2][#3]{#4}}

%% file: Macro/Logics.tex
\newcommandx{\BF}[5][1=, 2=, 3=, 4=, 5=]
  {\txtargname{BF#5{\small\argint{$[$}{#1}{$]$}}}[#2][#3]{#4}}

\newcommandx{\QBF}
  {{\txtname{Q}}\BF}

\newcommandx{\FOL}[5][1=, 2=, 3=, 4=, 5=]
  {\txtargname{FOL#5{\small\argint{$[$}{#1}{$]$}}}[#2][#3]{#4}}

\newcommandx{\SOL}[5][1=, 2=, 3=, 4=, 5=]
  {\txtargname{SOL#5{\small\argint{$[$}{#1}{$]$}}}[#2][#3]{#4}}

\newcommandx{\TL}[5][1=, 2=, 3=, 4=, 5=]
  {\txtargname{TL#5{\small\argint{$[$}{#1}{$]$}}}[#2][#3]{#4}}

\newcommandx{\PL}[5][1=, 2=, 3=, 4=, 5=]
  {\txtargname{PL#5{\small\argint{$[$}{#1}{$]$}}}[#2][#3]{#4}}

\newcommandx{\ML}[5][1=, 2=, 3=, 4=, 5=]
  {\txtargname{ML#5{\small\argint{$[$}{#1}{$]$}}}[#2][#3]{#4}}

\newcommandx{\MC}[5][1=, 2=, 3=, 4=, 5=]
  {\txtargname{$\mu$Calculus#5{\small\argint{$[$}{#1}{$]$}}}[#2][#3]{#4}}

\newcommandx{\LTL}[5][1=, 2=, 3=, 4=, 5=]
  {\txtargname{LTL#5{\small\argint{$[$}{#1}{$]$}}}[#2][#3]{#4}}

\newcommandx{\PTL}[5][1=, 2=, 3=, 4=, 5=]
  {\txtargname{PTL#5{\small\argint{$[$}{#1}{$]$}}}[#2][#3]{#4}}

\newcommandx{\CTL}[5][1=, 2=, 3=, 4=, 5=]
  {\txtargname{CTL#5{\small\argint{$[$}{#1}{$]$}}}[#2][#3]{#4}}

\newcommandx{\CTLP}[5][1=, 2=, 3=, 4=, 5=]
  {\txtargname{CTL$^{+}$#5{\small\argint{$[$}{#1}{$]$}}}[#2][#3]{#4}}

\newcommandx{\CTLS}[5][1=, 2=, 3=, 4=, 5=]
  {\txtargname{CTL$^{\star}$#5{\small\argint{$[$}{#1}{$]$}}}[#2][#3]{#4}}

\newcommandx{\STL}[5][1=, 2=, 3=, 4=, 5=]
  {\txtargname{STL#5{\small\argint{$[$}{#1}{$]$}}}[#2][#3]{#4}}

\newcommandx{\STLP}[5][1=, 2=, 3=, 4=, 5=]
  {\txtargname{STL$^{+}$#5{\small\argint{$[$}{#1}{$]$}}}[#2][#3]{#4}}

\newcommandx{\STLS}[5][1=, 2=, 3=, 4=, 5=]
  {\txtargname{STL$^{\star}$#5{\small\argint{$[$}{#1}{$]$}}}[#2][#3]{#4}}

\newcommandx{\ATL}[5][1=, 2=, 3=, 4=, 5=]
  {\txtargname{ATL#5{\small\argint{$[$}{#1}{$]$}}}[#2][#3]{#4}}

\newcommandx{\ATLP}[5][1=, 2=, 3=, 4=, 5=]
  {\txtargname{ATL$^{+}$#5{\small\argint{$[$}{#1}{$]$}}}[#2][#3]{#4}}

\newcommandx{\ATLS}[5][1=, 2=, 3=, 4=, 5=]
  {\txtargname{ATL$^{\star}$#5{\small\argint{$[$}{#1}{$]$}}}[#2][#3]{#4}}

\newcommandx{\SL}[5][1=, 2=, 3=, 4=, 5=]
  {\txtargname{SL#5{\small\argint{$[$}{#1}{$]$}}}[#2][#3]{#4}}

%% file: Macro/Complexity.tex
\newcommandx{\LogTime}[4][1=, 2=, 3=, 4=]
  {\txtargname{LogTime#4}[#2][#3]{#1}}
\newcommandx{\LogTimeE}[4][1=, 2=, 3=, 4=]
  {\LogTime[#1][#2][#3][#4]-\EComplexity}
\newcommandx{\LogTimeH}[4][1=, 2=, 3=, 4=]
  {\LogTime[#1][#2][#3][#4]-\HComplexity}
\newcommandx{\LogTimeC}[4][1=, 2=, 3=, 4=]
  {\LogTime[#1][#2][#3][#4]-\CComplexity}

\newcommandx{\LogSpace}[4][1=, 2=, 3=, 4=]
  {\txtargname{LogSpace#4}[#2][#3]{#1}}
\newcommandx{\LogSpaceE}[4][1=, 2=, 3=, 4=]
  {\LogSpace[#1][#2][#3][#4]-\EComplexity}
\newcommandx{\LogSpaceH}[4][1=, 2=, 3=, 4=]
  {\LogSpace[#1][#2][#3][#4]-\HComplexity}
\newcommandx{\LogSpaceC}[4][1=, 2=, 3=, 4=]
  {\LogSpace[#1][#2][#3][#4]-\CComplexity}

\newcommandx{\PTime}[4][1=, 2=, 3=, 4=]
  {\txtargname{PTime#4}[#2][#3]{#1}}
\newcommandx{\PTimeE}[4][1=, 2=, 3=, 4=]
  {\PTime[#1][#2][#3][#4]-\EComplexity}
\newcommandx{\PTimeH}[4][1=, 2=, 3=, 4=]
  {\PTime[#1][#2][#3][#4]-\HComplexity}
\newcommandx{\PTimeC}[4][1=, 2=, 3=, 4=]
  {\PTime[#1][#2][#3][#4]-\CComplexity}

\newcommand{\UPTime}
  {\UComplexity\PTime}

\newcommand{\CoUPTime}
  {\CoComplexity\UPTime}

\newcommand{\NPTime}
  {\NComplexity\PTime}

\newcommand{\CoNPTime}
  {\CoComplexity\NPTime}

\newcommandx{\PSpace}[4][1=, 2=, 3=, 4=]
  {\txtargname{PSpace#4}[#2][#3]{#1}}
\newcommandx{\PSpaceE}[4][1=, 2=, 3=, 4=]
  {\PSpace[#1][#2][#3][#4]-\EComplexity}
\newcommandx{\PSpaceH}[4][1=, 2=, 3=, 4=]
  {\PSpace[#1][#2][#3][#4]-\HComplexity}
\newcommandx{\PSpaceC}[4][1=, 2=, 3=, 4=]
  {\PSpace[#1][#2][#3][#4]-\CComplexity}

\newcommandx{\ExpTime}[4][1=, 2=, 3=, 4=]
  {\txtargname{ExpTime#4}[#2][#3]{#1}}
\newcommandx{\ExpTimeE}[4][1=, 2=, 3=, 4=]
  {\ExpTime[#1][#2][#3][#4]-\EComplexity}
\newcommandx{\ExpTimeH}[4][1=, 2=, 3=, 4=]
  {\ExpTime[#1][#2][#3][#4]-\HComplexity}
\newcommandx{\ExpTimeC}[4][1=, 2=, 3=, 4=]
  {\ExpTime[#1][#2][#3][#4]-\CComplexity}

\newcommandx{\ExpSpace}[4][1=, 2=, 3=, 4=]
  {\txtargname{ExpSpace#4}[#2][#3]{#1}}
\newcommandx{\ExpSpaceE}[4][1=, 2=, 3=, 4=]
  {\ExpSpace[#1][#2][#3][#4]-\EComplexity}
\newcommandx{\ExpSpaceH}[4][1=, 2=, 3=, 4=]
  {\ExpSpace[#1][#2][#3][#4]-\HComplexity}
\newcommandx{\ExpSpaceC}[4][1=, 2=, 3=, 4=]
  {\ExpSpace[#1][#2][#3][#4]-\CComplexity}

\newcommandx{\NonElm}[4][1=, 2=, 3=, 4=]
  {\txtargname{NonElementary#4}[#2][#3]{#1}}
\newcommandx{\NonElmE}[4][1=, 2=, 3=, 4=]
  {\NonElm[#1][#2][#3][#4]-\EComplexity}
\newcommandx{\NonElmH}[4][1=, 2=, 3=, 4=]
  {\NonElm[#1][#2][#3][#4]-\HComplexity}
\newcommandx{\NonElmC}[4][1=, 2=, 3=, 4=]
  {\NonElm[#1][#2][#3][#4]-\CComplexity}

\newcommandx{\NonElmTime}[4][1=, 2=, 3=, 4=]
  {\txtargname{NonElementaryTime#4}[#2][#3]{#1}}
\newcommandx{\NonElmTimeE}[4][1=, 2=, 3=, 4=]
  {\NonElmTime[#1][#2][#3][#4]-\EComplexity}
\newcommandx{\NonElmTimeH}[4][1=, 2=, 3=, 4=]
  {\NonElmTime[#1][#2][#3][#4]-\HComplexity}
\newcommandx{\NonElmTimeC}[4][1=, 2=, 3=, 4=]
  {\NonElmTime[#1][#2][#3][#4]-\CComplexity}

\newcommandx{\NonElmSpace}[4][1=, 2=, 3=, 4=]
  {\txtargname{NonElementarySpace#4}[#2][#3]{#1}}
\newcommandx{\NonElmSpaceE}[4][1=, 2=, 3=, 4=]
  {\NonElmSpace[#1][#2][#3][#4]-\EComplexity}
\newcommandx{\NonElmSpaceH}[4][1=, 2=, 3=, 4=]
  {\NonElmSpace[#1][#2][#3][#4]-\HComplexity}
\newcommandx{\NonElmSpaceC}[4][1=, 2=, 3=, 4=]
  {\NonElmSpace[#1][#2][#3][#4]-\CComplexity}

\newcommandx{\DLHier}[4][2=, 3=, 4=]
  {\mthargset[0]{\Delta#4}[#1][#3]{#2}}
\newcommandx{\DLHierE}[4][2=, 3=, 4=]
  {\DLHier{#1}[#2][#3][#4]-\EComplexity}
\newcommandx{\DLHierH}[4][2=, 3=, 4=]
  {\DLHier{#1}[#2][#3][#4]-\HComplexity}
\newcommandx{\DLHierC}[4][2=, 3=, 4=]
  {\DLHier{#1}[#2][#3][#4]-\CComplexity}

\newcommandx{\ELHier}[4][2=, 3=, 4=]
  {\mthargset[0]{\Sigma#4}[#1][#3]{#2}}
\newcommandx{\ELHierE}[4][2=, 3=, 4=]
  {\ELHier{#1}[#2][#3][#4]-\EComplexity}
\newcommandx{\ELHierH}[4][2=, 3=, 4=]
  {\ELHier{#1}[#2][#3][#4]-\HComplexity}
\newcommandx{\ELHierC}[4][2=, 3=, 4=]
  {\ELHier{#1}[#2][#3][#4]-\CComplexity}

\newcommandx{\ULHier}[4][2=, 3=, 4=]
  {\mthargset[0]{\Pi#4}[#1][#3]{#2}}
\newcommandx{\ULHierE}[4][2=, 3=, 4=]
  {\ULHier{#1}[#2][#3][#4]-\EComplexity}
\newcommandx{\ULHierH}[4][2=, 3=, 4=]
  {\ULHier{#1}[#2][#3][#4]-\HComplexity}
\newcommandx{\ULHierC}[4][2=, 3=, 4=]
  {\ULHier{#1}[#2][#3][#4]-\CComplexity}

\newcommandx{\DBHier}[4][2=, 3=, 4=]
  {\mthargset[3]{\Delta#4}[#1][#3]{#2}}
\newcommandx{\DBHierE}[4][2=, 3=, 4=]
  {\DBHier{#1}[#2][#3][#4]-\EComplexity}
\newcommandx{\DBHierH}[4][2=, 3=, 4=]
  {\DBHier{#1}[#2][#3][#4]-\HComplexity}
\newcommandx{\DBHierC}[4][2=, 3=, 4=]
  {\DBHier{#1}[#2][#3][#4]-\CComplexity}

\newcommandx{\EBHier}[4][2=, 3=, 4=]
  {\mthargset[3]{\Sigma#4}[#1][#3]{#2}}
\newcommandx{\EBHierE}[4][2=, 3=, 4=]
  {\EBHier{#1}[#2][#3][#4]-\EComplexity}
\newcommandx{\EBHierH}[4][2=, 3=, 4=]
  {\EBHier{#1}[#2][#3][#4]-\HComplexity}
\newcommandx{\EBHierC}[4][2=, 3=, 4=]
  {\EBHier{#1}[#2][#3][#4]-\CComplexity}

\newcommandx{\UBHier}[4][2=, 3=, 4=]
  {\mthargset[3]{\Pi#4}[#1][#3]{#2}}
\newcommandx{\UBHierE}[4][2=, 3=, 4=]
  {\UBHier{#1}[#2][#3][#4]-\EComplexity}
\newcommandx{\UBHierH}[4][2=, 3=, 4=]
  {\UBHier{#1}[#2][#3][#4]-\HComplexity}
\newcommandx{\UBHierC}[4][2=, 3=, 4=]
  {\UBHier{#1}[#2][#3][#4]-\CComplexity}

\newcommandx{\DPolHier}[4][2=, 3=, 4=]
  {\DLHier{#1}[#2][\argb{\mathrm{P}}{#3}][#4]}
\newcommandx{\DPolHierE}[4][2=, 3=, 4=]
  {\DPolHier{#1}[#2][#3][#4]-\EComplexity}
\newcommandx{\DPolHierH}[4][2=, 3=, 4=]
  {\DPolHier{#1}[#2][#3][#4]-\HComplexity}
\newcommandx{\DPolHierC}[4][2=, 3=, 4=]
  {\DPolHier{#1}[#2][#3][#4]-\CComplexity}

\newcommandx{\EPolHier}[4][2=, 3=, 4=]
  {\ELHier{#1}[#2][\argb{\mathrm{P}}{#3}][#4]}
\newcommandx{\EPolHierE}[4][2=, 3=, 4=]
  {\EPolHier{#1}[#2][#3][#4]-\EComplexity}
\newcommandx{\EPolHierH}[4][2=, 3=, 4=]
  {\EPolHier{#1}[#2][#3][#4]-\HComplexity}
\newcommandx{\EPolHierC}[4][2=, 3=, 4=]
  {\EPolHier{#1}[#2][#3][#4]-\CComplexity}

\newcommandx{\UPolHier}[4][2=, 3=, 4=]
  {\ULHier{#1}[#2][\argb{\mathrm{P}}{#3}][#4]}
\newcommandx{\UPolHierE}[4][2=, 3=, 4=]
  {\UPolHier{#1}[#2][#3][#4]-\EComplexity}
\newcommandx{\UPolHierH}[4][2=, 3=, 4=]
  {\UPolHier{#1}[#2][#3][#4]-\HComplexity}
\newcommandx{\UPolHierC}[4][2=, 3=, 4=]
  {\UPolHier{#1}[#2][#3][#4]-\CComplexity}

\newcommandx{\DAriHier}[4][2=, 3=, 4=]
  {\DLHier{#1}[#2][\argb{0}{#3}][#4]}
\newcommandx{\DAriHierE}[4][2=, 3=, 4=]
  {\DAriHier{#1}[#2][#3][#4]-\EComplexity}
\newcommandx{\DAriHierH}[4][2=, 3=, 4=]
  {\DAriHier{#1}[#2][#3][#4]-\HComplexity}
\newcommandx{\DAriHierC}[4][2=, 3=, 4=]
  {\DAriHier{#1}[#2][#3][#4]-\CComplexity}

\newcommandx{\EAriHier}[4][2=, 3=, 4=]
  {\ELHier{#1}[#2][\argb{0}{#3}][#4]}
\newcommandx{\EAriHierE}[4][2=, 3=, 4=]
  {\EAriHier{#1}[#2][#3][#4]-\EComplexity}
\newcommandx{\EAriHierH}[4][2=, 3=, 4=]
  {\EAriHier{#1}[#2][#3][#4]-\HComplexity}
\newcommandx{\EAriHierC}[4][2=, 3=, 4=]
  {\EAriHier{#1}[#2][#3][#4]-\CComplexity}

\newcommandx{\UAriHier}[4][2=, 3=, 4=]
  {\ULHier{#1}[#2][\argb{0}{#3}][#4]}
\newcommandx{\UAriHierE}[4][2=, 3=, 4=]
  {\UAriHier{#1}[#2][#3][#4]-\EComplexity}
\newcommandx{\UAriHierH}[4][2=, 3=, 4=]
  {\UAriHier{#1}[#2][#3][#4]-\HComplexity}
\newcommandx{\UAriHierC}[4][2=, 3=, 4=]
  {\UAriHier{#1}[#2][#3][#4]-\CComplexity}

\newcommandx{\DAnaHier}[4][2=, 3=, 4=]
  {\DLHier{#1}[#2][\argb{1}{#3}][#4]}
\newcommandx{\DAnaHierE}[4][2=, 3=, 4=]
  {\DAnaHier{#1}[#2][#3][#4]-\EComplexity}
\newcommandx{\DAnaHierH}[4][2=, 3=, 4=]
  {\DAnaHier{#1}[#2][#3][#4]-\HComplexity}
\newcommandx{\DAnaHierC}[4][2=, 3=, 4=]
  {\DAnaHier{#1}[#2][#3][#4]-\CComplexity}

\newcommandx{\EAnaHier}[4][2=, 3=, 4=]
  {\ELHier{#1}[#2][\argb{1}{#3}][#4]}
\newcommandx{\EAnaHierE}[4][2=, 3=, 4=]
  {\EAnaHier{#1}[#2][#3][#4]-\EComplexity}
\newcommandx{\EAnaHierH}[4][2=, 3=, 4=]
  {\EAnaHier{#1}[#2][#3][#4]-\HComplexity}
\newcommandx{\EAnaHierC}[4][2=, 3=, 4=]
  {\EAnaHier{#1}[#2][#3][#4]-\CComplexity}

\newcommandx{\UAnaHier}[4][2=, 3=, 4=]
  {\ULHier{#1}[#2][\argb{1}{#3}][#4]}
\newcommandx{\UAnaHierE}[4][2=, 3=, 4=]
  {\UAnaHier{#1}[#2][#3][#4]-\EComplexity}
\newcommandx{\UAnaHierH}[4][2=, 3=, 4=]
  {\UAnaHier{#1}[#2][#3][#4]-\HComplexity}
\newcommandx{\UAnaHierC}[4][2=, 3=, 4=]
  {\UAnaHier{#1}[#2][#3][#4]-\CComplexity}

\newcommandx{\DBorHier}[4][2=, 3=, 4=]
  {\DBHier{#1}[#2][\argb{\mathrm{B}}{#3}][#4]}
\newcommandx{\DBorHierE}[4][2=, 3=, 4=]
  {\DBorHier{#1}[#2][#3][#4]-\EComplexity}
\newcommandx{\DBorHierH}[4][2=, 3=, 4=]
  {\DBorHier{#1}[#2][#3][#4]-\HComplexity}
\newcommandx{\DBorHierC}[4][2=, 3=, 4=]
  {\DBorHier{#1}[#2][#3][#4]-\CComplexity}

\newcommandx{\EBorHier}[4][2=, 3=, 4=]
  {\EBHier{#1}[#2][\argb{\mathrm{B}}{#3}][#4]}
\newcommandx{\EBorHierE}[4][2=, 3=, 4=]
  {\EBorHier{#1}[#2][#3][#4]-\EComplexity}
\newcommandx{\EBorHierH}[4][2=, 3=, 4=]
  {\EBorHier{#1}[#2][#3][#4]-\HComplexity}
\newcommandx{\EBorHierC}[4][2=, 3=, 4=]
  {\EBorHier{#1}[#2][#3][#4]-\CComplexity}

\newcommandx{\UBorHier}[4][2=, 3=, 4=]
  {\UBHier{#1}[#2][\argb{\mathrm{B}}{#3}][#4]}
\newcommandx{\UBorHierE}[4][2=, 3=, 4=]
  {\UBorHier{#1}[#2][#3][#4]-\EComplexity}
\newcommandx{\UBorHierH}[4][2=, 3=, 4=]
  {\UBorHier{#1}[#2][#3][#4]-\HComplexity}
\newcommandx{\UBorHierC}[4][2=, 3=, 4=]
  {\UBorHier{#1}[#2][#3][#4]-\CComplexity}

\newcommand{\EComplexity}
  {{\txtname{easy}}}

\newcommand{\HComplexity}
  {{\txtname{hard}}}

\newcommand{\CComplexity}
  {{\txtname{complete}}}

\newcommand{\UComplexity}
  {{\txtname{U}}}

\newcommand{\NComplexity}
  {{\txtname{N}}}

\newcommand{\CoComplexity}
  {{\txtname{Co}}}

%% file: Figures.tex
\usepackage{tikz}
\usetikzlibrary{arrows,shapes,calc, positioning}

\usepackage{wrapfig}

\tikzstyle{every node} =
  [draw = none, fill = white, thin]
\tikzstyle{every edge} +=
  [black, thick]

\tikzstyle{noall} =
  [draw = none, fill = none]
\tikzstyle{nodraw} =
  [draw = none, fill = white]
\tikzstyle{nofill} =
  [draw = black, fill = none]

\tikzstyle{cnode} =
  [circle, draw = black]
\tikzstyle{ssnode} =
  [regular polygon, regular polygon sides = 4, inner sep=0pt, draw = black!50]
\tikzstyle{snode} =
  [regular polygon, inner sep=1pt, regular polygon sides = 4,  draw = black!50]
\tikzstyle{lnode} =
  [diamond, inner sep=2pt, draw = black!75]
\tikzstyle{llnode} =
  [diamond, inner sep=0pt,draw = black!75]
\tikzstyle{pnode} =
  [regular polygon, regular polygon sides = 5, draw = black]

\tikzstyle{hassenode} =
  [inner sep = 0.5em, draw = black!75]
\tikzstyle{selectedhassenode} =
  [inner sep = 0.5em, draw = black, very thick]
\tikzstyle{nonselectedhassenode} =
  [inner sep = 0.5em, draw = black!75, dashed]

\tikzstyle{hasseedge} =
  [draw = black!75]
\tikzstyle{selectedhasseedge} =
  [very thick]
\tikzstyle{nonselectedhasseedge} =
  [draw = black!75, dashed]

\tikzstyle{hassenodex} =
  [inner sep = 0.5em, draw = black, shape = newhassenode]
\tikzstyle{selectedhassenodex} =
  [inner sep = 0.5em, draw = black!75, very thick, shape = newhassenode]
\tikzstyle{smallhassenodex} =
  [inner sep = 0.5em, draw = black!75,  shape = smallhassenode]
\tikzstyle{smalltophassenodex} =
  [inner sep = 0.5em, draw = black!75, shape = smalltophassenode]

\makeatletter

\pgfdeclareshape{newhassenode}
  {
  \inheritanchorborder[from=rectangle]
  \inheritanchor[from=rectangle]{north}
  \inheritanchor[from=rectangle]{west}
  \inheritanchor[from=rectangle]{east}
  \inheritanchor[from=rectangle]{south}
    \savedanchor\northeast
    {%
    \pgfmathsetlength\pgf@x{\pgfshapeminwidth}%
    \pgfmathsetlength\pgf@y{\pgfshapeminheight}%
    \pgf@x=0.6\pgf@x%
    \pgf@y=0.5\pgf@y%
    }
  \savedanchor\southwest
    {%
    \pgfmathsetlength\pgf@x{\pgfshapeminwidth}%
    \pgfmathsetlength\pgf@y{\pgfshapeminheight}%
    \pgf@x=-0.6\pgf@x%
    \pgf@y=-0.4\pgf@y%
    }
  \anchor{center}{\pgfpointorigin}
  \anchor{text}
    {%
    \pgfpointorigin%
    \advance\pgf@x by -.5\wd\pgfnodeparttextbox%
    \advance\pgf@y by -.5\ht\pgfnodeparttextbox%
    \advance\pgf@y by +.5\dp\pgfnodeparttextbox%
    }
  \anchor{N}
    {%
    \pgf@process{\northeast}%
    \pgf@x=0\pgf@x%
    \pgf@y=1\pgf@y%
    }
  \anchor{S}
    {%
    \pgf@process{\southwest}%
    \pgf@x=0\pgf@x%
    \pgf@y=1\pgf@y%
    }
  \backgroundpath
    {
    \pgfpathrectanglecorners{\southwest}{\northeast}
    }
  }

\pgfdeclareshape{smallhassenode}
  {
  \inheritanchorborder[from=rectangle]
  \inheritanchor[from=rectangle]{north}
  \inheritanchor[from=rectangle]{west}
  \inheritanchor[from=rectangle]{east}
  \inheritanchor[from=rectangle]{south}
    \savedanchor\northeast
    {%
    \pgfmathsetlength\pgf@x{\pgfshapeminwidth}%
    \pgfmathsetlength\pgf@y{\pgfshapeminheight}%
    \pgf@x=0.3\pgf@x%
    \pgf@y=0.3\pgf@y%
    }
  \savedanchor\southwest
    {%
    \pgfmathsetlength\pgf@x{\pgfshapeminwidth}%
    \pgfmathsetlength\pgf@y{\pgfshapeminheight}%
    \pgf@x=-0.3\pgf@x%
    \pgf@y=-0.4\pgf@y%
    }
  \anchor{center}{\pgfpointorigin}
  \anchor{text}
    {%
    \pgfpointorigin%
    \advance\pgf@x by -.5\wd\pgfnodeparttextbox%
    \advance\pgf@y by -.5\ht\pgfnodeparttextbox%
    \advance\pgf@y by +.5\dp\pgfnodeparttextbox%
    }
  \anchor{N}
    {%
    \pgf@process{\northeast}%
    \pgf@x=0\pgf@x%
    \pgf@y=1\pgf@y%
    }
  \anchor{S}
    {%
    \pgf@process{\southwest}%
    \pgf@x=0\pgf@x%
    \pgf@y=1\pgf@y%
    }
  \backgroundpath
    {
    \pgfpathrectanglecorners{\southwest}{\northeast}
    }
  }

\pgfdeclareshape{smalltophassenode}
  {
  \inheritanchorborder[from=rectangle]
  \inheritanchor[from=rectangle]{north}
  \inheritanchor[from=rectangle]{west}
  \inheritanchor[from=rectangle]{east}
  \inheritanchor[from=rectangle]{south}
    \savedanchor\northeast
    {%
    \pgfmathsetlength\pgf@x{\pgfshapeminwidth}%
    \pgfmathsetlength\pgf@y{\pgfshapeminheight}%
    \pgf@x=0.5\pgf@x%
    \pgf@y=0.3\pgf@y%
    }
  \savedanchor\southwest
    {%
    \pgfmathsetlength\pgf@x{\pgfshapeminwidth}%
    \pgfmathsetlength\pgf@y{\pgfshapeminheight}%
    \pgf@x=-0.5\pgf@x%
    \pgf@y=-0.4\pgf@y%
    }
  \anchor{center}{\pgfpointorigin}
  \anchor{text}
    {%
    \pgfpointorigin%
    \advance\pgf@x by -.5\wd\pgfnodeparttextbox%
    \advance\pgf@y by -.5\ht\pgfnodeparttextbox%
    \advance\pgf@y by +.5\dp\pgfnodeparttextbox%
    }
  \anchor{N}
    {%
    \pgf@process{\northeast}%
    \pgf@x=0\pgf@x%
    \pgf@y=1\pgf@y%
    }
  \anchor{S}
    {%
    \pgf@process{\southwest}%
    \pgf@x=0\pgf@x%
    \pgf@y=1\pgf@y%
    }
  \backgroundpath
    {
    \pgfpathrectanglecorners{\southwest}{\northeast}
    }
  }

\makeatother

\AfterEndPreamble
  {

  \newcommand{\figcorfam}
    {
    \begin{wrapfigure}[11]{R}{0.55\textwidth}
    \begin{center}
      \vspace{-4.25em}
      \footnotesize
      \scalebox{0.60}[0.65]
        {
        \begin{tikzpicture}[node distance =7.5em, bend angle = 15]

          \tikzset{every loop/.style = {max distance = 1.5em}}

          \node [lnode]
                (a0)
                []
                {$\alpha_{0}/5$};
          \node [snode]
                (a1)
                [node distance = 8.5em, right of = a0]
                {$\alpha_{1}/6$};
          \node [lnode]
                (a2)
                [node distance = 8.5em,right of = a1]
                {$\alpha_{2}/7$};
          \node [snode]
                (a3)
                [node distance = 8.5em, right of = a2]
                {$\alpha_{3}/8$};
          \node [lnode]
                (a4)
                [node distance = 8.5em,right of = a3]
                {$\alpha_{4}/9$};
          \node [lnode]
                (b0)
                [below left of = a0, xshift=1.5em]
                {$\beta_{0}/0$};
          \node [snode]
                (g0)
                [below of = b0]
                {$\gamma_{0}/0$};
          \node [snode]
                (b1)
                [below left of = a1, , xshift=.5em]
                {$\beta_{1}/1$};
          \node [lnode]
                (g1)
                [below of = b1]
                {$\gamma_{1}/1$};
          \node [lnode]
                (b2)
                [below left of = a2, xshift=.5em]
                {$\beta_{2}/2$};
          \node [snode]
                (g2)
                [below of = b2]
                {$\gamma_{2}/2$};
          \node [snode]
                (b3)
                [below left of = a3, xshift=.5em]
                {$\beta_{3}/3$};
          \node [lnode]
                (g3)
                [below of = b3]
                {$\gamma_{3}/3$};
          \node [lnode]
                (b4)
                [below left of = a4, xshift=.5em]
                {$\beta_{4}/4$};
          \node [snode]
                (g4)
                [below of = b4]
                {$\gamma_{4}/4$};

          \path[->]
            (a0) edge [bend right]
                      (b0)
            (a1) edge [bend right]
                      (b1)
            (a2) edge [bend right]
                      (b2)
            (a3) edge [bend right]
                      (b3)
            (a4) edge [bend right]
                      (b4)
            (b1) edge [bend right]
                      (a0)
            (b2) edge [bend right]
                      (a1)
            (b3) edge [bend right]
                      (a2)
            (b4) edge [bend right]
                      (a3)
            (b0) edge [bend right]
                      (g0)
            (b1) edge [bend right]
                      (g1)
            (b2) edge [bend right]
                      (g2)
            (b3) edge [bend right]
                      (g3)
            (b4) edge [bend right]
                      (g4)
            (g0) edge [bend right]
                      (b0)
            (g1) edge [bend right]
                      (b1)
            (g2) edge [bend right]
                      (b2)
            (g3) edge [bend right]
                      (b3)
            (g4) edge [bend right]
                      (b4)
            (g0) edge [loop left]
                      ()
            (g1) edge [loop left]
                      ()
            (g2) edge [loop left]
                      ()
            (g0) edge [loop left]
                      ()
            (g3) edge [loop left]
                      ()
            (g4) edge [loop left]
                      ()
            (g0) edge [bend angle = 35, bend right, out=-90, in=-120,
                      looseness=1.5]
                      (a1)
            (g1) edge [bend angle = 35, bend right, , out=-90, in=-120,
                      looseness=1.5]
                      (a2)
            (g2) edge [bend angle = 35, bend right, out=-90, in=-120,
                      looseness=1.5]
                      (a3)
            (g3) edge [bend angle = 35, bend right, out=-90, in=-120,
                      looseness=1.5]
                      (a4)
            ;

          \begin{scope}
            [very thick, draw = gray!75, fill = gray!75, fill opacity = 0.25,
            draw opacity = 0.40]

            \filldraw
              ($(a0) + (-0.90, 0.90)$)
                to [out = 180, in = 90]
              ($(g0) + (-0.90, 0)$)
                to [out = 270, in = 180]
              ($(g2) + (0, -1.50)$)
                to [out = 0, in = 270]
              ($(a2) + (0.40, -1.50)$)
                to [out = 90, in = 315]
              ($(a2) + (0.90, 0.40)$)
                to [out = 135, in = 0]
              ($(a0) + (-0.90, 0.90)$)
              ;

          \end{scope}

        \end{tikzpicture}
        }
        \vspace{-4.25em}
        \captionof{figure}{\label{fig:corfam} Game $\GamName[\CSym][2]$ of the
        core family.}
    \end{center}
    \end{wrapfigure}
    }

  \newcommand{\figrectree}
    {
    \begin{wrapfigure}[21]{R}{0.55\textwidth}
     \vspace{-7.75em}
      \parbox{0.450\textwidth}
        {
        \centering
        \footnotesize
        \scalebox{1.1}[1]
          {
          \begin{tikzpicture}
            [node distance = 8.5em, bend angle = 40, minimum size = 8em]

            \node [hassenodex]
                  (G)
                  []
                  {\usebox{\boxfigGIbase}};
            \node [noall]
                  (Gl)
                  [below of = G, yshift = 4em]
                  {$\GamName[\varepsilon][1]$};
            \node [noall]
                  (X)
                  [node distance = 10em, below of = G]
                  {};
            \node [hassenodex]
                  (GLH)
                  [left of = X, xshift = 2em]
                  {\usebox{\boxfigGIlh}};
            \node [noall]
                  (GLHl)
                  [below of = GLH, yshift = 4.30em, xshift = 1em]
                  {$\der{\GamName}[\LSym][1]$};
            \node [hassenodex]
                  (GRH)
                  [right of = X, xshift = -2em]
                  {\usebox{\boxfigGIrh}};
            \node [noall]
                  (GRHl)
                  [below of = GRH, yshift = 4.30em, xshift = -1em]
                  {$\der{\GamName}[\RSym][1]$};
            \node [hassenodex]
                  (GL)
                  [below of = GLH, yshift = -1em]
                  {\usebox{\boxfigGIl}};
            \node [noall]
                  (GLl)
                  [below of = GL, yshift = 4em]
                  {$\GamName[\LSym][1]$};
            \node [hassenodex]
                  (GR)
                  [below of = GRH, yshift = -1em]
                  {\usebox{\boxfigGIr}};
            \node [noall]
                  (GRl)
                  [below of = GR, yshift = 4em]
                  {$\GamName[\RSym][1]$};
            \node [noall]
                  (XFL)
                  [node distance = 9em, below of = GL]
                  {};
            \node [smalltophassenodex]
                  (GLLH)
                  [left of = XFL, xshift = 5em]
                  {\usebox{\boxfigGIllh}};
            \node [noall]
                  (GLLHl)
                  [below of = GLLH, yshift = 4.5em]
                  {$\der{\GamName}^{1}_{\LSym\LSym}$};
            \node [smallhassenodex]
                  (GLRH)
                  [right of = XFL, xshift = -5em]
                  {\usebox{\boxfigGIlrh}};
            \node [noall]
                  (GLRHl)
                  [below of = GLRH, yshift = 4.5em]
                  {$\der{\GamName}^{1}_{\LSym\RSym}$};
            \node [noall]
                  (XFR)
                  [node distance = 9em, below of = GR]
                  {};
            \node [smallhassenodex]
                  (GRLH)
                  [left of = XFR, xshift = 5em]
                  {\usebox{\boxfigGIrlh}};
            \node [noall]
                  (GRLHl)
                  [below of = GRLH, yshift = 4.5em]
                  {$\der{\GamName}^{1}_{\RSym\LSym}$};
            \node [smallhassenodex]
                  (GRRH)
                  [right of = XFR, xshift = -6em]
                  {$\emptyset$};
            \node [noall]
                  (GRRHl)
                  [below of = GRRH, yshift = 4.5em]
                  {$\der{\GamName}^{1}_{\RSym\RSym}$};

            \path[-]
              (G.S)   edge  [hasseedge]
                            (GLH.N)
                      edge  [hasseedge]
                            (GRH.N)
              (GLH.S) edge  [hasseedge]
                            (GL.N)
              (GRH.S) edge  [hasseedge]
                            (GR.N)
              (GL.S)  edge  [hasseedge]
                            (GLLH.N)
                      edge  [hasseedge]
                            (GLRH.N)
              (GR.S)  edge  [hasseedge]
                            (GRLH.N)
                      edge  [hasseedge]
                            (GRRH.N)
              ;
          \end{tikzpicture}
          }
          \vspace{-5em}
          \caption{\label{fig:rectre} The induced subgame tree $\IndSubTre[][1]$
          of $\GamName[][1]$.}
        }
    \end{wrapfigure}
    }

  \newsavebox{\boxfigGIbase}
  \savebox{\boxfigGIbase}
    {
    \scalebox{0.43}[0.43]
      {
      \begin{tikzpicture}[node distance = 4.5em, bend angle = 15]

          \tikzset{every loop/.style = {max distance = 1.5em}}

          \node [lnode]
                (a0)
                []
                {{\large $\alpha_{0}$}};
          \node [opacity=0]
                (ghost)
                [left of = a0, yshift = 3em]
                {};
          \node [snode]
                (a1)
                [node distance = 6em, right of = a0]
                {$\alpha_{1}$};
          \node [lnode]
                (a2)
                [node distance = 6em,right of = a1]
                {$\alpha_{2}$};
          \node [lnode]
                (b0)
                [below left of = a0]
                {$\beta_{0}$};
          \node [snode]
                (g0)
                [below of = b0]
                {$\gamma_{0}$};
          \node [snode]
                (b1)
                [below left of = a1]
                {$\beta_{1}$};
          \node [lnode]
                (g1)
                [below of = b1]
                {$\gamma_{1}$};
          \node [lnode]
                (b2)
                [below left of = a2]
                {$\beta_{2}$};
          \node [snode]
                (g2)
                [below of = b2]
                {$\gamma_{2}$};

          \path[->]
            (a0) edge  [bend right]
                      (b0)
            (a1) edge  [bend right]
                      (b1)
            (a2) edge  [bend right]
                      (b2)
            (b1)    edge  [bend right]
                      (a0)
            (b2)    edge  [bend right]
                      (a1)
            (b0)    edge  [bend right]
                      (g0)
            (b1)    edge  [bend right]
                      (g1)
            (b2)    edge  [bend right]
                      (g2)
            (g0)    edge  [bend right]
                      (b0)
            (g1)    edge  [bend right]
                      (b1)
            (g2)    edge  [bend right]
                      (b2)
            (g0)    edge  [loop left]
                      ()
            (g1)    edge  [loop left]
                      ()
            (g2)    edge  [loop left]
                      ()
            (g0)    edge  [loop left]
                      ()
            (g0)    edge  [bend angle = 35, bend right, out=-90, in=-120,
looseness=1.5]
                      (a1)
            (g1)    edge  [bend angle = 35, bend right, , out=-90, in=-120,
looseness=1.5]
                      (a2)
            ;

        \end{tikzpicture}
    }
  }

\newsavebox{\boxfigGIlh}
  \savebox{\boxfigGIlh}
    {
    \scalebox{0.40}[0.40]
      {
      \begin{tikzpicture}[node distance = 4.5em, bend angle = 15]

          \tikzset{every loop/.style = {max distance = 1.5em}}

          \node [lnode]
                (a0)
                []
                {$\alpha_{0}$};
          \node [opacity=0]
                (ghost)
                [left of = a0, yshift = 2.8em]
                {};
          \node [snode]
                (a1)
                [node distance = 6em, right of = a0]
                {$\alpha_{1}$};
          \node [lnode]
                (b0)
                [below left of = a0]
                {$\beta_{0}$};
          \node [snode]
                (g0)
                [below of = b0]
                {$\gamma_{0}$};
          \node [snode]
                (b1)
                [below left of = a1]
                {$\beta_{1}$};
          \node [lnode]
                (g1)
                [below of = b1]
                {$\gamma_{1}$};
          \node [lnode]
                (b2)
                [below right of = a1]
                {$\beta_{2}$};
          \node [snode]
                (g2)
                [below of = b2]
                {$\gamma_{2}$};

          \path[->]
            (a0) edge  [bend right]
                      (b0)
            (a1) edge  [bend right]
                      (b1)
            (b1)    edge  [bend right]
                      (a0)
            (b2)    edge  [bend right]
                      (a1)
            (b0)    edge  [bend right]
                      (g0)
            (b1)    edge  [bend right]
                      (g1)
            (b2)    edge  [bend right]
                      (g2)
            (g0)    edge  [bend right]
                      (b0)
            (g1)    edge  [bend right]
                      (b1)
            (g2)    edge  [bend right]
                      (b2)
            (g0)    edge  [loop left]
                      ()
            (g1)    edge  [loop left]
                      ()
            (g2)    edge  [loop left]
                      ()
            (g0)    edge  [loop left]
                      ()
            (g0)    edge  [bend angle = 35, bend right, out=-90, in=-120,
looseness=1.5]
                      (a1)
            ;

        \end{tikzpicture}
    }
  }

\newsavebox{\boxfigGIrh}
  \savebox{\boxfigGIrh}
    {
    \scalebox{0.40}[0.40]
      {
      \begin{tikzpicture}[node distance = 4.5em, bend angle = 15]

          \tikzset{every loop/.style = {max distance = 1.5em}}

          \node [lnode]
                (a0)
                []
                {$\alpha_{0}$};
          \node [opacity=0]
                (ghost)
                [left of = a0, yshift = 2.8em]
                {};
          \node [snode]
                (a1)
                [node distance = 6em, right of = a0]
                {$\alpha_{1}$};
          \node [lnode]
                (b0)
                [below left of = a0]
                {$\beta_{0}$};
          \node [snode]
                (g0)
                [below of = b0]
                {$\gamma_{0}$};
          \node [snode]
                (b1)
                [below left of = a1]
                {$\beta_{1}$};

          \path[->]
            (a0) edge  [bend right]
                      (b0)
            (a1) edge  [bend right]
                      (b1)
            (b1)    edge  [bend right]
                      (a0)
            (b0)    edge  [bend right]
                      (g0)
            (g0)    edge  [bend right]
                      (b0)
            (g0)    edge  [loop left]
                      ()
            (g0)    edge  [loop left]
                      ()
            (g0)    edge  [bend angle = 35, bend right, out=-90, in=-120,
looseness=1.5]
                      (a1)
            ;

        \end{tikzpicture}
    }
  }

\newsavebox{\boxfigGIl}
  \savebox{\boxfigGIl}
    {
    \scalebox{0.40}[0.40]
      {
      \begin{tikzpicture}[node distance = 4.5em, bend angle = 15]

          \tikzset{every loop/.style = {max distance = 1.5em}}

          \node [lnode]
                (a0)
                []
                {$\alpha_{0}$};
          \node []
                (a1)
                [node distance = 6em, right of = a0]
                {};
          \node [lnode]
                (b0)
                [below left of = a0]
                {$\beta_{0}$};
          \node [snode]
                (g0)
                [below of = b0]
                {$\gamma_{0}$};
          \node [snode]
                (b1)
                [below left of = a1]
                {$\beta_{1}$};
          \node [lnode]
                (g1)
                [below of = b1]
                {$\gamma_{1}$};
          \node []
                (b2)
                [below right of = a1]
                {};
          \node [snode]
                (g2)
                [below of = b2]
                {$\gamma_{2}$};
          \node []
                (ghost)
                [below of = g2, yshift = 1em]
                {};

          \path[->]
            (a0) edge  [bend right]
                      (b0)
            (b1)    edge  [bend right]
                      (a0)
            (b0)    edge  [bend right]
                      (g0)
            (b1)    edge  [bend right]
                      (g1)
            (g0)    edge  [bend right]
                      (b0)
            (g1)    edge  [bend right]
                      (b1)
            (g0)    edge  [loop left]
                      ()
            (g1)    edge  [loop left]
                      ()
            (g2)    edge  [loop left]
                      ()
            (g0)    edge  [loop left]
                      ()
            ;

        \end{tikzpicture}
    }
  }

\newsavebox{\boxfigGIr}
  \savebox{\boxfigGIr}
    {
    \scalebox{0.40}[0.40]
      {
      \begin{tikzpicture}[node distance = 4.5em, bend angle = 15]

          \tikzset{every loop/.style = {max distance = 1.5em}}

          \node [lnode]
                (a0)
                []
                {$\alpha_{0}$};
          \node [lnode]
                (b0)
                [below left of = a0]
                {$\beta_{0}$};
          \node [snode]
                (g0)
                [below of = b0]
                {$\gamma_{0}$};
          \node [snode]
                (b1)
                [below right of = a0]
                {$\beta_{1}$};
          \node []
                (ghost)
                [below of = g0, yshift = 1em]
                {};

          \path[->]
            (a0) edge  [bend right]
                      (b0)
            (b1)    edge  [bend right]
                      (a0)
            (b0)    edge  [bend right]
                      (g0)
            (g0)    edge  [bend right]
                      (b0)
            (g0)    edge  [loop left]
                      ()
            (g0)    edge  [loop left]
                      ()
            ;

        \end{tikzpicture}
    }
  }

\newsavebox{\boxfigGIllh}
  \savebox{\boxfigGIllh}
    {
    \scalebox{0.40}[0.40]
      {
      \begin{tikzpicture}[node distance = 4.5em, bend angle = 15]

          \tikzset{every loop/.style = {max distance = 1.5em}}

          \node []
                (a0)
                []
                {};
          \node []
                (a1)
                [node distance = 6em, right of = a0]
                {};
          \node [lnode]
                (b0)
                [below left of = a0]
                {$\beta_{0}$};
          \node [snode]
                (g0)
                [below of = b0]
                {$\gamma_{0}$};
          \node []
                (b1)
                [below left of = a1]
                {};
          \node [lnode]
                (g1)
                [below of = b1]
                {$\gamma_{1}$};
          \node []
                (b2)
                [below right of = a1]
                {};
          \node [snode]
                (g2)
                [below of = b2]
                {$\gamma_{2}$};

          \path[->]
            (b0)    edge  [bend right]
                      (g0)
            (g0)    edge  [bend right]
                      (b0)
            (g0)    edge  [loop left]
                      ()
            (g1)    edge  [loop left]
                      ()
            (g2)    edge  [loop left]
                      ()
            (g0)    edge  [loop left]
                      ()
            ;

        \end{tikzpicture}
    }
  }

\newsavebox{\boxfigGIlrh}
  \savebox{\boxfigGIlrh}
    {
    \hspace{-1em}
    \scalebox{0.40}[0.40]
      {
      \begin{tikzpicture}[node distance = 4.5em, bend angle = 15]

          \tikzset{every loop/.style = {max distance = 1.5em}}

          \node []
                (a0)
                []
                {};
          \node [snode]
                (b1)
                [below right of = a0]
                {$\beta_{1}$};
          \node [lnode]
                (g1)
                [below of = b1]
                {$\gamma_{1}$};

          \path[->]
            (b1)    edge  [bend right]
                      (g1)
            (g1)    edge  [bend right]
                      (b1)
            (g1)    edge  [loop left]
                      ()
            ;

        \end{tikzpicture}
    }
  }

\newsavebox{\boxfigGIrlh}
  \savebox{\boxfigGIrlh}
    {
    \hspace{0.5em}
    \scalebox{0.40}[0.40]
      {
      \begin{tikzpicture}[node distance = 4.5em, bend angle = 15]

          \tikzset{every loop/.style = {max distance = 1.5em}}

          \node []
                (a0)
                []
                {};
          \node [lnode]
                (b0)
                [below left of = a0]
                {$\beta_{0}$};
          \node [snode]
                (g0)
                [below of = b0]
                {$\gamma_{0}$};

          \path[->]
            (b0)    edge  [bend right]
                      (g0)
            (g0)    edge  [bend right]
                      (b0)
            (g0)    edge  [loop left]
                      ()
            ;

        \end{tikzpicture}
    }
  }

  \newcommand{\figsccfam}
    {
    \begin{wrapfigure}[10]{R}{0.425\textwidth}
    \vspace{-6em}
    \begin{center}
      {
        \footnotesize
        \scalebox{0.75}[0.75] {
  \begin{tikzpicture}
    \draw [draw=none] (4,6.6) -- (8,4.4) node (a1) [llnode,pos=0.34]
{$\delta^{0}_{1,2}$} node (a2) [ssnode,pos=0.66] {$\delta^{1}_{1,2}$};
          \draw [draw=none] (8,4.4) -- (6.2,0) node (a3) [ssnode,pos=0.34]
{$\delta^{1}_{0,1}$} node (a4) [llnode,pos=0.66] {$\delta^{0}_{0,1}$};
          \draw [draw=none] (6.2,0) -- (1.8,0) node (a5) [llnode,midway]
{$\delta^{0}_{0,4}$};
          \draw [draw=none] (1.8,0) -- (0,4.4) node (a6) [llnode,pos=0.34]
{$\delta^{0}_{3,4}$} node (a7) [ssnode,pos=0.66] {$\delta^{1}_{3,4}$};
          \draw [draw=none] (0,4.4) -- (4,6.6) node (a8) [ssnode,pos=0.34]
{$\delta^{1}_{2,3}$} node (a9) [llnode,pos=0.66] {$\delta^{0}_{2,3}$};
    \draw [draw=none] (4,6.6) -- (6.2,0) node (b1) [llnode,pos=.475]
{$\delta^{0}_{0,2}$};
          \draw [draw=none] (4,6.6) -- (1.8,0) node (b2) [llnode,pos=.475]
{$\delta^{0}_{2,4}$};
          \draw [draw=none] (8,4.4) -- (0,4.4) node (b3) [ssnode,midway]
{$\delta^{1}_{1,3}$};
          \draw [draw=none] (8,4.4) -- (1.8,0) node (b4) [ssnode,near start]
{$\delta^{1}_{1,4}$} node (b5) [llnode,pos=.535] {$\delta^{0}_{1,4}$};
          \draw [draw=none] (0,4.4) -- (6.2,0) node (b6) [ssnode,near start]
{$\delta^{1}_{0,3}$} node (b7) [llnode,pos=.535] {$\delta^{0}_{0,3}$};
          \draw (4,6.6) node (g2) [snode, very thick, draw = black]
{$\gamma_{2}$};
          \draw (8,4.4) node (g1) [lnode, very thick, draw = black]
{$\gamma_{1}$};
          \draw (6.2,0) node (g0) [snode, very thick, draw = black]
{$\gamma_{0}$};
          \draw (1.8,0) node (g4) [snode, very thick, draw = black]
{$\gamma_{4}$};
          \draw (0,4.4) node (g3) [lnode, very thick, draw = black]
{$\gamma_{3}$};
          \draw [<->] (a1) -- (a2);
          \draw [<->] (g2) -- (a1);
          \draw [<->] (a2) -- (g1);
          \draw [<->] (a3) -- (a4);
          \draw [<->] (g1) -- (a3);
          \draw [<->] (a4) -- (g0);
          \draw [<->] (g0) -- (a5);
          \draw [<->] (a5) -- (g4);
          \draw [<->] (a6) -- (a7);
          \draw [<->] (g3) -- (a7);
          \draw [<->] (a6) -- (g4);
          \draw [<->] (a8) -- (a9);
          \draw [<->] (g2) -- (a9);
          \draw [<->] (a8) -- (g3);
          \draw [<->] (g2) -- (b1);
          \draw [<->] (b1) -- (g0);
          \draw [<->] (g2) -- (b2);
          \draw [<->] (b2) -- (g4);
          \draw [<->] (g3) -- (b3);
          \draw [<->] (b3) -- (g1);
          \draw [<->] (b6) -- (b7);
          \draw [<->] (g3) -- (b6);
          \draw [<->] (b7) -- (g0);
          \draw [<->] (b4) -- (b5);
          \draw [<->] (g1) -- (b4);
          \draw [<->] (b5) -- (g4);
  \end{tikzpicture}
        }
      }
    \vspace{-1.55em}
    \captionof{figure}{\label{fig:sccfam} Clique of $\GamName[][2]$.}
    \end{center}
  \end{wrapfigure}
  }

  }

%% file: Algorithms.tex
\usepackage[ruled,vlined]{algorithm2e}

\DontPrintSemicolon

\SetKw{Signature}{signature}
\SetKwFor{Function}{function}{}{}

\SetKwFor{Let}{let}{in}{}

\AfterEndPreamble
  {

  \newcommand{\algrec}
    {
    \begin{algorithm}[H]
      \caption{\label{alg:rec} Recursive algorithm.}
      \Signature{$\solFun \colon \PG \cto[\GamName] \pow{\PosSet[\GamName]}
        \times \pow{\PosSet[\GamName]}$} \;
      \Function{$\solFun(\GamName)$}
        {
        \nl $(\GamName[\LSym], \wp) \gets \fFun[\LSym](\GamName)$ \;
        \nl $(\WinSet[\LSym][0], \WinSet[\LSym][1]) \gets
            \solFun(\GamName[\LSym])$ \;
        \nl \eIf{$\preFun[\GamName][\dual{\wp}](\WinSet[\LSym][\dual{\wp}])
            \setminus \WinSet[\LSym][\dual{\wp}] = \emptyset$}
          {
          \nl $(\WinSet[][\wp], \WinSet[][\dual{\wp}]) \gets (\PosSet[\GamName]
              \!\setminus\! \WinSet[\LSym][\dual{\wp}],
              \WinSet[\LSym][\dual{\wp}])$ \;
          }
          {
          \nl $\GamName[\RSym] \gets \fFun[\RSym](\GamName,
              \WinSet[\LSym][\dual{\wp}], \dual{\wp})$ \;
          \nl $(\WinSet[\RSym][0], \WinSet[\RSym][1]) \gets
              \solFun(\GamName[\RSym])$ \;
          \nl $(\WinSet[][\wp], \WinSet[][\dual{\wp}]) \gets
              (\WinSet[\RSym][\wp], \PosSet[\GamName] \!\setminus\!
              \WinSet[\RSym][\wp])$ \;
          }
        \nl \Return $(\WinSet[][0], \WinSet[][1])$ \;
        }
    \end{algorithm}
    }

  \newcommand{\algsubcal}
    {
    \begin{minipage}[t]{\textwidth}
      \vspace{-1.00em}
      \begin{algorithm}[H]
        \caption{\label{alg:lefsubcal} Left-subgame function.}
        \Signature{$\fFun[\LSym] \colon \PG \to \PG \times \{ 0, 1 \}$} \;
        \Function{$\fFun[\LSym](\GamName)$}
          {
          \nl $\wp \gets \prtFun(\GamName) \bmod{2}$ \;
          \nl $\GamName[][\star] \gets \GamName \setminus
              \atrFun[\GamName][\wp](\prtFun[\GamName][-1](\prtFun(\GamName)))$
              \;
          \nl \Return $(\GamName[][\star], \wp)$ \;
          }
      \end{algorithm}
    \end{minipage}
    \begin{minipage}[t]{\textwidth}
      \vspace{0.7em}
      \begin{algorithm}[H]
        \caption{\label{alg:rigsubcal} Right-subgame function.\!\!}
        \Signature{$\fFun[\RSym] \colon \PG \ctimes[\GamName]\,
          \pow{\PosSet[\GamName]} \!\times \{ 0, 1 \} \to \PG$\!\!} \;
        \Function{$\fFun[\RSym](\GamName, \WSet, \wp)$}
          {
          \nl $\GamName[][\star] \gets \GamName \setminus
              \atrFun[\GamName][\wp](\WSet)$ \;
          \nl \Return $\GamName[][\star]$ \;
          }
      \end{algorithm}
    \end{minipage}
    }

  }

%% file: Abstract.tex
%%****************************************************************************%%
%%                                                                            %%
%% Robust Worst Cases for Divide-et-Impera Algorithms for Parity Games        %%
%%                                                                            %%
%% Abstract.tex                                                               %%
%%                                                                            %%
%% Revision 0                                                                 %%
%%                                                                            %%
%% Copyright (C) 2017, Massimo Benerecetti, Daniele Dell'Erba, and            %%
%%                     Fabio Mogavero.                                        %%
%% All rights reserved.                                                       %%
%%                                                                            %%
%%****************************************************************************%%

% Begin of file Abstract.tex

\begin{abstract}

  The McNaughton-Zielonka \divideetimpera algorithm is the simplest and most
  flexible approach available in the literature for determining the winner in a
  parity game.
  Despite its theoretical exponential worst-case complexity and the negative
  reputation as a poorly effective algorithm in practice, it has been shown to
  rank among the best techniques for the solution of such games.
  Also, it proved to be resistant to a lower bound attack, even more than the
  strategy improvements approaches, and only recently a family of games on
  which the algorithm requires exponential time has been provided by Friedmann.
  An easy analysis of this family shows that a simple memoization technique can
  help the algorithm solve the family in polynomial time.
  The same result can also be achieved by exploiting an approach based on the
  dominion-decomposition techniques proposed in the literature.
  These observations raise the question whether a suitable combination of
  dynamic programming and game-decomposition techniques can improve on the
  exponential worst case of the original algorithm.
  In this paper we answer this question negatively, by providing a robustly
  exponential worst case, showing that no possible intertwining of the above
  mentioned techniques can help mitigating the exponential nature of the
  \divideetimpera approaches.

\end{abstract}

% End of file Abstract.tex

%%% Local Variables:
%%% mode: latex
%%% TeX-master: "Article"
%%% End:

%% file: Introduction.tex
%%****************************************************************************%%
%%                                                                            %%
%% Robust Worst Cases for Divide-et-Impera Algorithms for Parity Games        %%
%%                                                                            %%
%% Introduction.tex                                                           %%
%%                                                                            %%
%% Revision 0                                                                 %%
%%                                                                            %%
%% Copyright (C) 2017, Massimo Benerecetti, Daniele Dell'Erba, and            %%
%%                     Fabio Mogavero.                                        %%
%% All rights reserved.                                                       %%
%%                                                                            %%
%%****************************************************************************%%

% Begin of file Introduction.tex

\begin{section}{Introduction}

  \emph{Parity games}~\cite{Mos91} are perfect-information two-player turn-based
  games of infinite duration, usually played on finite directed graphs.
  Their vertices, labeled by natural numbers called \emph{priorities}, are
  assigned to one of two players, named \emph{Even} and \emph{Odd} or, simply,
  $0$ and $1$, respectively.
  A play in the game is an infinite sequence of moves between vertices and it is
  said to be winning for player $0$ (\resp, $1$), if the maximal priority
  encountered infinitely often along the play is \emph{even} (\resp, odd).
  These games have been extensively studied in the attempt to find efficient
  solutions to the problem of determining the winner.
  From a complexity theoretic perspective, this decision problem lies in \NPTime
  $\cap$ \CoNPTime~\cite{EJS01}, since it is \emph{memoryless
  determined}~\cite{Mos91,EJ91,Mar75,Mar85}.
  It has been even proved to belong to \UPTime $\cap$ \CoUPTime~\cite{Jur98}
  and, very recently, to be solvable in quasi-polynomial time~\cite{CJKLS17}.
  They are the simplest class of games in a wider family with similar
  complexities and containing, \eg, \emph{mean payoff games}~\cite{EM79,GKK90},
  \emph{discounted payoff games}~\cite{ZP96}, and \emph{simple stochastic
  games}~\cite{Con92}.
  In fact, polynomial time reductions exist from parity games to the latter
  ones.
  However, despite being the most likely class among those games to admit a
  polynomial-time solution, the answer to the question whether such a solution
  exists still remains elusive.
  The effort devoted to provide efficient solutions stems primarily from the
  fact that many problems in formal verification and synthesis can be
  reformulated in terms of solving parity games.
  Emerson, Jutla, and Sistla~\cite{EJS01} have shown that computing winning
  strategies for these games is linear-time equivalent to solving the modal \MC
  model checking problem~\cite{EL86}.
  Parity games also play a crucial role in automata
  theory~\cite{Mos84,EJ91,KV98}, where they can be applied to solve the
  complementation problem for alternating automata~\cite{GTW02} and the
  emptiness of the corresponding nondeterministic tree automata~\cite{KV98}.
  These automata, in turn, can be used to solve the satisfiability and model
  checking problems for expressive logics, such as the modal~\cite{Wil01} and
  alternating~\cite{SF06,AHK02} \MC, \ATLS~\cite{AHK02,Sch08}, Strategy
  Logic~\cite{CHP10,MMV10a,MMPV12,MMPV14,MMPV17}, Substructure Temporal
  Logic~\cite{BMM13,BMM15}, and fixed-point extensions of guarded first-order
  logics~\cite{BG04}.
  \\\indent
  Previous exponential solutions essentially divide into two families.
  The first one collects procedures that attempt to directly build winning
  strategies for the two players on the entire game.
  To such family belongs the \emph{Small Progress Measure} approach by
  Jurdzi\'nski~\cite{Jur00}, which exploits the connection between the notions
  of progress measures~\cite{KK91} and winning strategies.
  A second approach in same vein is the \emph{Strategy Improvement} algorithm by
  Jurdzi\'nski and V\"oge~\cite{VJ00}, based on the idea of iteratively
  improving an initial, non necessarily winning, strategy.
  \\\indent
  The second family gathers, instead, the approaches based on decomposing the
  solution of a game into the analysis of its subgames.
  To this family belong the so called \divideetimpera approaches led by the
  \emph{Recursive} algorithm proposed by Zielonka~\cite{Zie98}, which adapts to
  parity games an earlier algorithm proposed by McNaughton for Muller
  games~\cite{McN93}.
  Intuitively, it decomposes the input game into subgames and solves them
  recursively.
  Using the Recursive algorithm as a back-end, and in the attempt to obtain a
  better upper bound, the \emph{Dominion Decomposition}~\cite{JPZ06,JPZ08} and
  the \emph{Big Step}~\cite{Sch07} approaches were devised.
  Both share the idea of intertwining the recursive calls of the back-end with a
  preprocessing phase, applied to the current subgame, in search of a
  sufficiently small dominion for some player $\wp$, \ie, a set of positions
  from where $\wp$ wins without ever exiting the set.
  The first technique does so by means of a brute force search, while the second
  one exploits a suitable variation of the Small Progress Measure procedure.
  A different direction has been followed recently within the
  decomposition-based family, that leads to a novel solution technique based on
  the notion of \emph{priority promotion}~\cite{BDM16, BDM16a, BDM16b}.
  The approach relies on a new procedure that finds dominions of arbitrary size,
  which proved to be quite efficient in practice and exhibits the best space
  complexity among the known solution algorithms, even better than the recently
  introduced quasi-linear space algorithms~\cite{JL17,FJSSW17}.
  \\\indent
  The literature also suggests several heuristics to tune parity game solvers.
  One of the most successful ones is that of decomposing the game into
  strongly-connected components (SCCs, for short) and solving it SCC-wise.
  \emph{SCC-decomposition}, together with some other minor techniques such as
  removal of self-cycles and priority compression, can significantly improve the
  solution process, as empirically demonstrated in~\cite{FL09}.
  The same authors also show that, against the negative reputation as far as
  performances are concerned, the Recursive algorithm often stands out as the
  best solver among those proposed in the literature, particularly when paired
  with the SCC-decomposition heuristic.
  Despite having a quite straightforward exponential upper bound, this algorithm
  has resisted an exponential lower bound for more than ten years, until
  Friedmann~\cite{Fri11a} devised an indexed family of games that forces the
  algorithm to execute a number of recursive calls that grows exponentially with
  the index.
  The family is also resilient to the SCC-decomposition technique, since each
  subgame passed to a recursive call always forms a single SCC.
  On a closer look, however, the games proposed there force an exponential
  behavior by requiring the algorithm to repeatedly solve a small number of
  subgames, actually only a linear number of them.
  As a consequence, all those games are amenable to a polynomial-time solution,
  by simply providing the algorithm with a suitable memoization mechanism that
  prevents it from wasting computational resources on solving already solved
  subgames.
  For different reasons, also a dominion decomposition approach can break the
  lower bound easily, as most of the subgames of a game in the family contain a
  dominion of constant size.
  %
  % [This observations shifts the question of the existence of an exponential
  % lower bound from the original version of the recursive algorithm to any
  % variant of the algorithm that intertwines memoization, scc-decomposition
  % and dominion decomposition techniques.]
  %
  \\\indent
  These observations raise the question whether the Recursive algorithm admits
  an exponential lower bound robust enough to be resilient to a suitable
  intertwining with memoization, SCC-decomposition, and dominion decomposition
  techniques.
  The difficulty here is that such a robust worst case should induce an
  exponential number of different subgames to prevent memoization from being of
  any help.
  At the same time, each of those subgames must contain a single SCC and only
  dominions of sufficiently large size to prevent both SCC-decomposition and
  dominion decomposition techniques from simplifying the game.
  In this paper, we answer positively to the question, by providing a robust, and
  harder, worst case family that meets all the above requirements, thereby
  shading some light on the actual power of aforementioned techniques and
  sanctioning that no combination of them can indeed help improving the
  exponential lower bound of the \divideetimpera approaches.

  A recent breakthrough~\cite{CJKLS17} by Calude \etal proposes a succinct
  reduction from parity to reachability games based on a clever encoding of the
  sequences of priorities a player finds along a play.
  This allows for a mere quasi-polynomial blow up in the size of the underlying
  graph and sets the basis of the fixed-parameter tractability \wrt the number
  of priorities.
  The approach has been then considerably refined in~\cite{FJSSW17}, where these
  encodings are modeled as progress measures.
  A similar technique is also used in~\cite{JL17}.
  Despite the theoretical relevance of this new idea, preliminary
  experiments~\cite{BDM?} seem to suggest that the practical impact of the
  result does not match the theoretical one, as all exponential algorithms
  outperform, often by orders of magnitude, the current implementations of the
  quasi-polynomial ones, which do not scale beyond few hundred vertices.
  This evaluation is consistent with the fact that the new techniques
  essentially amount to clever and succinct encodings embedded within a brute
  force search, which makes matching quasi-polynomial worst cases quite easy to
  find.
  These observations suggest that the road to a polynomial solution may need to
  take another direction.
  Our work is, therefore, intended to evaluate the weaknesses of classic
  exponential algorithms, in the same vein of~\cite{NS17,NS17a}, where the
  authors study the pitfalls of existing exponential algorithms for graphs
  isomorphism, in spite of the fact that a quasi-polynomial, but impractical,
  algorithm exists~\cite{Bab16}.
  We believe that a better understanding of the different issues of the known
  approaches may lead to progress in the quest for a polynomial algorithm.

\end{section}

% End of file Introduction.tex

%%% Local Variables:
%%% mode: latex
%%% TeX-master: "Article"
%%% End:

%% file: Preliminaries.tex
%%****************************************************************************%%
%%                                                                            %%
%% Robust Worst Cases for Divide-et-Impera Algorithms for Parity Games        %%
%%                                                                            %%
%% Preliminaries.tex                                                          %%
%%                                                                            %%
%% Revision 0                                                                 %%
%%                                                                            %%
%% Copyright (C) 2017, Massimo Benerecetti, Daniele Dell'Erba, and            %%
%%                     Fabio Mogavero.                                        %%
%% All rights reserved.                                                       %%
%%                                                                            %%
%%****************************************************************************%%

% Begin of file Preliminaries.tex

\begin{section}{Parity Games}
  \label{sec:pargam}

  Let us first briefly recall the notation and basic definitions concerning
  parity games that expert readers can simply skip.
  We refer to~\cite{AG11}\cite{Zie98} for a comprehensive presentation of the
  subject.
  \\\indent
  A two-player turn-based \emph{arena} is a tuple $\ArnName =
  \tuplec{\PosSet[][0]}{\PosSet[][1]}{\MovRel}$, with $\PosSet[][0] \cap
  \PosSet[][1] = \emptyset$ and $\PosSet \defeq \PosSet[][0] \cup \PosSet[][1]$,
  such that $\tupleb{\PosSet}{\MovRel}$ is a finite directed graph.
  $\PosSet[][0]$ (\resp, $\PosSet[][1]$) is the set of positions of player $0$
  (\resp, $1$) and $\MovRel \subseteq \PosSet \times \PosSet$ is a left-total
  relation describing all possible moves.
  A \emph{path} in $\VSet \subseteq \PosSet$ is an infinite sequence $\pthElm
  \in \PthSet(\VSet)$ of positions in $\VSet$ compatible with the move relation,
  \ie, $(\pthElm_{i}, \pthElm_{i + 1}) \in \MovRel$, for all $i \in \SetN$.
  A positional \emph{strategy} for player $\wp \in \{ 0, 1 \}$ on $\VSet
  \subseteq \PosSet$ is a function $\strElm[\wp] \in \StrSet[][\wp](\VSet)
  \subseteq (\VSet \cap \PosSet[][\wp]) \to \VSet$, mapping each $\wp$-position
  $\posElm \in \VSet \cap \PosSet[][\wp]$ to position $\strElm[\wp](\posElm) \in
  \VSet$ compatible with the move relation, \ie, $(\posElm,
  \strElm[\wp](\posElm)) \in \MovRel$.
  By $\StrSet[][\wp](\VSet)$ we denote the set of all $\wp$-strategies on
  $\VSet$.
  A \emph{play} in $\VSet \subseteq \PosSet$ from a position $\posElm \in \VSet$
  \wrt a pair of strategies $(\strElm[0], \strElm[1]) \in \StrSet[][0](\VSet)
  \times \StrSet[][1](\VSet)$, called \emph{$((\strElm[0], \strElm[1]),
  \posElm)$-play}, is a path $\pthElm \in \PthSet(\VSet)$ such that $\pthElm[0]
  = \posElm$ and, for all $i \in \SetN$, if $\pthElm_{i} \in \PosSet[][0]$, then
  $\pthElm_{i + 1} = \strElm[][0](\pthElm_{i})$ else $\pthElm_{i + 1} =
  \strElm[][1](\pthElm_{i})$.
  \\\indent
  A \emph{parity game} is a tuple $\GamName =
  \tuplec{\ArnName}{\PrtSet}{\prtFun}$, where $\ArnName$ is an arena, $\PrtSet
  \subset \SetN$ is a finite set of priorities, and $\prtFun : \PosSet \to
  \PrtSet$ is a \emph{priority function} assigning a priority to each position.
  The priority function can be naturally extended to games and paths as follows:
  $\prtFun(\GamName) \defeq \max[\posElm \in \PosSet] \, \prtFun(\posElm)$; for
  a path $\pthElm \in \PthSet$, we set $\prtFun(\pthElm) \defeq \limsup_{i \in
  \SetN} \prtFun(\pthElm_{i})$.
  A set of positions $\VSet \subseteq \PosSet$ is a $\wp$-\emph{dominion}, with
  $\wp \in \{ 0, 1 \}$, if there exists a $\wp$-strategy $\strElm[\wp] \in
  \StrSet[][\wp](\VSet)$ such that, for all $\dual{\wp}$-strategies
  $\strElm[\dual{\wp}] \in \StrSet[][\dual{\wp}](\VSet)$ and positions $\posElm
  \in \VSet$, the induced $((\strElm[0], \strElm[1]), \posElm)$-play $\pthElm$
  has priority of parity $\wp$, \ie, $\prtFun(\pthElm) \equiv_{2} \wp$.
  In other words, $\strElm[\wp]$ only induces on $\VSet$ plays whose maximal
  priority visited infinitely often has parity $\wp$.
  The \emph{winning region} for player $\wp \in \{ 0, 1 \}$ in game $\GamName$,
  denoted by $\WinSet[\GamName][\wp]$, is the maximal set of positions that is
  also a $\wp$-dominion in $\GamName$. Since parity games are determined
  games~\cite{EJ91}, meaning that from each position one of the two players
  wins, the two winning regions of a game $\GamName$ form a partition of its
  positions, \ie, $\WinSet[\GamName][0] \cup \WinSet[\GamName][1] =
  \PosSet[\GamName]$.
  By $\GamName \!\setminus\! \VSet$ we denote the maximal subgame of $\GamName$
  with set of positions $\PosSet'$ contained in $\PosSet \!\setminus\! \VSet$
  and move relation $\MovRel'$ equal to the restriction of $\MovRel$ to
  $\PosSet'$.
  The $\wp$-predecessor of $\VSet$, in symbols $\preFun[][\wp](\VSet) \defeq
  \set{ \posElm \in \PosSet[][\wp] }{ \MovRel(\posElm) \cap \VSet \neq \emptyset
  } \cup \set{ \posElm \in \PosSet[][\dual{\wp}] }{ \MovRel(\posElm) \subseteq
  \VSet }$, collects the positions from which player $\wp$ can force the game to
  reach some position in $\VSet$ with a single move.
  The $\wp$-attractor $\atrFun[][\wp](\VSet)$ generalizes the notion of
  $\wp$-predecessor $\preFun[][\wp](\VSet)$ to an arbitrary number of moves.
  Thus, it corresponds to the least fix-point of that operator.
  When $\VSet = \prdFun[][\wp](\VSet)$, player $\wp$ cannot force any position
  outside $\VSet$ to enter this set.
  For such a $\VSet$, the set of positions of the subgame $\GamName \setminus
  \VSet$ is precisely $\PosSet \setminus \VSet$.
  When confusion cannot arise, we may abuse the notation and write $\GamName$ to
  mean its set of positions $\PosSet[\GamName]$.

\end{section}

% End of file Preliminaries.tex

%%% Local Variables:
%%% mode: latex
%%% TeX-master: "Article"
%%% End:

%% file: SectionI.tex
%%****************************************************************************%%
%%                                                                            %%
%% Robust Worst Cases for Divide-et-Impera Algorithms for Parity Games        %%
%%                                                                            %%
%% SectionI.tex                                                               %%
%%                                                                            %%
%% Revision 0                                                                 %%
%%                                                                            %%
%% Copyright (C) 2017, Massimo Benerecetti, Daniele Dell'Erba, and            %%
%%                     Fabio Mogavero.                                        %%
%% All rights reserved.                                                       %%
%%                                                                            %%
%%****************************************************************************%%

% Begin of file SectionI.tex

\begin{section}{The Recursive Algorithm}
  \label{sec:recalg}

  \vspace{0.50em}
  \noindent
  \begin{tabular}{l|l}
    \begin{minipage}[t]{0.475\textwidth}
      \vspace{-1.00em}
      \algrec
    \end{minipage} &
    \begin{minipage}[t]{0.475\textwidth}
      \algsubcal
    \end{minipage}
  \end{tabular}
  \vspace{0.25em}

  The Recursive procedure, reported in Algorithm~\ref{alg:rec} and proposed by
  Zielonka~\cite{Zie98} in an equivalent version, solves a parity game
  $\GamName$ by decomposing it into two subgames, each of which is, then, solved
  recursively.
  Intuitively, the procedure works as follows.
  Algorithm~\ref{alg:rec}, by means of Algorithm~\ref{alg:lefsubcal}, starts by
  collecting all the positions that are forced to pass through a position with
  maximal priority $\pElm \defeq \prtFun(\GamName)$ in that game.
  This first step results in computing the set $\ASet \defeq
  \atrFun[\GamName][\wp](\prtFun[\GamName][-1](\pElm))$, \ie, the attractor to
  the set $\prtFun[\GamName][-1](\pElm)$ of positions with priority $\pElm$ \wrt
  player $\wp \defeq \pElm \bmod 2$.
  The subgame $\GamName[\LSym]$ is, then, obtained from $\GamName$ by removing
  $\ASet$ from it and solved recursively.
  The result is a partitioning of the positions of $\GamName[\LSym]$ into two
  winning regions, $\WinSet[\LSym][0]$ and $\WinSet[\LSym][1]$, one per player.
  At this point, the algorithm checks whether the subgame $\GamName[\LSym]$ is
  completely won by $\wp$ or, more generally, if the adversary $\dual{\wp}$
  cannot force any other position in $\GamName$ into its own winning region
  $\WinSet[\LSym][\dual{\wp}]$ in one move.
  In other words, none of the winning positions of the adversary $\dual{\wp}$
  can attract something outside that region, \ie,
  $\preFun[\GamName][\dual{\wp}](\WinSet[\LSym][\dual{\wp}]) \setminus
  \WinSet[\LSym][\dual{\wp}] = \emptyset$.
  If this is the case, the entire game $\GamName$ is solved.
  Indeed, the positions of $\GamName$ winning for $\wp$ are all its positions
  except, possibly, for those won by $\dual{\wp}$ in subgame $\GamName[\LSym]$
  (see Line~$4$ of Algorithm~\ref{alg:rec}).
  If, on the other hand, the above condition does not hold, the winning region
  of $\dual{\wp}$ can be extended with some other positions in $\GamName$.
  Let $\BSet \defeq \atrFun[\GamName][\dual{\wp}](\WinSet[\LSym][\dual{\wp}])$
  be the set collecting all such positions.
  Observe that all the positions in $\BSet$ are certainly winning for
  $\dual{\wp}$ in the entire game, as, from each such position, $\dual{\wp}$ can
  force entering its own winning region $\WinSet[\LSym][\dual{\wp}]$, from which
  its opponent $\wp$ cannot escape.
  The residual subgame $\GamName[\RSym]$, obtained by removing $\BSet$ form
  $\GamName$, as computed by Algorithm~\ref{alg:rigsubcal}, may now contain
  positions winning for either player, and, therefore, still needs to be solved
  recursively (see Line~$6$ of Algorithm~\ref{alg:rec}).
  All the positions of $\GamName[\RSym]$ that turn out to be winning for $\wp$
  in that game, namely $\WinSet[\RSym][\wp]$, are, then, all and only those
  positions winning for $\wp$ in the entire game $\GamName$, while the remaining
  ones are winning for $\dual{\wp}$ (see Line~$7$ of Algorithm~\ref{alg:rec}).

  As shown by Friedmann in~\cite{Fri11a}, the algorithm admits a worst case
  family of games $\{ \GamName[][k] \}_{k = 1}^{\omega}$ that requires a number
  of recursive calls exponential in $k$.
  The reason is essentially the following.
  Each game $\GamName[][k]$ of that family contains all $\GamName[][j]$, with
  $1 \leq j < k$, as subgames.
  Each recursive call that receives as input one such subgame $\GamName[][j]$
  requires to eventually solve both $\GamName[][j - 1]$ and $\GamName[][j - 2]$.
  As a consequence, the number of recursive calls performed by the algorithm on
  game $\GamName[][k]$ can be put in correspondence with a Fibonacci sequence.
  This proves that their number grows at least as fast as the sequence of the
  Fibonacci numbers, namely that their number is $\AOmega{((1 + \sqrt{5}) /
  2)^{k}}$.
  The very reason that makes this family exponential also makes it amenable to a
  polynomial-time solution.
  It suffices to endow the Recursive algorithm with a memoization mechanism
  that, for each solved game $\GamName$, records the triple $(\GamName,
  \WinSet[\GamName][0], \WinSet[\GamName][1])$.
  Each recursive call can, then, directly extract the winning regions of a
  subgame that is already contained in the collection, thus preventing the
  procedure from solving any subgame more than once.
  Not only does the resulting procedure make Friedman worst case vain, but it
  also speeds up the solution of games significantly, as long as the number of
  repeated subgames remains relatively small, \eg, linear in the size of the
  original game, which is often the case in practice.

\end{section}

% End of file SectionI.tex

%%% Local Variables:
%%% mode: latex
%%% TeX-master: "Article"
%%% End:

%% file: SectionII.tex
%%****************************************************************************%%
%%                                                                            %%
%% Robust Worst Cases for Divide-et-Impera Algorithms for Parity Games        %%
%%                                                                            %%
%% SectionII.tex                                                              %%
%%                                                                            %%
%% Revision 0                                                                 %%
%%                                                                            %%
%% Copyright (C) 2017, Massimo Benerecetti, Daniele Dell'Erba, and            %%
%%                     Fabio Mogavero.                                        %%
%% All rights reserved.                                                       %%
%%                                                                            %%
%%****************************************************************************%%

% Begin of file SectionII.tex

\begin{section}{Memoization Resilient Games}
  \label{sec:memresgam}

  As mentioned above, the main requirement for an exponential worst case family
  for the memoized version of the Recursive algorithm is to contain games that
  force the procedure to solve an exponential number of different subgames.
  In this section we shall focus primarily on this problem, by showing that such
  a family exists.
  More generally, we identify a \emph{core family} $\{ \GamName[\CSym][k] \}_{k
  = 1}^{\omega}$ of games enjoying that specific property.
  For each $k \in \mathbb{N}_+$, game $\GamName[\CSym][k]$ contains $2 k +
  1$ gadgets, each one formed by three positions $\alpha_{i}, \beta_{i}$, and
  $\gamma_{i}$, for $i \in \numcc{0}{2k}$.
  The positions $\beta_{i}$ and $\gamma_{i}$, in gadget $i$, share the same
  priority $i$ and opposite owners, namely player $i \bmod{2}$ for $\beta_{i}$
  and $(i + 1) \bmod{2}$ for $\gamma_{i}$.
  The position $\alpha_{i}$ in the gadget has the same owner as the
  corresponding $\beta_{i}$.
  These positions are leading ones, having higher priorities than all the
  $\beta$'s and $\gamma$'s of the other gadgets.
  Positions within gadget $i$ are connected as follows: $\alpha_{i}$ can only
  move to $\beta_{i}$; $\beta_{i}$ can only move to $\gamma_{i}$; $\gamma_{i}$
  can choose either to move to $\beta_{i}$ or to stay in $\gamma_{i}$ itself.
  Two adjacent gadgets, of indexes $i$ and $i + 1$, are connected by only two
  moves: one from $\gamma_{i}$ to $\alpha_{i + 1}$ and one from $\beta_{i + 1}$
  to $\alpha_{i}$.
  Figure~\ref{fig:corfam} depicts game $\GamName[\CSym][2]$.
  The gray portion in the figure represents game $\GamName[\CSym][1]$, where all
  the priorities of all its $\alpha$'s have been increased by $2$, so as to
  comply with the requirement that the $\alpha$'s have the higher priorities.
  Indeed, in general, given an index $k$, game $\GamName[\CSym][k + 1]$ is
  obtained from $\GamName[\CSym][k]$ by increasing by two units the priorities
  of each $\alpha_{i}$ in the gadgets of $\GamName[\CSym][k]$ and adding two
  new gadgets with indexes $2k + 1$ and $2(k + 1)$ connected as in figure.
  The games in the core family are formally described by the following
  definition.

  \begin{definition}[Core Family]
    \label{def:expcorfam}
    The \emph{core family} $\{ \GamName[\CSym][k] \}_{k = 1}^{\omega}$, where
    $\GamName[\CSym][k] \defeq \tuplec{\ArnName}{\PrtSet}{\prtFun}$, $\AName
    \defeq \tuplec{\PosSet[][0]}{\PosSet[][1]}{\MovRel}$, and $\PrtSet \defeq
    \numcc{0}{4k}$, is defined as follows.
    For any index $k \geq 1$, the set of positions $\PosSet \defeq \set{
    \alpha_{i}, \beta_{i}, \gamma_{i} }{ 0 \leq i \leq 2k }$ of
    $\GamName[\CSym][k]$ is divided into three categories:
    \begin{itemize}
      \item
        $\alpha_{i}$ belongs to player $\wp \defeq i \bmod 2$, \ie, $\alpha_{i}
        \in \PosSet[][\wp]$, and has priority $\prtFun(\alpha_{i}) \defeq 2k + i
        + 1$;
      \item
        $\beta_{i}$ belongs to player $\wp \defeq i \bmod 2$, \ie, $\beta_{i}
        \in \PosSet[][\wp]$, and has priority $\prtFun(\beta_{i}) \defeq i$;
      \item
        $\gamma_{i}$ belongs to player $\dual{\wp} \defeq (i + 1) \bmod 2$, \ie,
        $\gamma_{i} \in \PosSet[][\dual{\wp}]$, and has priority
        $\prtFun(\gamma_{i}) \defeq i$.
    \end{itemize}
    Moreover, the moves from positions $\alpha_{i}$, $\beta_{i}$, and
    $\gamma_{i}$, with $0 \leq i \leq 2k$, are prescribed as follows:
    \begin{itemize}
      \item
        $\alpha_{i}$ has a unique move to $\beta_{i}$, \ie, $\MovRel(\alpha_{i})
        = \{ \beta_{i} \}$;
      \item
        $\beta_{i}$ has one move to $\gamma_{i}$ and one to $\alpha_{i - 1}$, if
        $i > 0$, \ie, $\MovRel(\beta_{i}) = \{ \gamma_{i} \} \cup \set{
        \alpha_{i - 1} }{ i > 0 }$;
      \item
        $\gamma_{i}$ has one move to $\gamma_{i}$ itself, one to $\beta_{i}$,
        and, if $i < 2k$, one to $\alpha_{i + 1}$, \ie, $\MovRel(\gamma_{i}) =
        \{ \beta_{i}, \gamma_{i} \} \cup \set{ \alpha_{i + 1} }{ i < 2k }$.
    \end{itemize}
  \end{definition}

  \figcorfam

  It is not hard to verify that, for any $k \in \mathbb{N}_+$, the game
  $\GamName[\CSym][k]$ is completely won by player $0$ and contains precisely
  $6k + 3$ positions and $12k + 4$ moves.
  As we shall see later in detail, the solution of each such games requires
  Algorithm~\ref{alg:rec} to solve an exponential number of different subgames.

  These core games form the backbone for a more general framework, consisting of
  an entire class of game families, with the property that each of them remains
  resilient to memoization techniques.
  Essentially, each game in any such family extends a core game.
  In order to define such a wider class, let us first establish what counts as a
  suitable extension of a core.
  Clearly, for a game $\GamName$ to be an extension of $\GamName[\CSym][k]$, for
  some index $k \in \mathbb{N}_+$, it must contain $\GamName[\CSym][k]$ as
  a subgame.
  However, in order to prevent the Recursive algorithm from disrupting the core
  while processing $\GamName$, we have to enforce some additional requirements.
  In particular, we need the algorithm to behave on $\GamName$ virtually in the
  same way as it does on the core subgame.
  To this end, we require that all positions $\alpha_{i}$ still have the maximal
  priorities as in the core.
  Moreover, all positions $\alpha_{i}$ and $\beta_{i}$ cannot have additional
  moves in $\GamName$ \wrt those contained in the core.
  Finally, if $\gamma_{i}$ can escape to some position $\posElm$ outside the
  core, then $\posElm$ does not have a higher priority, it has a move back to
  $\gamma_{i}$, and belongs to the opponent player \wrt $\gamma_{i}$.
  This condition ensures that no $\gamma_{i}$ can decide to escape the core
  without being bounced back immediately by the opponent.
  The following definition makes the notion of extension precise.

  \begin{definition}[Core Extension]
    \label{def:corext}
    An arbitrary parity game $\GamName \in \PG$ is a \emph{core extension} of
    $\GamName[\CSym][k]$, for a given index $k \in \mathbb{N}_+$, if the
    following four conditions hold:
    \begin{enumerate}
      \item\label{def:corext(cor)}
        $\GamName \setminus \PSet = \GamName[\CSym][k]$, where $\PSet \defeq
        \set{ \posElm \in \GamName }{ \posElm \not\in \GamName[\CSym][k] }$;
      \item\label{def:corext(prt)}
        $\prtFun[\GamName](\posElm) < \prtFun[\GamName](\alpha_{0})$, for all
        $\posElm \in \PSet$;
      \item\label{def:corext(mov)}
        $\set{ \alpha_{i}, \beta_{i} \in \GamName }{ 0 \leq i \leq 2k } \cap
        (\MovRel(\PSet) \cup \MovRel^{-1}(\PSet)) = \emptyset$;
      \item\label{def:corext(sep)}
        $\posElm \in \PosSet[][\wp]$, $\gamma_{i} \in \MovRel(\posElm)$,
        and $\prtFun(\posElm) \leq i$, for all $\posElm \in \PSet \cap
        \MovRel(\gamma_{i})$, $i \in \numcc{0}{2k}$, and $\wp \defeq i \bmod 2$.
    \end{enumerate}
    We shall denote with $\PGC \subseteq \PG$ the set of all core extensions of
    $\GamName[\CSym][k]$, for any index $k \in \mathbb{N}_+$.
  \end{definition}

  We can now define the abstract notion of worst-case family that extends the
  core family, while still preserving the same essential properties that we are
  going to prove shortly.

  \begin{definition}[Worst-Case Family]
    \label{def:expworcasfam}
    A family of parity games $\{ \GamName[][k] \}_{k = 1}^{\omega}$ is a
    \emph{worst-case family} if $\GamName[][k]$ is a core extension of
    $\GamName[\CSym][k]$, for every index $k \in \mathbb{N}_+$.
  \end{definition}

  \figrectree
  In order to prove that any worst-case family requires an exponential number of
  different subgames to be solved, we shall characterize a suitable subtree of
  the recursion tree generated by the algorithm, when called on one of the games
  in the family.
  Starting from the root, which contains the original game $\GamName[][k]$, we
  fix specific observation points in the recursion tree that are identified by
  sequences in the set $\wElm \in \{ \LSym, \RSym \}^{\leq k}$, where $\LSym$
  (\resp, $\RSym$) denotes the recursive call on the left (\resp, right)
  subgame.
  Each sequence $\wElm$ identifies two subgames, $\der{\GamName}[\wElm][k]$ and
  $\GamName[\wElm][k]$, of $\GamName[][k]$ that correspond to the input subgames
  of two successive nested calls.
  In the analysis of the recursion tree, we shall only take into account the
  left subgame $\GamName[\wElm][k]$ of each $\der{\GamName}[\wElm][k]$, thus
  disregarding its right subtree as it is inessential to the argument.
  An example of the resulting subgame tree for game $\GamName[\CSym][1]$ of the
  core family is depicted in Figure~\ref{fig:rectre}.
  According to Algorithm~\ref{alg:rec} on input $\GamName[\CSym][1] =
  \GamName[\varepsilon][1]$, the first (left) recursive call is executed on the
  subgame obtained by removing the $1$-attractor to the positions with maximal
  priority, in this case $\alpha_{9}$, which only contains $\alpha_{9}$.
  Therefore, the left subgame coincides precisely with
  $\der{\GamName}[\LSym][1]$.
  The second call is executed on the game obtained by removing the $0$-attractor
  in $\GamName[\varepsilon][1]$ to the winning region for player $0$ of the left
  subgame $\der{\GamName}[\LSym][1]$.
  In this case, that winning region is precisely $\{ \beta_{2}, \gamma_{2} \}$,
  and its $0$-attractor is $\{ \alpha_{2}, \gamma_{1}, \beta_{2}, \gamma_{2}\}$.
  As consequence, the subgame passed to the right-hand call precisely coincides
  with $\der{\GamName}[\RSym][1]$.
  The rest of the subtree is generated applying the same reasoning.
  The following definition generalizes this notion to a game of any worst-case
  family and characterizes the portion of the recursion tree we are interested
  in analyzing.

  \begin{definition}[Induced Subgame Tree]
    \label{def:indsubtre}
    Given a worst-case family $\{ \GamName[][k] \}_{k = 1}^{\omega}$, the
    \emph{induced subgame tree} $\IndSubTre[][k] \defeq \{ \GamName[\wElm][k]
    \}_{\wElm \in \{ \LSym, \RSym \}^{\leq k}} \cup \{ \der{\GamName}[\wElm][k]
    \}_{\wElm \in \{ \LSym, \RSym \}^{\leq k + 1}}^{\wElm \neq \varepsilon}$
    \wrt an index $k \in \mathbb{N}_+$ is defined inductively on the
    structure of the sequence $\wElm \in \{ \LSym, \RSym \}^{\leq k}$ as
    follows, where $\GamName[\varepsilon][k] \defeq \GamName[][k]$ and $z \defeq
    2(k - \card{\wElm})$:
    \begin{enumerate}
      \item\label{def:indsubtre(lef)}
        $\der{\GamName}[\wElm \LSym][k] \defeq \GamName[\wElm][k] \setminus
        \atrFun[ {\GamName[\wElm][k]} ][1](\{ \alpha_{z} \})$;
      \item\label{def:indsubtre(rig)}
        $\der{\GamName}[\wElm \RSym][k] \defeq \GamName[\wElm][k] \setminus
        \atrFun[ {\GamName[\wElm][k]} ][0](\WinSet[ {\der{\GamName}[\wElm
        \LSym][k]} ][0]\!)$;
      \item\label{def:indsubtre(hat)}
        $\GamName[\wElm][k] \defeq \der{\GamName}[\wElm][k] \setminus \atrFun[
        {\der{\GamName}[\wElm][k]} ][0](\{ \alpha_{z + 1} \})$, if $\wElm \neq
        \varepsilon$.
    \end{enumerate}
  \end{definition}

  Before proceeding with proving the main result of this section, we need some
  additional properties of the induced subgame tree of any worst-case family.
  The following lemma states some invariant of the elements contained in the
  tree of a games for $\GamName[][k]$, extending the core $\GamName[\CSym][k]$,
  that will be essential to the result.
  In particular, they ensure that all of them are subgames of $\GamName[][k]$
  and that, depending on the identifying sequence $\wElm$, they contain the
  required leading positions $\alpha_{i}$ of the core.
  In addition, it states two important properties of every left child in the
  tree, \ie, those elements identified by a sequence $\wElm$ ending with
  $\LSym$.
  Both of them will be instrumental in proving that all the subgames in the tree
  are indeed different and to assess their number, as we shall see in
  Lemmas~\ref{lmm:gamdif} and~\ref{lmm:indsubtrecar}.
  The first one ensures that each such game necessarily contain a specific
  position $\gamma_{i}$, with the index $i$ depending on $\wElm$.
  The second one characterizes the winning region for player $0$ of the
  left-child subgames $\der{\GamName}[\wElm\LSym][k]$.
  It states that, in each such game, the winning positions for player $0$
  contained in the corresponding core $\GamName[\CSym][k]$ are all its
  $\beta_{2j}$ and $\gamma_{2j}$, with even index greater than the maximal index
  of a leading position $\alpha_{x}$ in that game.
  Indeed, as soon as the higher positions $\alpha_{i}$, with $i \in \numcc{x +
  1}{k}$, are removed from the game, each residual corresponding $\gamma_{2j}$,
  possibly together with its associated $\beta_{2j}$, is necessarily contained
  in an independent $0$-dominion.

  In the sequel, by $\lst{\wElm}$ we denoted the last position of a non-empty
  sequence $\wElm \in \{ \LSym, \RSym \}^{+}$.

  \begin{lemma}
    \label{lmm:recinv}
    For any index $k \!\in\! \mathbb{N}_+$ and sequence $\wElm \!\in\! \{
    \LSym, \RSym \}^{\leq k + 1}$, let $z \!\defeq\! 2(k - \card{\wElm})$.
    Then, the following properties hold, where $\card{\wElm} \leq k$, in the
    first three items, and $\wElm \neq \varepsilon$, in the remaining ones:
    \begin{enumerate}
      \item\label{lmm:recinv(sub)}
        $\GamName[\wElm][k]$ is a subgame of $\GamName[][k]$;
      \item\label{lmm:recinv(alp)}
        $\alpha_{j} \in \GamName[\wElm][k]$ \iff $j \in \numcc{0}{z}$;
      \item\label{lmm:recinv(gam)}
        $\gamma_{j} \in \GamName[\wElm][k]$, for all $j \in \numcc{0}{z}$, and
        $\gamma_{z + 1} \in \GamName[\wElm][k]$, if $\wElm \neq \varepsilon$ and
        $\lst{\wElm} = \LSym$;
      \item\label{lmm:recinv(hatsub)}
        $\der{\GamName}[\wElm][k]$ is a subgame of $\GamName[][k]$;
      \item\label{lmm:recinv(hatalp)}
        $\alpha_{j} \in \der{\GamName}[\wElm][k]$ \iff $j \in \numcc{0}{z + 1}$;
      \item\label{lmm:recinv(hatgam)}
        $\gamma_{j} \!\in\! \der{\GamName}[\wElm][k]$, for all $j \!\in\!
        \numcc{0}{z}$, and $\gamma_{z + 1} \!\in\! \der{\GamName}[\wElm][k]$, if
        $\card{\wElm} \leq k$, and $\gamma_{0} \!\in\!
        \der{\GamName}[\wElm][k]$, otherwise, when $\lst{\wElm} \!=\! \LSym$;
      \item\label{lmm:recinv(hatwin)}
        $\WinSet[ {\der{\GamName}[\wElm][k]} ][0]\! \cap \GamName[\CSym][k] =
        \set{ \beta_{z + 2j}, \gamma_{z + 2j} \in \der{\GamName}[\wElm][k] }{ j
        \in \numcc{1}{\card{\wElm}} }$, if $\lst{\wElm} = \LSym$.
    \end{enumerate}
  \end{lemma}

  Finally, the next lemma simply establishes that all the subgames contained in
  the induced subgame tree of Definition~\ref{def:indsubtre} are indeed
  generated by the Recursive algorithm when called with input game
  $\GamName[][k]$ of some worst-case family.

  \begin{lemma}
    \label{lmm:recsub}
    For any index $k \in \mathbb{N}_+$ and sequence $\wElm \in \{ \LSym,
    \RSym \}^{\leq k}$, the following properties hold:
    \begin{enumerate}
      \item\label{lmm:recsub(lef)}
        $\fFun[\LSym](\GamName[\wElm][k]) = (\der{\GamName}[\wElm \LSym][k],
        1)$;
      \item\label{lmm:recsub(rig)}
        $\fFun[\RSym](\GamName[\wElm][k], \WinSet[ {\der{\GamName}[\wElm
        \LSym][k]} ][0], 0) = \der{\GamName}[\wElm \RSym][k]$;
      \item\label{lmm:recsub(hat)}
        $\fFun[\LSym](\der{\GamName}[\wElm][k]) = (\GamName[\wElm][k], 0)$, if
        $\wElm \neq \varepsilon$.
    \end{enumerate}
  \end{lemma}

  We are now ready for the main result of this section, namely that the induced
  subgame tree contains elements which are all different from each other and
  whose number is exponential in the index $k$.
  We split the result into two lemmas.
  The first one simply states that any subgame in the left subtree of some
  $\GamName[\wElm][k]$ is different from any other subgame in the right subtree.
  The idea is that for for any subgame in the tree, all subgames of its left
  subtree contain at least one position, a specific position $\gamma_{i}$ with
  $i$ depending on $\wElm$, that is not contained in any subgame of its right
  subtree.

  \begin{lemma}
    \label{lmm:gamdif}
    For all indexes $k \in \mathbb{N}_+$ and sequences $\wElm, \vElm \in
    \{ \LSym, \RSym \}^{*}$, with $\ell \defeq \card{\wElm} + \card{\vElm} \leq
    k$ and $z \defeq 2(k - \card{\wElm}) - 1$, the following properties hold:
    \begin{enumerate}
      \item\label{lmm:gamdif(lef)}
        $\gamma_{z} \in \GamName[\wElm \LSym \vElm][k]$ and $\gamma_{z} \in
        \der{\GamName}[\wElm \LSym \vElm][k]$, if $\ell < k$, and $\gamma_{0}
        \in \der{\GamName}[\wElm \LSym \vElm][k]$, otherwise;
      \item\label{lmm:gamdif(rig)}
        $\gamma_{z} \not\in \GamName[\wElm \RSym \vElm][k]$ and $\gamma_{z}
        \not\in \der{\GamName}[\wElm \RSym \vElm][k]$, if $\ell < k$, and
        $\gamma_{0} \not\in \der{\GamName}[\wElm \RSym \vElm][k]$, otherwise.
    \end{enumerate}
  \end{lemma}
  \begin{proof}
    First observe that, if $\ell < k$, by Items~\ref{lmm:recinv(gam)}
    and~\ref{lmm:recinv(hatgam)} of Lemma~\ref{lmm:recinv}, position
    $\gamma_{z}$ belongs to both $\GamName[\wElm][k]$ and
    $\der{\GamName}[\wElm][k]$, since $0 < z < 2(k - \card{\wElm})$.
    Let us consider Item~\ref{lmm:gamdif(lef)} of the current lemma first and
    show that the every position $\gamma_{z}$ belongs to all the descendants of
    $\GamName[\wElm][k]$ in its left subtree.
    The proof proceeds by induction on the length of the sequence $\vElm$ and
    recall that, due to Item~\ref{lmm:recinv(alp)}
    (\resp,~\ref{lmm:recinv(hatalp)}) of Lemma~\ref{lmm:recinv}, the position
    with maximal priority in $\GamName[\wElm][k]$ (\resp, in
    $\der{\GamName}[\wElm][k]$) is $\alpha_{z}$ (\resp, $\alpha_{z + 1}$).
    Assume for the base case that $\card{\vElm} = 0$.
    The thesis becomes $\gamma_{z} \in \GamName[\wElm \LSym][k]$ and $\gamma_{z}
    \in \der{\GamName}[\wElm \LSym][k]$.
    By Item~\ref{def:indsubtre(lef)} of Definition~\ref{def:indsubtre},
    $\der{\GamName}[\wElm \LSym][k]$ is defined as $\GamName[\wElm][k] \setminus
    \ASet$, where $\ASet = \atrFun[ {\GamName[\wElm][k]} ][1](\{ \alpha_{z + 1}
    \})$.
    By Definition~\ref{def:corext} of core extension
    (Items~\ref{def:corext(cor)},~\ref{def:corext(mov)},
    and~\ref{def:corext(sep)}), a move entering $\alpha_{z + 1}$ may only come
    from $\beta_{z + 2}$, if it is present in the subgame, which is always the
    case unless $\wElm = \varepsilon$, or from $\gamma_{z}$, whose owner is the
    opponent player $0$ and cannot be attracted.
    Hence, $\ASet = \{ \alpha_{z + 1}, \beta_{z + 2} \}$, if $\wElm \neq
    \varepsilon$, and $\ASet = \{ \alpha_{z + 1} \}$, otherwise.
    Similarly, by Item~\ref{def:indsubtre(hat)}, $\GamName[\wElm \LSym][k]$ is
    defined as $\der{\GamName}[\wElm \LSym][k] \setminus \ASet$, where $\ASet =
    \atrFun[ {\der{\GamName}[\wElm \LSym][k]} ][0](\{ \alpha_{z} \})$.
    For the same observations as in the previous case, we have that $\ASet = \{
    \alpha_{z}, \beta_{z+1} \}$.
    Hence, no position $\gamma_{i}$ is removed from either games and the thesis
    immediately follows.
    Assume now $\card{\vElm} > 0$, let $\vElm \defeq \vElm' \cdot \xElm$, with
    $\xElm \in \{ \LSym, \RSym \}$, and set $\trn{\wElm} \defeq \wElm \LSym
    \vElm'$.
    By the inductive hypothesis, $\gamma_{z} \in \der{\GamName}[\trn{\wElm}][k]$
    and $\gamma_{z} \in \GamName[\trn{\wElm}][k]$.
    We have two cases, depending on whether $\xElm = \LSym$ or $\xElm = \RSym$.
    Let $r \defeq 2(k - \card{\trn{\wElm}})$.
    If $\xElm = \LSym$, according to Item~\ref{def:indsubtre(lef)} of
    Definition~\ref{def:indsubtre}, $\der{\GamName}[\trn{\wElm} \LSym][k]$ is
    obtained from $\GamName[\trn{\wElm}][k]$ by removing $\ASet = \atrFun[
    {\GamName[\trn{\wElm}][k]} ][1](\{ \alpha_{r} \}) = \{ \alpha_{r}, \beta_{r
    + 1} \}$.
    Similarly, by Item~\ref{def:indsubtre(hat)} of
    Definition~\ref{def:indsubtre}, $\GamName[\trn{\wElm} \LSym][k]$ is defined
    as $\der{\GamName}[\trn{\wElm} \LSym][k] \setminus \ASet$, where $\ASet =
    \atrFun[ {\der{\GamName}[\trn{\wElm} \LSym][k]} ][0](\{ \alpha_{r - 1} \})$,
    by observing that $\card{\trn{\wElm} \LSym} = \card{\trn{\wElm}} + 1$ and,
    therefore, $2(k - \card{\trn{\wElm} \LSym}) + 1 = r - 1$.
    In both cases the thesis follows immediately.
    Let us now consider the case with $\xElm = \RSym$.
    According to Item~\ref{def:indsubtre(rig)} of
    Definition~\ref{def:indsubtre}, $\der{\GamName}[\trn{\wElm} \RSym][k]$ is
    obtained by removing the set $\ASet = \atrFun[ {\GamName[\trn{\wElm}][k]}
    ][0](\WinSet[ {\der{\GamName}[\trn{\wElm} \LSym][k]} ][0])$ from
    $\GamName[\trn{\wElm}][k]$.
    Position $\gamma_{z}$ has odd index and cannot belong to $\WinSet[
    {\der{\GamName}[\trn{\wElm} \LSym][k]} ][0]$, which, by
    Item~\ref{lmm:recinv(hatwin)} of Lemma~\ref{lmm:recinv}, only contains,
    among the positions from the core, those $\beta_{i}$ and $\gamma_{i}$, with
    $i \geq 2(k - \card{\wElm}) > z$ and even.
    Since, $\der{\GamName}[\trn{\wElm} \LSym][k]$ is a subgame of a core
    extension, it holds that $\gamma_{z}$ is owned by player $0$ and can only
    have a move leading to $\alpha_{z + 1}$, which is not in the subgame, or to
    a position outside the core and owned by player $1$.
    As a consequence, it cannot end up in $\atrFun[ {\GamName[\trn{\wElm}][k]}
    ][0](\WinSet[ {\der{\GamName}[\trn{\wElm} \LSym][k]} ][0])$ and the thesis
    holds.
    Finally, recall that $\GamName[\trn{\wElm} \RSym][k] =
    \der{\GamName}[\trn{\wElm} \RSym][k] \setminus \ASet$, where $\ASet =
    \atrFun[ {\der{\GamName}[\trn{\wElm} \RSym][k]} ][0](\{ \alpha_{\hat{r}}
    \})$, with $\hat{r} \defeq 2(k - \card{\trn{\wElm} \RSym}) + 1$.
    In the considered subgame, $\alpha_{\hat{r}}$ has incoming moves only from
    $\beta_{\hat{r} + 1}$ and $\gamma_{\hat{r} - 1}$.
    However, $\hat{r} - 1 = 2(k - \card{\trn{\wElm} \RSym}) < 2(k -
    \card{\wElm}) - 1 = z$.
    Moreover, $\hat{r} + 1$ is an even index, and thus $\beta_{\hat{r} + 1}$ is
    not contained in $\der{\GamName}[\trn{\wElm} \RSym][k]$, being in $\WinSet[
    {\der{\GamName}[\trn{\wElm} \LSym][k]} ][0]$ as shown above.
    As a consequence, $\ASet = \{ \alpha_{\hat{r}} \}$ and the thesis follows.
    In addition, when $\card{w} = k$, Item~\ref{lmm:recinv(gam)} of
    Lemma~\ref{lmm:recinv} tells us that $\gamma_{0} \in \GamName[\wElm][k]$.
    If $\ell = k$, instead, we have that $\gamma_{0}$ belongs to
    $\der{\GamName}[\wElm \LSym][k]$, due to Item~\ref{lmm:recinv(hatgam)} of
    Lemma~\ref{lmm:recinv}.
    This ends the proof of Item~\ref{lmm:gamdif(lef)} of the lemma.
    As to Item~\ref{lmm:gamdif(rig)} of the lemma, first observe that, if
    $\card{\wElm} < k$, then $\gamma_{z} \not\in \der{\GamName}[\wElm
    \RSym][k]$.
    Indeed, position $\gamma_{z} \in \atrFun[ {\GamName[\wElm][k]} ][0](\WinSet[
    {\der{\GamName}[\wElm \LSym][k]} ][0])$, as shown above.
    Since this set is removed from $\GamName[\wElm][k]$ to obtain
    $\der{\GamName}[\wElm \RSym][k]$, the thesis holds for $\der{\GamName}[\wElm
    \RSym][k]$.
    Moreover, every descendant of $\der{\GamName}[\wElm \RSym][k]$ in the
    subgame tree is obtained only by removing positions.
    As a consequence, none of them can contain position $\gamma_{z}$.
    In case $\card{\wElm} = k$, instead, it suffices to observe that, according
    to Item~\ref{lmm:recinv(hatwin)} of Lemma~\ref{lmm:recinv}, $\gamma_{0} \in
    \WinSet[ {\der{\GamName}[\wElm \LSym][k]} ][0]$, hence it cannot be
    contained in $\der{\GamName}[\wElm \RSym][k]$.
  \end{proof}

  The main result asserting the exponential size of the induced subtrees of any
  worst-case family is given by the next lemma.
  This follows by observing that the number of nodes in the induced tree is
  exponential in $k$ and by showing that the subgames associated with any two
  nodes in the tree $\IndSubTre[][k]$ are indeed different.

  \begin{lemma}
    \label{lmm:indsubtrecar}
    $\card{\IndSubTre[][k]} = 3(2^{k + 1} - 1)$, for any $k \in
    \mathbb{N}_+$.
  \end{lemma}
  \begin{proof}
    To prove that the size of $\IndSubTre[][k]$ is as stated, we first need to
    show that all the elements contained in the subgame trees are different,
    namely that, for each $\wElm \neq \wElm'$, the subgames
    $\GamName[\wElm][k]$, $\GamName[\wElm'][k]$, $\der{\GamName}[\wElm][k]$, and
    $\der{\GamName}[\wElm'][k]$ are pairwise different.
    Let us start by showing that $\GamName[\wElm][k] \neq
    \der{\GamName}[\wElm][k]$, for each $\wElm \neq \varepsilon$.
    By Item~\ref{lmm:recinv(hatalp)} of Lemma~\ref{lmm:recinv}, position
    $\alpha_{2(k - \card{\wElm}) + 1} \in \der{\GamName}[\wElm][k]$ and, by
    Item~\ref{def:indsubtre(hat)} of Definition~\ref{def:indsubtre}, this
    position is removed from $\der{\GamName}[\wElm][k]$ to obtain
    $\GamName[\wElm][k]$.
    Hence, those two subgames cannot be equal.
    Let us consider now two subgames, each associated with one of the sequences
    $\wElm$ and $\wElm'$.
    There are two possible cases: either \emph{(i)}~$\wElm$ is a strict prefix
    of $\wElm'$, \ie, the one subgame is a descendant of the other in the
    subgame tree, or \emph{(ii)}~$\wElm$ and $\wElm'$ share a common longest
    prefix $\trn{\wElm}$ that is different from both, \ie, the two subgames lie
    in two distinct subtrees of the subgame associated with $\trn{\wElm}$.
    In case \emph{(i)} we have that $\wElm' = \wElm \vElm$, for some $\vElm \neq
    \varepsilon$.
    An easy induction on the length of $\vElm$ can prove that if $\GamName \in
    \{ \GamName[\wElm][k], \der{\GamName}[\wElm][k] \}$ and each $\GamName' \in
    \{ \GamName[\wElm'][k], \der{\GamName}[\wElm'][k] \}$ are the subgames
    associated with $\wElm$ and $\wElm'$, respectively, then the second is a
    strict subgame of the first, \ie, $\GamName' \subset \GamName$.
    Indeed, Definition~\ref{def:indsubtre} together with
    Items~\ref{lmm:recinv(alp)},~\ref{lmm:recinv(hatalp)},
    and~\ref{lmm:recinv(hatwin)} of Lemma~\ref{lmm:recinv} ensure that, at each
    step downward along a path in the tree starting from $\GamName$, whether we
    proceed on the left or the right branch, at least one position is always
    removed from the current subgame.
    In case \emph{(ii)}, instead, Item~\ref{lmm:gamdif(lef)} of
    Lemma~\ref{lmm:gamdif} tells us that there is at least one position,
    $\gamma_{z}$ in the lemma, contained in all the subgames of the left
    subtree, while Item~\ref{lmm:gamdif(rig)} states that the same position is
    not contained in any subgame of the right subtree.
    Therefore, each of the two subgames associated with $\wElm$ must be
    different from either of the two subgames associated with $\wElm'$.

    Finally, to prove the statement of the lemma, it suffices to observe that
    the number of sequences of length at most $k$ over the alphabet $\{ \LSym,
    \RSym \}$ are precisely $2^{k + 1} - 1$ and with each such sequence $\wElm$
    a subgame $\GamName[\wElm][k]$ is associated.
    As a consequence, the set $\{ \GamName[\wElm][k] \}_{\wElm \in \{ \LSym,
    \RSym \}^{\leq k}}$ contains $2^{k + 1} - 1$ different elements.
    Moreover, each such subgame has two children in $\{ \der{\GamName}[\wElm][k]
    \}_{\wElm \in \{ \LSym, \RSym \}^{\leq k + 1}}^{\wElm \neq \varepsilon}$.
    We can, then, conclude that the size of $\IndSubTre[][k]$ is precisely
    $3(2^{k + 1} - 1)$.
  \end{proof}

  As a consequence of Lemma~\ref{lmm:indsubtrecar}, we can obtain a stronger
  lower bound on the execution time of the Recursive algorithm.
  Indeed, the result holds regardless of whether the algorithm is coupled with a
  memoization technique.

  \begin{theorem}[Exponential Worst Case]
    \label{thm:expworcas}
    The number of distinct recursive calls executed by the Recursive algorithm,
    with or without memoization, on a game with $\nElm$ positions is
    $\AOmega{2^{\frac{n}{6}}}$ in the worst case.
  \end{theorem}
  \begin{proof}
    To prove the theorem, it suffices to consider a game $\GamName[\CSym][k]$
    belonging to the core family.
    Indeed, Lemma~\ref{lmm:recsub} states that the induced subgame tree of
    $\GamName[\CSym][k]$ is a subset of the recursion tree induced by the
    Recursive algorithm executed on that game.
    Therefore, according to Lemma~\ref{lmm:indsubtrecar}, the algorithm performs
    at least $3 (2^{k + 1} - 1)$ calls, each on a different subgame.
    By Definition~\ref{def:expcorfam}, game $\GamName[\CSym][k]$ has $n = 6k +
    3$ positions and, therefore, we have $k = \frac{n}{6} - \frac{1}{2}$.
    As a consequence, the number of recursive calls is bounded from below by $3
    (2^{\frac{n}{6} + \frac{1}{2}} - 1) = \AOmega{2^{\frac{n}{6}}}$.
  \end{proof}

\end{section}

% End of file SectionII.tex

%%% Local Variables:
%%% mode: latex
%%% TeX-master: "Article"
%%% End:

%% file: SectionIII.tex
%%****************************************************************************%%
%%                                                                            %%
%% Robust Worst Cases for Divide-et-Impera Algorithms for Parity Games        %%
%%                                                                            %%
%% SectionIII.tex                                                             %%
%%                                                                            %%
%% Revision 0                                                                 %%
%%                                                                            %%
%% Copyright (C) 2017, Massimo Benerecetti, Daniele Dell'Erba, and            %%
%%                     Fabio Mogavero.                                        %%
%% All rights reserved.                                                       %%
%%                                                                            %%
%%****************************************************************************%%

% Begin of file SectionIII.tex

\begin{section}{SCC-Decomposition Resilient Games}
  \label{sec:sccdecresgam}

  \figsccfam
  The previous section provides a class of parametric families of parity games
  over which a dynamic-programming approach cannot help improving the asymptotic
  exponential behavior of the classic Recursive algorithm.
  However, it is not hard to observe that an SCC-decomposition of the underlying
  game graph, if applied by each recursive call as described in~\cite{FL09},
  would disrupt the recursive structure of the core family $\{
  \GamName[\CSym][k] \}_{k = 1}^{\omega}$ and, consequently, break the
  exponential worst-case.
  This is due to the fact that the subgames $\GamName[\wElm][k]$ in the induced
  subgame tree get decomposed into distinct SCCs, which can then be solved as
  independent subgames and memoized.
  In other words, the Recursive algorithm extended with memoization and
  SCC-decomposition can easily solve the core family.
  A concrete instance of this behavior can be observed by looking at the two
  games $\der{\GamName}[\LSym\LSym][1]$ and $\der{\GamName}[\RSym\LSym][1]$ of
  Figure~\ref{fig:rectre}.
  The first one is formed by three distinct components, one of which exactly
  corresponds to $\der{\GamName}[\RSym\LSym][1]$.
  Therefore, a solution of these components immediately implies that the leaves
  in the induced subgame tree could not be considered distinct games \wrt to the
  behavior of the combined algorithm anymore.
  This behavior can, however, be prevented by introducing a suitable extension
  of the core family, which complies with the requirements of
  Definition~\ref{def:corext}.
  The basic idea is to connect all the pairs of positions $\gamma_{i}$ and
  $\gamma_{j}$ together in a clique-like fashion, by means of additional
  positions, denoted $\delta_{\{ i, j \}}^{\wp}$ with $\wp \in \{ 0, 1 \}$,
  whose owners $\wp$ are chosen so as to preserve the exponential behavior on
  the underlying core family.
  With more detail, if $i \equiv_{2} j$, there is a unique connecting position
  $\delta_{\{ i, j \}}^{\wp}$ of parity $\wp \equiv_{2} i$, the opposite of that
  of $\gamma_{i}$ and $\gamma_{j}$.
  If, on the other hand, $i \not\equiv_{2} j$, two mutually connected positions,
  $\{ \delta_{\{ i, j \}}^{0}, \delta_{\{ i, j \}}^{1} \}$, separate
  $\gamma_{i}$ and $\gamma_{j}$.
  Figure~\ref{fig:sccfam} depicts the extension of the core game
  $\GamName[\CSym][2]$, where, besides the positions $\gamma{i}$, only the
  additional positions and their moves are shown.
  The complete formalization of the new family follows.

  \begin{definition}[SCC Family]
    \label{def:sccfam}
    The \emph{SCC family} $\{ \GamName[\SSym][k] \}_{k = 1}^{\omega}$, where
    $\GamName[\SSym][k] \defeq \tuplec{\ArnName}{\PrtSet}{\prtFun}$, $\AName
    \defeq \tuplec{\PosSet[][0]}{\PosSet[][1]}{\MovRel}$, $\PrtSet \defeq
    \numcc{0}{4k}$, and $\PosSet \defeq \GamName[\CSym][k] \cup \PSet$, is
    defined, for any index $k \geq 1$, as follows:
    \begin{enumerate}
      \item\label{def:sccf(pos)}
        $\PSet \defeq \set{ \delta_{\{ i, j \}}^{\wp} }{ \wp \in \{ 0, 1 \}
        \land i, j \in \numcc{0}{2k} \land i \neq j \land (i \equiv_{2} j
        \rightarrow \wp \equiv_{2} i) }$;
      % \item
      %   $\PSet \defeq \set{ \delta_{\{ i, j \}}^{\wp} }{ i \equiv_{2} j
      %   \equiv_{2} \wp \land i \neq j \land i, j \in \numcc{0}{2k} } \cup
      %   \set{ \delta_{\{ i, j \}}^{\wp} }{ i \not\equiv_{2} j \land i, j \in
      %   \numcc{0}{2k} }$, and
      % \item
      %   $\PSet \defeq \set{ \delta_{\{ i, j \}}^{\wp} }{ \wp \in \{ 0, 1 \}
      %   \land i, j \in \numcc{0}{2k} \land i \neq j \land (\wp \equiv_{2} i
      %   \equiv_{2} j \vee i \not\equiv_{2} j) }$, and
      \item\label{def:sccf(mov1)}
        $(\gamma_{i}, \delta_{\ISet}^{\wp}), (\delta_{\ISet}^{\wp}, \gamma_{i})
        \in \MovRel$ \iff $i \in \ISet$ and $i \equiv_{2} \wp$, for $i \in
        \numcc{0}{2k}$ and $\delta_{\ISet}^{\wp} \in \PSet$;
      \item\label{def:sccf(mov2)}
        $(\delta_{\ISet}^{\wp}, \delta_{\ISet}^{\dual{\wp}}) \in \MovRel$ \iff
        $i \not\equiv_{2} j$, for $\delta_{\ISet}^{\wp} \in \PSet$ and $\ISet =
        \{ i, j \}$;
      \item\label{def:sccf(prt)}
        $\delta_{\ISet}^{\wp} \in \PosSet[][\wp]$ and
        $\prtFun(\delta_{\ISet}^{\wp}) \defeq 0$, for $\delta_{\ISet}^{\wp} \in
        \PSet$;
      \item\label{def:sccf(cor)}
        $\GamName[\SSym][k] \setminus \PSet = \GamName[\CSym][k]$.
    \end{enumerate}
  \end{definition}

  Intuitively, in Item~\ref{def:sccf(pos)}, $\PSet$ denotes the set of
  additional positions of $\GamName[\SSym][k]$ \wrt to the core family game
  $\GamName[\CSym][k]$, which is, indeed, a proper subgame, as stated in
  Item~\ref{def:sccf(cor)}.
  Item~\ref{def:sccf(mov1)}, instead, formalizes the moves connecting the
  additional $\delta_{\ISet}^{\wp}$ positions with the $\gamma_{i}$ of the core,
  while Item~\ref{def:sccf(mov2)} describes the mutual connection between the
  $\delta$ positions that share the same doubleton of indexes $\ISet$.
  Finally, Item~\ref{def:sccf(prt)} associates each $\delta_{\ISet}^{\wp}$ with
  its corresponding owner $\wp$ and priority $0$.
  The following lemma proves that such a parity-game family is indeed a
  worst-case family.

  \begin{lemma}
    \label{lmm:sccfam}
    The SCC family $\{ \GamName[\SSym][k] \}_{k = 1}^{\omega}$ is a worst-case
    family.
  \end{lemma}
  \begin{proof}
    To prove that the \emph{SCC family} of Definition~\ref{def:sccfam} is a
    worst-case family, we need to show that each game $\GamName[\SSym][k]$ is a
    core extension of $\GamName[\CSym][k]$, \ie, that it complies with the
    Definition~\ref{def:corext}.
    Items~\ref{def:sccf(pos)} and~\ref{def:sccf(cor)} of
    Definition~\ref{def:sccfam} imply Item~\ref{def:corext(cor)} of
    Definition~\ref{def:corext}, since the set $\PSet$ does not contain any
    position of the core and, in addition, $\GamName[\CSym][k]$ is a subgame of
    $\GamName[\SSym][k]$, as all the positions and moves of the core are
    contained in $\GamName[\SSym][k]$.
    Item~\ref{def:corext(prt)} of Definition~\ref{def:corext} follows from
    Item~\ref{def:sccf(prt)} of Definition~\ref{def:sccfam}.
    Indeed, by Definition~\ref{def:expcorfam}, $\prtFun(\alpha_{i}) \defeq 2k +
    i + 1$, for $i \in [0,2k]$ and $k\geq 1$, while all the additional positions
    in $\PSet$ have priority $0$.
    By Items~\ref{def:sccf(mov1)} and~\ref{def:sccf(mov2)} of
    Definition~\ref{def:sccfam}, there are no moves connecting positions
    $\delta_{\ISet}$ with positions $\alpha_i$ or $\beta_i$, for any $i \in
    \numcc{0}{2k}$, hence Item~\ref{def:corext(mov)} of
    Definition~\ref{def:corext} is satisfied.
    Finally, we need to show that whenever a $\gamma_{i}$ has a move to a
    position $\posElm$ in $\PSet$, then $\posElm$ does not have higher priority,
    belongs to the opponent of $\gamma_{i}$, and has a move back to $\gamma_{i}$
    (Item~\ref{def:corext(sep)} of Definition~\ref{def:corext}).
    This property is enforced by Items~\ref{def:sccf(mov1)}
    and~\ref{def:sccf(prt)} of Definition~\ref{def:sccfam}.
    Indeed, by the latter, position $\delta_{\ISet}^{\wp}$ is owned by player
    $\wp$.
    Moreover, by the former, $\gamma_{i}$, whose owner is $(i+1)\bmod 2$, can
    only have a move to $\delta_{\{i,j\}}^{\wp}$ if $\delta_{\{i,j\}}^{\wp}$ has
    a move back to $\gamma_{i}$ and $\wp = i\bmod 2$.
    Hence, the two positions belong to opposite players.
    Since, in addition, all positions $\delta$ have priority $0$, the
    requirement is satisfied.
  \end{proof}

  Due to the clique-like structure of the new family, it is not hard to see that
  every game in the induced subgame tree forms a single SCC.
  This guarantees that the intertwining of SCC-decomposition and memoization
  cannot prevent an exponential worst-case behavior of the Recursive algorithm
  on this family.

  \begin{lemma}
    \label{lmm:sinsccsccfam}
    Each game in the induced subgame tree $\IndSubTre[][k]$ of the SCC family
    $\{ \GamName[\SSym][k] \}_{k = 1}^{\omega}$, for an arbitrary index $k \in
    \mathbb{N}_+$, forms a single SCC.
  \end{lemma}
  \begin{proof}
    Let $\GamName[\wElm][k] \in \IndSubTre[][k]$ (\resp,
    $\der{\GamName}[\wElm][k] \in \IndSubTre[][k]$) be a game in the induced
    subgame tree.
    By induction on the structure of the string $\wElm$, it is not hard to see
    that, for all indexes $i, j \in \numcc{0}{2k}$ with $i \neq j$, it holds
    that $\gamma_{i}, \gamma_{j} \in \GamName[\wElm][k]$ (\resp, $\gamma_{i},
    \gamma_{j} \in \der{\GamName}[\wElm][k]$) iff the positions $\delta_{\{ i, j
    \}}^{\wp} \in \PSet$, with $\wp \in \{ 0, 1 \}$, belong to
    $\GamName[\wElm][k]$ (\resp, $\der{\GamName}[\wElm][k]$), as well.
    For the base case $\GamName[\varepsilon][k] = \GamName[][k]$, the thesis
    trivially follows from Definition~\ref{def:sccfam}.
    For the inductive case $\der{\GamName}[\wElm \xElm][k] = \GamName[\wElm][k]
    \setminus \ASet$ (\resp, $\GamName[\wElm \xElm][k] =
    \der{\GamName}[\wElm][k] \setminus \ASet$), let us assume, as inductive
    hypothesis, that the statement holds for $\GamName[\wElm][k]$ (\resp,
    $\der{\GamName}[\wElm][k]$).
    By Definition~\ref{def:indsubtre}, the set $\ASet$ is computed as the
    attractor to some set of positions $\BSet$ such that either
    \emph{(i)}~$\gamma_{i}, \gamma_{j}, \delta_{\{ i, j \}}^{\wp} \not\in
    \ASet$ or \emph{(ii)}~$\delta_{\{ i, j \}}^{\wp} \in \ASet$ and at least one
    between $\gamma_{i}$ and $\gamma_{j}$ belongs to $\ASet$, for all $i, j \in
    \numcc{0}{2k}$.
    Case~\emph{(i)} arises when $\BSet = \{ \alpha_{2(k - \card{\wElm})} \}$
    (\resp, $\BSet = \{ \alpha_{2(k - \card{\wElm}) + 1} \}$), while
    Case~\emph{(ii)} when $\BSet = \WinSet[0](\der{\GamName}[\wElm \LSym][k])$.
    Consequently, the required property on $\wElm \xElm$ immediately follows
    from the inductive hypothesis on $\wElm$, since, if a position $\delta_{\{ i,
    j \}}^{\wp}$ is removed from the game, also one between $\gamma_{i}$ and
    $\gamma_{j}$ is removed as well and \viceversa.
    Now, let $\GamName$ be an arbitrary game in $\IndSubTre[][k]$.
    Thanks to the topology of the games in the SCC family and to the property
    proved above, it easy to see that all positions in $\XSet = \{ \beta_{i},
    \gamma_{i}, \delta_{\ISet}^{\wp} \in \GamName \}$ form a strongly connected
    subgame.
    Indeed, two positions $\beta_{i}$ and $\gamma_{i}$ are mutually reachable
    due to the two moves $(\beta_{i}, \gamma_{i}), (\gamma_{i}, \beta_{i}) \in
    \MovRel$.
    Moreover, two arbitrary positions $\gamma_{i}$ and $\gamma_{j}$, with $i \neq
    j$, are mutually reachable via the positions $\delta_{\{ i, j \}}^{\wp}$.
    Also, there are no isolated positions $\delta_{\{ i, j \}}^{\wp}$.
    Finally, to prove that $\GamName$ is indeed a single SCC, it remains just to
    show that the positions in $\YSet = \{ \alpha_{i} \in \GamName \}$ can reach
    and can be reached by those in $\XSet$.
    The first part is implied by Item~\ref{lmm:recinv(sub)} of
    Lemma~\ref{lmm:recinv}, since every $\alpha_{i}$ has only a move to
    $\beta_{i}$, which needs to belong to $\GamName$ in order for this to be a
    game.
    Now, due to the same observation, all positions $\alpha_{i}$, but possibly
    the last one $\alpha_{m}$ with maximal index $m$ in $\GamName$, are
    reachable by $\beta_{i + 1}$.
    Finally, $\alpha_{m}$ can be reached by $\gamma_{m - 1}$, which necessarily
    belongs to $\GamName$ due to Item~\ref{lmm:recinv(gam)} of the same lemma.
  \end{proof}

  Putting everything together, we obtain the following structural, although non
  asymptotic, strengthening of Theorem~\ref{thm:expworcas}.

  \begin{theorem}[SCC-Decomposition Worst Case]
    \label{thm:expsccdecworcas}
    The number of distinct recursive calls executed by the Recursive algorithm
    with SCC decomposition on a game with $\nElm$ positions is
    $\AOmega{2^{\sqrt{n / 3}}}$ in the worst case.
  \end{theorem}
  \begin{proof}
    Due to Lemma~\ref{lmm:sccfam}, the \emph{SCC family} is an exponential
    worst-case family.
    As shown in the proof of Theorem~\ref{thm:expworcas}, the Recursive
    algorithm performs at least $3(2^{k + 1} - 1)$ calls to solve a game of $\{
    \GamName[][k] \}_{k = 1}^{\omega}$, and therefore, of $\{ \GamName[\SSym][k]
    \}_{k = 1}^{\omega}$.
    As consequence of Lemma~\ref{lmm:sinsccsccfam}, the number of calls cannot
    be affected by an SCC-decomposition technique, since there is an exponential
    number of subgames, the ones in the induced subgame tree of
    $\GamName[\SSym][k]$, each of which forms a single SCC.
    Moreover, a core game $\GamName[\CSym][k]$ has $6k + 3$ positions, while the
    number of additional positions $\PSet$ in the extension $\GamName[\CSym][k]$
    is given by $3k^{2} + 2k$, which follows from Definition~\ref{def:sccfam}.
    Indeed, for every pair of positions $\gamma_{i}, \gamma_{j}$, with $i, j \in
    \numcc{0}{2k}$ and $i \neq j$, there is a single position $\delta_{\{ i, j
    \}}^{\wp}$, if $i \equiv_{2} j$, and two such positions, otherwise.
    Hence, $\GamName[\SSym][k]$ has $n \defeq 3k^{2} + 8k +3 $ positions, from
    which we obtain that $k = \frac{\sqrt{3n + 7} - 4}{3}$.
    As a consequence, the number of recursive calls is bounded from below by
    $3(2^{\frac{\sqrt{3n + 7} - 1}{3}} - 1) = \AOmega{2^{\sqrt{n / 3}}}$.
    %FORMULE ESATTE
    %  \begin{equation}
    %    \sum\limits_{i=0}^{2k-1} \sum\limits_{j=i+1}^{2k}=
    %    \begin{cases}
    %      1, & \text{if}\ i\equiv_{2}j \\
    %      2, & \text{otherwise}
    %    \end{cases}
    %  \end{equation}
    %  \begin{equation}
    %    \sum\limits_{i=0}^{2k-1}=
    %    \begin{cases}
    %      3k-\frac{3}{2}i, & \text{if}\ i\equiv_{2}j \\
    %      3k-\frac{3}{2}i+\frac{1}{2}, & \text{otherwise}
    %    \end{cases}
    %  \end{equation}
    %  \begin{equation}
    %    6k^2-3k^2+\frac{3}{2}k+\sum\limits_{i=0}^{2k-1}=
    %    \begin{cases}
    %      0, & \text{if}\ i\equiv_{2}j \\
    %      \frac{1}{2}, & \text{otherwise}
    %    \end{cases}
    %  \end{equation}
  \end{proof}

\end{section}

% End of file SectionIII.tex

%%% Local Variables:
%%% mode: latex
%%% TeX-master: "Article"
%%% End:

%% file: SectionIV.tex
%%****************************************************************************%%
%%                                                                            %%
%% Robust Worst Cases for Divide-et-Impera Algorithms for Parity Games        %%
%%                                                                            %%
%% SectionIV.tex                                                              %%
%%                                                                            %%
%% Revision 0                                                                 %%
%%                                                                            %%
%% Copyright (C) 2017, Massimo Benerecetti, Daniele Dell'Erba, and            %%
%%                     Fabio Mogavero.                                        %%
%% All rights reserved.                                                       %%
%%                                                                            %%
%%****************************************************************************%%

% Begin of file SectionIV.tex

\begin{section}{Dominion-Decomposition Resilient Games}
  \label{sec:domdecresgam}

  A deeper analysis of the SCC family reveals that the size of the smallest
  dominion for player $0$ in Game $\GamName[\SSym][k]$ (there are no dominions
  for player $1$, being the game completely won by its opponent) is of size $2(k
  + 1)$.
  This observation, together with the fact that the game has $3k^{2} + 8k + 3$
  positions, immediately implies that the proposal of~\cite{JPZ06,JPZ08} of a
  brute force search for dominions of size at most $\ceil{\!\sqrt{3k^{2} + 8k +
  3}\,} < 2(k + 1)$ cannot help improving the solution process on these games.
  We can prove an even stronger result, since this kind of search cannot reduce
  the running time in any of the subgames of the induced tree.
  The reason is that the smallest dominion in each such subgame that contains at
  least a position in the core has size linear \wrt $k$, as reported by the
  following lemma.

  \begin{lemma}
    \label{lmm:sizdomsccfam}
    For all $k \in \mathbb{N}_+$, let $\GamName$ be a subgame in the
    induced subgame tree $\IndSubTre[][k]$ of $\GamName[\SSym][k]$, and $\DSet
    \subseteq \GamName$ the smallest dominion such that $\DSet \cap
    \GamName[\CSym][k] \neq \emptyset$.
    Then, $\card{\DSet} \geq k + 1$.
  \end{lemma}
  \begin{proof}
    We start by proving that any such dominion $\DSet$ must necessarily contain
    at least a position $\gamma_{i}$, from some $i \in \mathbb{N}_+$.
    Assume, by contradiction, that it does not.
    Then, $\DSet \subseteq \set{ \alpha_{i}, \beta_{i} \in \GamName }{ i \in
    \numcc{0}{2k} } \cup \PSet$.
    % We need to prove that every dominion containing a position
    % $\alpha,\beta$ or $\gamma$, has size no lower than $k+1$.  Take an
    % arbitrary set $\XSet$ of positions $\alpha$ and $\beta$, this set does not
    % constitute a subgame.
    We can prove that $\DSet$ is not a game.
    By assumption, $\DSet$ must contain at least one position $\alpha_{i}$ or
    $\beta_{i}$, otherwise $\DSet \cap \GamName[\CSym][k] = \emptyset$.
    Let $j$ be the smallest index such that $\alpha_{j} \in \DSet$ or $\beta_{j}
    \in \DSet$.
    Then, at least one of those two positions has only moves leading outside
    $\DSet$.
    Hence $\DSet$ is not a game and, \afortiori, cannot be a dominion.

    Therefore, $\DSet$ must contain at least a position $\gamma_{i}$.
    By Definition~\ref{def:sccfam}, $\gamma_{i}$ is connected in
    $\GamName[\SSym][k]$ to precisely $2k$ positions $\delta_{\{ i, j
    \}}^{\wp}$, where $j \in \numcc{0}{2k}$, $j \neq i$, and whose owner $\wp
    \equiv_{2} i$ is the adversary of the owner of $\gamma_{i}$.
    Let us consider the $k$ indexes $j \in \numcc{0}{2k}$ such that $i
    \not\equiv_{2} j$.
    For each such $j$, we have two cases, depending on whether $\gamma_{j}$
    belongs to $\GamName$ or not.
    If it does, then so does $\delta_{\{ i, j \}}^{\wp}$.
    If it does not, then it must have been removed by some application of
    Item~\ref{def:indsubtre(rig)} of Definition~\ref{def:indsubtre}.
    If the involved attractor \wrt player $0$ does not attract $\delta_{\{ i, j
    \}}^{\dual{\wp}}$, then it cannot attract $\delta_{\{ i, j \}}^{\wp}$
    either, as it has no moves to $\gamma_{j}$.
    If, on the other hand, $\delta_{\{ i, j \}}^{\dual{\wp}}$ gets attracted, it
    must belong to player $0$.
    As a consequence, $\delta_{\{ i, j \}}^{\wp}$ belongs to player $1$ and is
    not attracted.
    In either case, we conclude hat $\delta_{\{ i, j \}}^{\wp}$ cannot be
    removed and, therefore, is still contained in $\GamName$.
    Since, in addition, all the $k$ positions $\delta^{\wp}_{\{ i, j \}}$, with
    $j\in[0,2k]$ and $j\not\equiv_2i$, are mutually connected to $\gamma_{i}$
    and their owner is the opponent of the one of $\gamma_{i}$, they must be
    contained in $\DSet$ as well.  Hence, $\DSet$ contains at least $k+1$
    positions.
    %
    % In addition, $\delta_{\{ i, j \}}^{\wp}$ is connected to $\gamma_{j}$ if
    % $i \equiv_{2} j$, to $\delta_{\{ i, j \}}^{\dual{\wp}}$ which is in turn
    % connected to $\gamma_{j}$, otherwise.
    %
    % Clearly for all $j$ such that $\gamma_{j}\in\GamName[w][k]$,
    % $\delta_{\{ i, j \}}^{\wp}\in\GamName[w][k]$. Suppose that for all $j$,
    % $\gamma_{j}\not\in\GamName[w][k]$.  For indexes $j$ such that
    % $i \equiv_{2} j$, $\gamma_{j}$ can attract $\delta_{\{ i, j \}}^{\wp}$ and
    % then remove it from the game, on the other hand, when $i \not\equiv_{2} j$,
    % $\gamma_{j}$ can attract only one of positions $\delta_{\{ i, j \}}^{\wp}$
    % and $\delta_{\{ i, j \}}^{\dual{\wp}}$. Therefore in this case at least $k$
    % position $\delta_{\{ i, j \}}\in\GamName[w][k]$.
    %
    % Hence, since all $\delta_{\{ i, j \}}$ are connected to $\gamma_{i}$,
    % every $\gamma_{i}$ belongs to a dominion with at least $k+1$ positions.
  \end{proof}

  The above observation allows us to obtain an exponential lower bound for the
  Recursive algorithm combined with memoization, SCC decomposition, and dominion
  decomposition techniques.
  Indeed, the brute-force procedure employed by the Dominion Decomposition
  algorithm of~\cite{JPZ06} needs at least time $\AOmega{2^{k + 1}}$ to find a
  dominion of size $k + 1$.
  For the sake of space and clarity of exposition, we postpone the formal
  treatment of the Big Step procedure to the extended version of this work,
  reporting here an informal description only.
  Intuitively, its exponential behavior on the SCC family follows from the
  following observations: \emph{(i)}~the algorithm only looks for dominions of
  size $d = \sqrt[3]{pn^{2}}$, where $p$ and $n$ are the numbers of priorities
  and positions, and \emph{(ii)} the update of the measure functions used by the
  procedure requires an exponential number of steps in $d$ before reaching a
  fixpoint.
  As a consequence, none of the dominion decomposition approaches, combined with
  memoization and SCC decomposition, can efficiently solve the SCC family.

  \begin{corollary}[Exponential Dominion-Decomposition Worst Case]
    \label{cor:expnavdomdecworcas}
    The solution time of the Recursive algorithm with memoization, SCC
    decomposition, and dominion decomposition on a game with $\nElm$ positions
    is $\AOmega{2^{\Theta(\sqrt{n})}}$ in the worst case.
  \end{corollary}

\end{section}

% End of file SectionIV.tex

%%% Local Variables:
%%% mode: latex
%%% TeX-master: "Article"
%%% End: